\documentclass{aa}
\pdfoutput=1

\usepackage{graphicx,amsmath}
\usepackage{txfonts}
\usepackage{pictex}
\usepackage{latexsym}
\usepackage{graphics}
\usepackage{afterpage}
\usepackage[colorlinks=true,
            urlcolor=blue,
            linkcolor=blue,
            citecolor=blue]
            {hyperref}
\usepackage{dcolumn}
\usepackage{booktabs}
\usepackage{subcaption}%
\usepackage{rotating}
\usepackage{blindtext}
\usepackage{tabu}
\usepackage{ulem}

\usepackage{units}
\usepackage{gensymb}

\newcommand{\arc}{\arcsec\hspace*{-1ex}.\hspace*{0.3ex}}

\makeatletter
 \def\hlinewd#1{%
   \noalign{\ifnum0=`}\fi\hrule \@height #1 \futurelet
    \reserved@a\@xhline}
\makeatother

\newcommand{\mcc}[1]{\multicolumn{1}{c}{#1}}

\begin{document} 

   \title{Emission-Line and Continuum Reverberation Mapping of the\\ NLS1 Galaxy WPVS\,48
   }

      \author{M.A.~Probst \inst{1},
           W.~Kollatschny \inst{1},
           M.W. Ochmann \inst{1},
           C.~Sobrino Figaredo \inst{2,3},
           D.~Chelouche \inst{2,3},
           M.~Haas \inst{4},
           S.~Kaspi \inst{5},
           L. Meerwart \inst{1},
           T.-O. Husser \inst{1} 
           }

   \institute{
          Institut f\"ur Astrophysik und Geophysik, Universit\"at G\"ottingen,
          Friedrich-Hund Platz 1, 37077 G\"ottingen, Germany \\
          \email{malte.probst@uni-goettingen.de}
          \and
          Department of Physics, Faculty of Natural Sciences, University of Haifa, Haifa 3498838, Israel
          \and
          Haifa Research Center for Theoretical Physics and Astrophysics, University of Haifa, Haifa 3498838, Israel
          \and
          Ruhr University Bochum, Faculty of Physics and Astronomy, Astronomical Institute (AIRUB), 44780 Bochum, Germany
          \and
          School of Physics \& Astronomy and the Wise Observatory, Tel-Aviv University, Tel-Aviv 6997801, Israel
          }

  \date{Received 01 April 2025 / Accepted 08 November 2025}

\abstract
 {
 WPVS\,48 is a nearby narrow-line Seyfert 1 galaxy without previous analysis of the broad-line region (BLR) by means of optical spectroscopic reverberation mapping.
 }
{
By studying the continuum and emission line variability of WPVS\,48, we aim to infer the BLR size as well as the mass of the central supermassive black hole (SMBH).
}
{
We analyse data from a dedicated optical spectroscopic reverberation mapping campaign of WPVS\,48 taken with the 10\,m Southern African Large Telescope (SALT) at 24 epochs over a period of 7 months between December 2013 and June 2014.
}
{ 
WPVS\,48 shows variability throughout the campaign.
We find a stratified BLR, where the variability amplitude of the integrated emission lines decreases with distance to the ionizing continuum source. Specifically, the variable emission of H$\alpha$, H$\beta$, H$\gamma$, \ion{He}{i}\,$\lambda5876$ originates at distances of $16.0^{+4.0}_{-2.0}$, $15.0^{+4.5}_{-1.9}$, $12.5^{+3.5}_{-2.5}$ and $14.0^{+2.5}_{-2.1}$ light-days, respectively, to the optical continuum at 5100\,\AA{}. The \ion{He}{ii}\,$\lambda 4686$ lag is $\lesssim 5$\,days. Based on the high S/N spectra, we identify variable emission of \ion{N}{iii}\,$\lambda4640$ and \ion{C}{iv}\,$\lambda4658$ in the line complex with \ion{He}{ii}\,$\lambda 4686$. We derive interband continuum delays increasing with wavelength up to $\sim 8$\,days. These delays are consistent with an additional diffuse continuum originating at the same distance as the variable Balmer emission. We derive a central black hole mass of $(1.3_{-0.6}^{+1.1})\times10^7M_{\odot}$ based on the integrated line-widths and distances of the BLR and discuss corrections for the inclination angle. This gives an Eddington ratio $L/L_{\text{Edd}}\approx 0.39$ without correction for inclination.
}
{}


  \titlerunning{Emission-Line and Continuum Variability of WPVS\,48}
  \authorrunning{M.A. Probst et al.}

  \maketitle

%
%**********************************************************************************
%
\section{Introduction} \label{sec:introduction}
%
%**********************************************************************************
%

Variability is widespread in active galactic nuclei (AGN) and was leveraged in the past $\sim$ 30 years to identify and map the innermost AGN structures -- namely the accretion disk (AD), the broad-line region (BLR), and the dusty torus (TOR) -- using methods such as reverberation mapping \citep[RM;][]{blandford82}. RM traces the lagging emissive response of the individual AGN components to the time-varying ionizing continuum radiation from the central source close to the supermassive black hole (SMBH). The AGN components are each associated with different wavelength bands, as emission from the AD and the BLR dominate in the UV to NIR, whereas emission from the dusty torus dominates in mid-IR \citep[for a review of different RM types, see][]{2021iSci...24j2557C}. 

In case of the BLR, the emissive response to the ionizing continuum is prominently observed in the variable emission lines. Regular and densely sampled observations of the ionizing X-ray/UV continuum, however, are difficult to acquire. The light curves of optical continua are often used instead and are highly correlated with the ionising UV continuum in RM studies that monitored continua from X-ray to optical bands \citep[e.g.][]{2015ApJ...806..129E, 2019ApJ...870..123E, 2024ApJ...973..152E, 2016ApJ...821...56F, 2020MNRAS.498.5399H}{}{}. While being a good proxy for the ionizing radiation, the variable continuum emission in the UV/optical itself may be complex, as a superposition of the illuminated AD and variable diffuse emission from the BLR is assumed \citep[e.g.][and references therein]{2016ApJ...821...56F, 2018ApJ...857...53C, 2019NatAs...3..251C, 2022MNRAS.509.2637N, 2024ApJ...973..152E}{}{}.
 
Here, we present the results of continuum and emission-line RM from a spectroscopic variability campaign on WPVS\,48, ($\alpha_{2000}=09^\text{h}59^\text{m}42.65^\text{s}, \delta_{2000} = -31^{\circ}12^{\prime}58.4^{\prime \prime}$) a local ($z=0.0372$, \citet{2022ApJS..261....4O}) Seyfert 1 galaxy.\footnote{\url{https://ned.ipac.caltech.edu/}} The host is a spiral galaxy and the nucleus luminosity was measured by \citet{1997MNRAS.292..273W} with $V = 14.78$\,mag. A photometric RM campaign in the optical and near-infrared was conducted in parallel to the spectroscopic campaign described in this work and has been presented by \citet{2018rnls.confE..57S, 2025ApJS..276...48S}. 

WPVS\,48 is more specifically classified as a narrow-line Seyfert 1 (NLS1), a sub-type of Seyfert 1 galaxies with narrow line-emission components originating from the BLR. Collective spectral properties of NLS1 galaxies are discussed in \citet{1985ApJ...297..166O} and \citet{1989ApJ...342..224G}, namely: First, the FWHM of the broad H$\beta$ component does not exceed $2000$\,km\,s$^{-1}$ to differentiate NLS1s from other type-I Seyfert galaxies with typical line widths on the order of several $1000$\,km\,s$^{-1}$ \citep[e.g.][and references therein]{2013MNRAS.433..622M, 2022ApJS..261...24L}{}{}, and, second, the flux ratio of [\ion{O}{iii}]\,$\lambda$5007/H$\beta \leq 3$. Further authors noted remarkably strong \ion{Fe}{ii} emission in spectra of NLS1s and included this as a third defining criterion \citep[e.g.][and references therein]{2001A&A...372..730V}{}{}. Since the first NLS1s were discovered, two scenarios are primarily discussed to explain the narrow emission lines in NLS1s: The first argues that the inferred low dispersion velocities in NLS1s are real, which would imply that NLS1s host comparatively low-mass SMBHs. The second scenario attributes the low velocities to the line-of-sight effect, that occurs when the BLR is primarily located on a plane, which is oriented at a low inclination angle $i$ to the line of sight.

While optical RM studies on timescales between several months and years exist for many AGN, e.g. NGC\,5548 \citep[][and references therein]{2002ApJ...581..197P, 2017ApJ...837..131P}, 3C\,120 \citep[][]{1998ApJ...501...82P, 2013ApJ...764...47G, 2014A&A...566A.106K}, NGC\,7603 \citep[][]{2000A&A...361..901K}, 3C\,390.3 \citep[][]{2010A&A...517A..42S} and HE\,1136-2304 \citep[][]{2018A&A...619A.168K}, such studies are scarce for the subtype of NLS1s. With this study on WPVS\,48, we expand on this relatively underexplored sample and compile key figures from existing RM campaigns on NLS1.

We extracted line profiles of multiple emission lines -- including low-intensity emission lines -- as well as light curves from several line-free continua and from line emission. We furthermore cross-correlated the light curves and inferred lags between the optical continuum and line emission. In Sect.\,\ref{sec:observations}, we describe the observations and the data reduction. In Sect.\,\ref{sec:results}, we present the analysis of the spectroscopic observations. We discuss the results in Sect.\,\ref{sec:discussion} and summarize the results in Sect.\,\ref{sec:conclusions}. This work covers the 1D analysis of this RM campaign, while the velocity-resolved analysis as well as the low-luminosity lines will be covered in a forthcoming paper (Paper II). Throughout this paper, we assume $\Lambda$CDM universe with a Hubble constant of $H_0 = 67.4$\,km\,s$^{-1}$\,Mpc$^{-1}$, $\Omega_{\Lambda} = 0.68$ and $\Omega_{\text{M}} = 0.32$ \citep[][]{2020A&A...641A...6P}, which results in a luminosity distance of $D_{\text{L}} =  170.2\,\text{Mpc} = 5.25 \times 10^{20}\,\text{cm}$ using the cosmology calculator from \citet{2006PASP..118.1711W}.

%**********************************************************************************
%
\section{Observations and data reduction}\label{sec:observations}
%
%**********************************************************************************
During the campaign at SALT, we acquired $24$ optical long-slit spectra of WPVS\,48 over a time span of $209$ days from December 2nd, 2013 to June 29th, 2014. % under the proposal code 2013-2-GU-001 (PI: Kollatschny).
This results in a mean and median sampling rate of $8.7$\,d and $7.2$\,d, respectively. The minimum and maximum separation between two epochs are $5$\,d and $16$\,d, respectively. A log of the observations is given in Table\,\ref{tab:spectroscopy_log}. We obtained galaxy spectra and necessary calibration images (flat-fields, Xe and Ar arc frames). All observations were taken under identical instrumental setup employing the Robert Stobie Spectrograph \citep[RSS;][]{kobulnicky03} and using the PG0900 grating as well as a 2x2 binning. To minimize differential refraction, the slit width was fixed to 2\arc0 projected onto the sky at an optimized projection angle. The elevation angle of the SALT telescope is fixed at $53^{\circ}$; all observations, thus, were taken with the same airmass. Finally, we used a aperture of 1\arc5 $\times$ 2\arc0 to extract the spectra. With this setup, we covered the wavelength range from 4350 to 7375\,\AA{} with a spectral resolution of $\sim 6.7$\,\AA{}. This corresponds to object rest-frame wavelengths of 4194 to 7110\,\AA{}. The two regions without spectral data are due to gaps between the three CCDs of the spectrograph. These regions range from 5356 to 5428 and 6398 to 6462\,\AA{} (5164 to 5234\,\AA{} and 6169 to 6231\,\AA{} in the rest frame), respectively. We performed the same standard reduction procedures for all observations using \textsc{iraf} packages including flat field correction, illumination correction, cosmics correction and background subtraction. The flux calibration was performed with standard star LTT\,4364.\\
%
%------------------------------------------------------------------------------
%
\begin{figure*}[!t]
\centering
\includegraphics[width=18.5cm,angle=0]{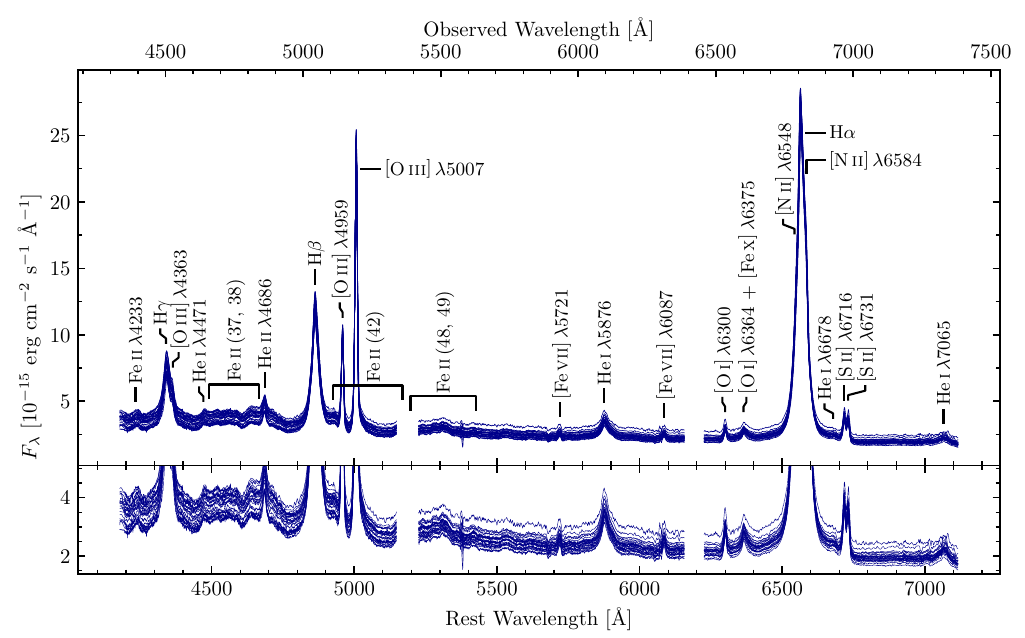}
\vspace*{-3mm} 
\caption{\textit{Top panel:} Reduced optical spectra of WPVS\,48 during the variability campaign from December 2013 to June 2014. The spectra were calibrated to the same absolute [\ion{O}{iii}]\,$\lambda5007$ flux of $244 \times 10^{-15} \rm\,erg\,cm^{-2}\,s^{-1}$. The identified emission features are labelled. \textit{Bottom panel:} Zoom on the continuum and the low-intensity emission lines. The conspicuous feature at $\sim 5490$\,{\AA} is a residual from night-sky emission.}
\label{fig:WPVS48_all_spectra.pdf}
\end{figure*}
%
%---------------------------------------------------------------------------
%                   
%
%------------------------------------------------------------------------------
%
 \begin{table}[t!]
 \caption{Log of spectroscopic observations of WPVS\,48 throughout the reverberation mapping campaign from late 2013 to mid 2014.}
 \centering
 \resizebox{0.48\textwidth}{!}{
     \begin{tabular}{rlcccl}
         \hline \hline
         \noalign{\smallskip}
         ID & Mod. JD & UT Date & Exp. Time    & Seeing             & Weather   \\
            &         &         &  \mcc{[s]}   & {\small FWHM}      &           \\
         \hline 
         \noalign{\smallskip}
          1 & 56628.53 & 2013-12-02 & $900$ & 3.5 - 5\,'' & thin clouds \\ 
          2 & 56640.51 & 2013-12-14 & $900$ & 1\,'' & thin clouds \\ 
          3 & 56647.50 & 2013-12-20 & $900$ & 1.6\,'' & clear \\ 
          4 & 56653.47 & 2013-12-26 & $900$ & 2.9\,'' & thin clouds \\ 
          5 & 56659.45 & 2014-01-01 & $900$ & 2.1\,'' & clear \\ 
          6 & 56670.42 & 2014-01-12 & $900$ & 2 - 2.6\,'' & clear \\ 
          7 & 56676.40 & 2014-01-18 & $900$ & 2.5\,'' & thin clouds \\ 
          8 & 56689.37 & 2014-01-31 & $900$ & 1.6\,'' & clouds \\ 
          9 & 56696.58 & 2014-02-08 & $900$ & 2.5\,'' & clear \\ 
         10 & 56704.33 & 2014-02-15 & $900$ & 1.6 - 1.8\,'' & clear \\ 
         11 & 56717.28 & 2014-02-28 & $900$ & 1.7\,'' & clear \\ 
         12 & 56722.52 & 2014-03-06 & $900$ & 1.5\,'' & clear \\ 
         13 & 56728.27 & 2014-03-11 & $900$ & 1.3\,'' & clouds \\ 
         14 & 56734.50 & 2014-03-17 & $900$ & 2.6\,'' & clear \\ 
         15 & 56745.47 & 2014-03-28 & $900$ & 1.4 - 1,7\,'' & clear \\ 
         16 & 56757.42 & 2014-04-09 & $900$ & 0.8\,'' & clear \\ 
         17 & 56773.39 & 2014-04-25 & $900$ & 1.2\,'' & clear \\ 
         18 & 56789.34 & 2014-05-11 & $874$ & 1.6\,'' & clouds \\ 
         19 & 56795.31 & 2014-05-17 & $874$ & 2.5\,'' & clear \\ 
         20 & 56801.30 & 2014-05-23 & $874$ & 2\,'' & thin clouds \\ 
         21 & 56809.30 & 2014-05-31 & $874$ & 2.5\,'' & clear \\ 
         22 & 56828.22 & 2014-06-19 & $874$ & 1.8\,'' & clear \\ 
         23 & 56833.23 & 2014-06-24 & $874$ & 1.6\,'' & clouds \\ 
         24 & 56838.21 & 2014-06-29 & $874$ & 1.5\,'' & thin clouds \\ 
         \hline 
     \end{tabular}}
 \label{tab:spectroscopy_log}
 \end{table}
%
%------------------------------------------------------------------------------
%
All flux-calibrated spectra are corrected for the telluric \ion{O$_2$}{B} absorption band and the \ion{H$_2$O}{} absorption band between $7080$\,{\AA} and $7450$\,{\AA} in the observer's frame using templates of the the two absorption bands and, further, for galactic reddening applying the extinction curve of \citet{cardelli89} and using a ratio $R$ of absolute extinction $A(v)$ to $E_{\rm B-V}=0.067$ \citep{schlafly11} of $3.1$. Subsequently, the spectra were intercalibrated to constant narrow-line fluxes to ensure high relative flux accuracy ($\lesssim 2$\,\%). The intercalibration was performed separately for the spectral regions on either side of $5855$\,{\AA} ($5645$\,{\AA} in the rest frame). The spectral region bluewards of $5855$\,{\AA} was intercalibrated with respect to the line fluxes of the narrow emission lines [\ion{O}{iii}]\,$\lambda\lambda4959,5007$ with respective fluxes of $(79.8 \pm 0.9)\, \times$\,10$^{-15}$\,erg\,cm$^{-2}$\,s$^{-1}$ and $(244.2 \pm 0.7)\,\times$\,10$^{-15}$\,erg\,cm$^{-2}$\,s$^{-1}$. The spectral region redwards of $5855$\,{\AA} was intercalibrated with respect to [\ion{S}{ii}]\,$\lambda\lambda6716,6731$ with the combined flux of $(43.2 \pm 1.2)\, \times$\,10$^{-15}$\,erg\,cm$^{-2}$\,s$^{-1}$ and [\ion{O}{iii}]\,$\lambda6300$ with a constant flux of $(10.4 \pm 0.4)\, \times$\,10$^{-15}$\,erg\,cm$^{-2}$\,s$^{-1}$. The rms spectrum furthermore shows no residual of the [\ion{N}{ii}]\,$\lambda\lambda6548,6584$ lines. We chose the spectrum from 2014-06-24 (ID 23 in Table \ref{tab:spectroscopy_log}) in both intercalibrations as a flux reference. 

We furthermore corrected for small wavelength shifts using the profiles of the aforementioned narrow lines. The wavelength accuracy of the intercalibration is optimal in the spectral regions close to [\ion{O}{iii}]\,$\lambda\lambda4959,5007$ and [\ion{S}{ii}]\,$\lambda\lambda6716,6731$. In regions further apart from the lines used for intercalibration, individual spectra may differ $\lesssim1$\,{\AA} from the optimal wavelength solution. To further increase the wavelength accuracy in the spectral region close to \ion{He}{i}\,$\lambda5876$, we applied shifts using the peaks of the coronal lines [\ion{Fe}{vii}]\,$\lambda5721$ and [\ion{Fe}{vii}]\,$\lambda6087$ as well as of the narrow line [\ion{N}{ii}]\,$\lambda5755$ as wavelength reference. Similarly, we used the single peak of \ion{Fe}{ii}\,$\lambda4233$ to improve the wavelength accuracy close to H$\gamma$.\footnote{We are able to perform this improvement as \ion{Fe}{ii}\,$\lambda4233$ is not variable in WPVS\,48. This is shown in the rms spectrum (see Fig.~\ref{fig:WPVS48_avg_rms.pdf}), which shows no variability in any \ion{Fe}{ii} line. In general, \ion{Fe}{ii} lines can be variable \citep[e.g.][]{2022AN....34310112G}.}

%**********************************************************************************
%
\section{Results}\label{sec:results}
%
%**********************************************************************************

%**********************************************************************************
%
\subsection{Optical spectral observations}\label{sec:optical_spectral_variations_results}
%
%**********************************************************************************
%
%---------------------------------------------------------------------------
%                              
\begin{figure*}[!t]
\centering
\includegraphics[width=18.5cm,angle=0]{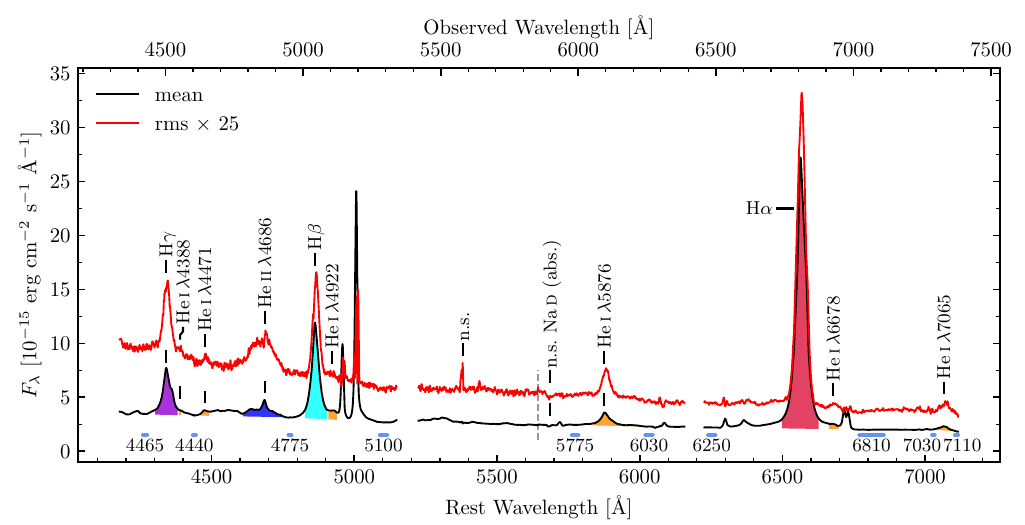}
\vspace*{-3mm} 
\caption{Combined mean spectrum (black) and rms spectrum (red) of WPVS\,48 from our campaign. The partition in the two intercalibrations (see Sect.~\ref{sec:observations} for details) is shown with a grey dashed line. The rms spectrum was scaled to allow for a direct comparison. The analysed continuum regions and the identical pseudo-continua used for linear continuum subtraction are highlighted with blue bars below the spectra. The line integration limits are marked by the shaded areas. All emission lines of \ion{He}{i} are displayed in orange.}
\label{fig:WPVS48_avg_rms.pdf}
\end{figure*}
%
%---------------------------------------------------------------------------
%  
We present all reduced, extinction-corrected and intercalibrated optical spectra in Fig.~\ref{fig:WPVS48_all_spectra.pdf}. The spectra are qualitatively equal and exhibit a variable continuum. We label the broad Balmer and \ion{He}{i} and \ion{He}{ii} emission lines, various narrow emission lines as well as the coronal lines [\ion{Fe}{vii}]\,$\lambda5721$, [\ion{Fe}{vii}]\,$\lambda6087$ and [\ion{Fe}{x}]\,$\lambda6375$ and further indicate the locations of the conspicuous \ion{Fe}{ii} multiplets (37, 38), (42) and (48, 49) \citep[for the individual transition wavelength see][and references therein]{2010ApJS..189...15K, 2022ApJS..258...38P}{}{}. A more detailed identification of the complex line blends in the proximity of the \ion{He}{ii}\,$\lambda4686$ is given in Sect.~\ref{sec:line_identification}.\\
              
Fig.~\ref{fig:WPVS48_avg_rms.pdf} shows the composite mean and rms spectrum from the intercalibrations (see Sect.~\ref{sec:observations}) shortwards and longwards of $5645$\,{\AA} respectively.\footnote{The presented mean and rms spectra are a combination of the blue and red intercalibrations merged at $5645$\,{\AA}.}\footnote{We note a [\ion{O}{iii}]\,$\lambda5007$ residual comparable in height to the profile of H$\beta$ in the rms spectrum. This high residual is produced by the skewed [\ion{O}{iii}]\,$\lambda5007$ profile in spectrum 15 (from 2014-03-28) likely caused by poor seeing conditions. If spectrum 15 is omitted, the [\ion{O}{iii}]\,$\lambda5007$ residual is smaller by the factor of $\sim 2$, whereas the rms profiles of the broad emission lines are not affected.} The division line between the two spectral regions is indicated by a grey dashed line. The rms spectrum is scaled by a factor of $\times\,25$ to facilitate a straightforward comparison. We highlight the line-free continuum regions analysed in this work with light blue lines below the spectrum as well as the Balmer and Helium lines with the corresponding integrated line flux. The spectral region used to measure the integrated line flux was chosen such that it encompasses the total variable component of the emission lines, which can be deduced from the rms spectrum. The specific continuum boundaries, integration limits and selected pseudo-continua are given in Table~\ref{tab:integrate_a_pc}. This work focuses on the Balmer as well as \ion{He}{i}\,$\lambda5876$ and \ion{He}{ii}\,$\lambda4686$. A detailed analysis on the remaining \ion{He}{i} lines will be given in a following publication (Probst et al. in prep.).
%
%------------------------------------------------------------------------------
%
\begin{table}[!t]
\centering
\caption{Continuum boundaries and line integration limits (2) as well as the sections used to fit the pseudo-continua (3).}
\begin{tabular}{lcc} 
\hline 
\hline 
\noalign{\smallskip}
\multicolumn{1}{c}{Cont./Line} &  Wavelength range &  Pseudo-continuum \\ 
\multicolumn{1}{c}{(1)} & \multicolumn{1}{c}{(2)} & \multicolumn{1}{c}{(3)} \\ 
\hline 
\noalign{\smallskip}
Cont. 4265 (4425) & $4260-4274$ &  \\
Cont. 4440 (4605) & $4435-4445$ &  \\
Cont. 4775 (4925) & $4770-4780$ &  \\
Cont. 5100 (5290) & $5090-5115$ &  \\
Cont. 5775 (5990) & $5763-5785$ &  \\
Cont. 6030 (6255) & $6020-6045$ &  \\
Cont. 6250 (6480) & $6240-6265$ &  \\
Cont. 6810 (7065) & $6770-6855$ &  \\
Cont. 7030 (7290) & $7025-7033$ &  \\
Cont. 7110 (7375) & $7105-7115$ &  \\
\hline 
\noalign{\smallskip}
H$\gamma$ & $4302-4382$ & $4260-4445$ \\
\text{}\ion{He}{ii}\,$\lambda4686$ & $4609-4765$ & $4435-4780$ \\
H$\beta$ & $4826-4905$ & $4770-5115$ \\
\text{}\ion{He}{i}\,$\lambda5876$ & $5835-5952$ & $5763-6045$ \\
H$\alpha$ & $6497-6630$ & $6240-6855$ \\
\hline 
\end{tabular}
\label{tab:integrate_a_pc}
\tablefoot{All wavelengths are given in the rest frame of the galaxy in units of \AA. In the case of the continua, the corresponding wavelengths in the observer's frame are given in brackets.}
\end{table}
%
%------------------------------------------------------------------------------
%

%**********************************************************************************
%
\subsubsection{Host galaxy contribution }\label{sec:host_galaxy_contribution_results}
\begin{figure}[h!]
    \centering
    \includegraphics[width=0.47\textwidth,angle=0]{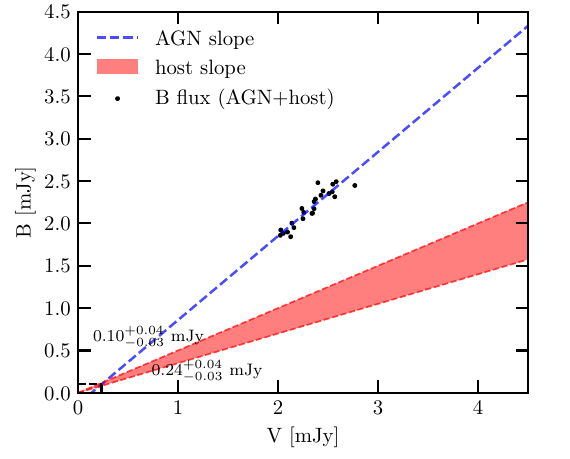}
    \caption{Extinction corrected B-V flux variations. The blue dashed line represents the best linear fit of the AGN slope of WPVS\,48 and the red shaded area host slopes of nearby AGN.}
    \label{fig:FVG_BV_WPVS48}
\end{figure}

\begin{figure}[h!]
    \centering
    \includegraphics[width=0.47\textwidth,angle=0]{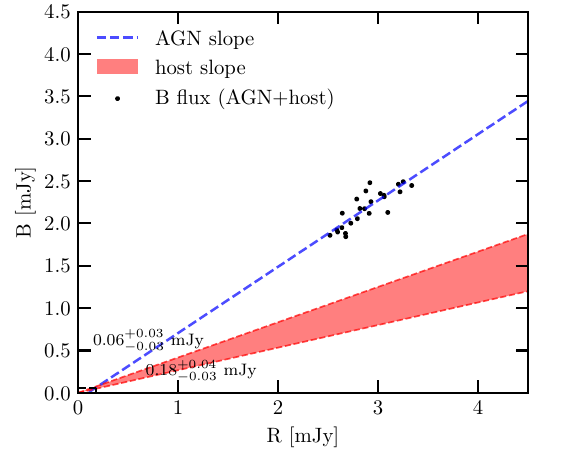}
    \caption{Extinction corrected B-R flux variations. The blue dashed line represents the best linear fit of the AGN slope of WPVS\,48 and the red shaded area host slopes of nearby AGN.}
    \label{fig:FVG_BR_WPVS48}
\end{figure}

%
%**********************************************************************************
We estimate the contribution of the host spectrum in the B, V and R band of the spectra applying the flux variation gradient (FVG) method \citep{1981AcA....31..293C, 1992MNRAS.257..659W, 2022A&A...657A.122K}. We adopt typical host slopes in the B-V and B-R colour indexes found by \citep{2010ApJ...711..461S} and \citep{2011A&A...535A..73H}, respectively. We determine the colour indices in WPVS\,48 using line-free continuum emission that has an effective wavelength (in the observed frame) close to the one of the corresponding filters employed by \citet{1992MNRAS.257..659W}, \citet{2010ApJ...711..461S} and \citet{2011A&A...535A..73H}. We hence select the continuum regions Cont. 4440, Cont. 5100 and Cont. 6250 for the B, V and R band, respectively. The three selected continuum regions have a respective effective wavelength of $4605$\,{\AA}, $5290$\,{\AA} and $6480$\,{\AA}. Figs.~\ref{fig:FVG_BV_WPVS48} and \ref{fig:FVG_BR_WPVS48} show the AGN slope and the typical host slope in the B-V and B-R colour plots. From the intersection of the slopes, we determine host fluxes of $0.08$\,mJy (mean of $0.06$ and $0.10$\,mJy), $0.24$\,mJy and $0.18$\,mJy in the B, V and R band, respectively, and thus a host contribution of $3-12\,\%$. The flux densities in units of mJy and erg$\,$cm$^{-2}\,$s$^{-1}\,$\AA$^{-1}$ for the combined as well as the separated AGN and host contributions in the three bands are given in Table \ref{tab:host_contribution}. 

No stellar signature is discernable in any of our taken spectra. Specifically, no stellar absorption lines are visibly superposed with the AGN emission. Hence, we did not subtract any possibly remaining host flux. A subtraction is neither necessary, as any constant host flux has no effect on the estimate of the BLR size by means of RM in the following analysis.

%
%------------------------------------------------------------------------------
%
%
%------------------------------------------------------------------------------
\begin{table}[!t] 
\centering 
\caption{Extinction-corrected B, V, and R values for the combined host galaxy plus AGN fluxes as well as for the host galaxy and AGN fluxes alone as determined by the FVG method.} 
\begin{tabular}{lccc} 
\hline 
\hline 
\noalign{\smallskip}
Flux component &  \textit{B} band &  \textit{V} band &  \textit{R} band \\ 
  & \multicolumn{3}{c}{[mJy]} \\ 
\hline 
\noalign{\smallskip}
Host+AGN (BvsR) & 1.84-2.49 &   & 2.52-3.34 \\ 
Host+AGN (BvsV) & 1.84-2.49 & 2.02-2.77 &   \\ 
\hline 
\noalign{\smallskip}
Host (BvsR) & 0.10 &   & 0.18 \\ 
Host (BvsV) & 0.06 & 0.24 &   \\ 
\hline 
\noalign{\smallskip}
AGN (BvsR) & 1.74-2.39 &   & 2.34-3.16 \\ 
AGN (BvsV) & 1.78-2.43 & 1.79-2.53 &   \\ 
\hline 
\hline 
\noalign{\smallskip}
  & \multicolumn{3}{c}{[$10^{-15}\,\text{erg}\,\text{cm}^{-2}\,\text{s}^{-1}\,\text{\AA}^{-1}$]} \\ 
\hline
\noalign{\smallskip}
Host+AGN (BvsR) & 2.8-3.79 &   & 1.93-2.56 \\ 
Host+AGN (BvsV) & 2.8-3.79 & 2.33-3.19 &   \\ 
\hline 
\noalign{\smallskip}
Host (BvsR) & 0.15 &   & 0.14 \\ 
Host (BvsV) & 0.09 & 0.27 &   \\ 
\hline 
\noalign{\smallskip}
AGN (BvsR) & 2.65-3.64 &   & 1.79-2.42 \\ 
AGN (BvsV) & 2.71-3.70 & 2.06-2.92 &   \\ 
\hline 
\label{tab:host_contribution} 
\end{tabular} 
\tablefoot{Given ranges correspond to the minimum and maximum flux.} 
\end{table}
%
%---------------------------------------------------------------------------
%     
%
%------------------------------------------------------------------------------
%

\subsubsection{Identification of Bowen Fluorescence near \ion{He}{ii}\,$\lambda$4686}\label{sec:line_identification}
\begin{figure}
\centering
\includegraphics[width=0.47\textwidth,angle=0]{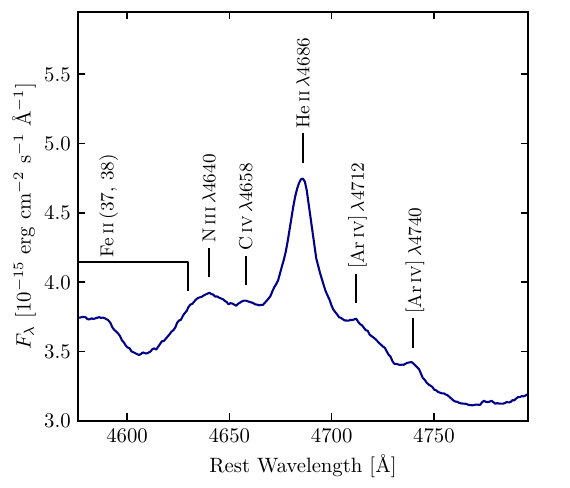}
\vspace*{-3mm} 
\caption{Detailed emission line identification in the mean spectrum close to \ion{He}{ii}\,$\lambda4686$. We identify \ion{N}{iii}\,$\lambda4640$ and \ion{C}{iv}\,$\lambda4658$ \citep{1982ApJ...261...64O} and further labelled the position of the \ion{Fe}{ii}\,$(37,\,38)$ emission bands \citep[for a template of \ion{Fe}{ii}, see][]{2022ApJS..258...38P}{}{} as well as [\ion{Ar}{iv}]\,$\lambda\lambda4712, 4740$.}
\label{fig:line_identification}
\end{figure}

Additionally to the line identification shown in Fig.~\ref{fig:WPVS48_all_spectra.pdf}, we analyse the spectral region close to \ion{He}{ii}\,$\lambda4686$ in more detail in Fig.~\ref{fig:line_identification}. Specifically, in the blue wing of \ion{He}{ii}\,$\lambda4686$, two broad peaks are conspicuous. The wavelength of the peaks match to \ion{N}{iii}\,$\lambda4640$ and \ion{C}{iv}\,$\lambda4658$, which together with \ion{He}{ii}\,$\lambda4686$ are often referred to as Wolf-Rayet feature and was interpreted in the past as the presence of a high number of Wolf-Rayet stars in the host galaxy \citep[e.g.][]{1982ApJ...261...64O}{}{}. The feature has also been seen in tidal disruption events \citep[][]{2015ApJ...815L...5G, 2018MNRAS.473.1130B, 2017MNRAS.466.4904B}{}{} as well as a new type of transient event in AGN defined by \citet{2019NatAs...3..242T} with distinctively strong \ion{He}{ii} emission. Specifically, the \ion{He}{ii} emission at $303.78$\,{\AA} produces the Bowen-fluorescence lines in multiple \ion{N}{iii} lines including \ion{N}{iii}\,$\lambda4640$ \citep[][and references therein]{1985ApJ...299..752N, 2019NatAs...3..242T}{}{}. To distinguish the Wolf-Rayet feature from emission of the \ion{Fe}{ii} emission, we mark the wavelength range of \ion{Fe}{ii}\,$(37,\,38)$ multiplets as given in the template of \citet{2022ApJS..258...38P}. Further, we mark two peaks in the red wing of \ion{He}{ii}\,$\lambda4686$, whose wavelengths match the [\ion{Ar}{iv}]\,$\lambda\lambda4712, 4740$ lines. 

%
%---------------------------------------------------------------------------
%                              
%
%---------------------------------------------------------------------------
%          

%**********************************************************************************
%
\subsection{Continuum Variability}\label{sec:Results_Continuum_Variability}
%
%**********************************************************************************

%**********************************************************************************
%
\subsubsection{Continuum Light Curves}\label{sec:Results_Continuum_Light_Curves}
%
%**********************************************************************************

\begin{figure*}
\centering
\includegraphics[width=17cm,angle=0]{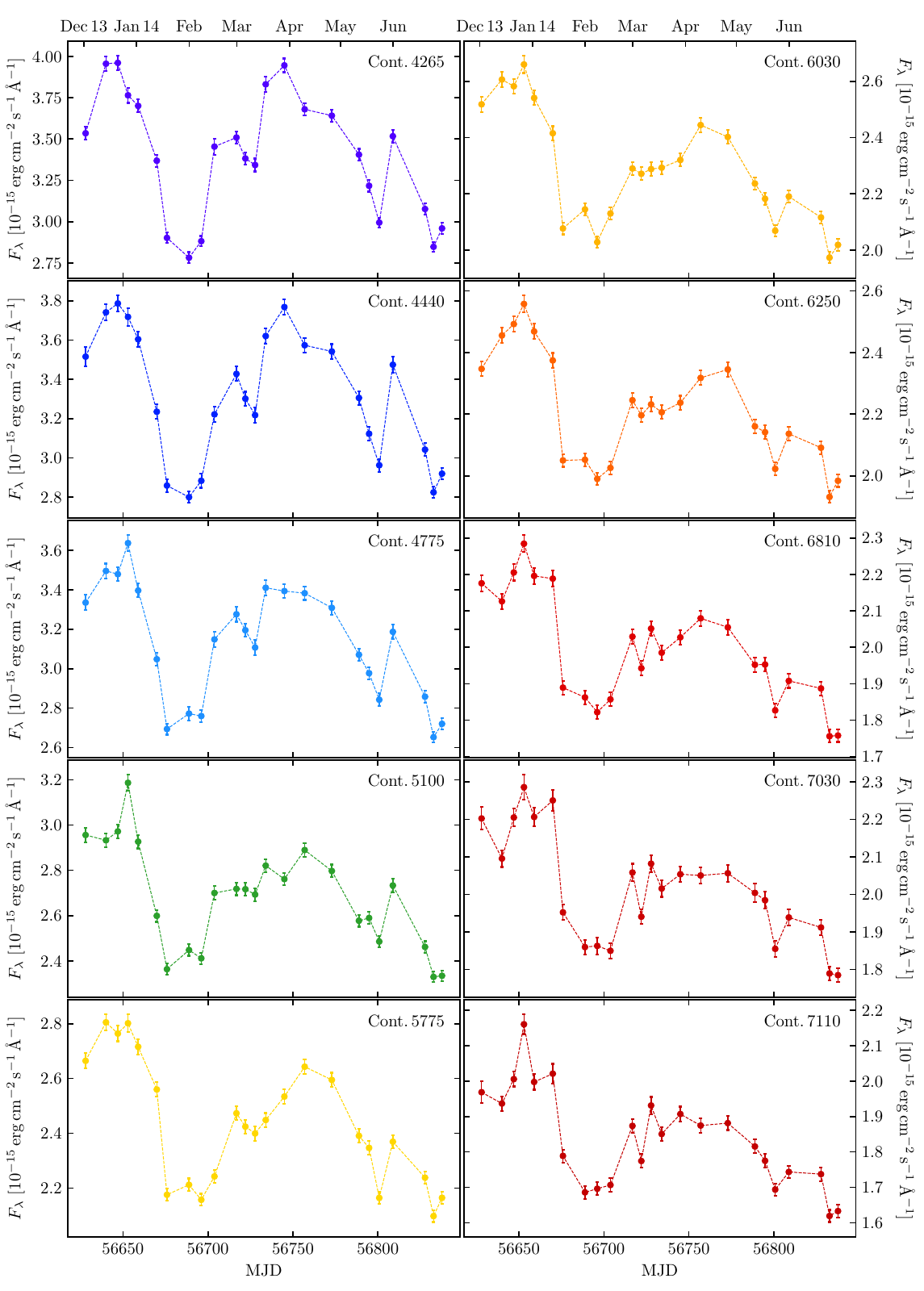}
\vspace*{-3mm} 
\caption{Optical continuum light curves measured within the integration limits given in Table~\ref{tab:integrate_a_pc}.}
\label{fig:WPVS48_cont_lightcurves.pdf}
\end{figure*}

We present the light curves of the continuum flux densities at $4265$\,{\AA}, $4440$\,{\AA}, $4775$\,{\AA}, $5100$\,{\AA}, $5760$\,{\AA}, $6030$\,{\AA}, $6250$\,{\AA}, $6810$\,{\AA}, $7030$\,{\AA}, $7110$\,{\AA} (rest wavelength) in Fig.~\ref{fig:WPVS48_cont_lightcurves.pdf}. The continuum bands are selected such they are not contaminated with emission lines. The selected continua are equivalent to line-free continuum bands identified in previous spectroscopic continuum RM campaigns \citep[e.g.][]{2001A&A...379..125K, 2018ApJ...857...53C, 2025ApJ...986..137Z}. The presented light curves are measured within the integration limits given in Table~\ref{tab:integrate_a_pc}. All measured light curves have a similar overall shape with a sequence of (1) a short duration of higher flux, (2) a steep decline and a short period of lower flux, (3) an increase in flux marking the beginning of a second, longer period of increased flux and (4) a second decline. Notably, the flux ratio of both periods of high flux is $\sim 1$ for the blue continua, whereas the flux of the second period is $\sim 10$ percent lower for the red continua. The flux densities of the measured continua are tabulated in \ref{tab:contLightcurves}.
%
%------------------------------------------------------------------------------
%
%
%------------------------------------------------------------------------------
%

We derive the minimum and maximum fluxes $F_{\text{min}}$, $F_{\text{max}}$, peak-to-peak amplitudes $R_{\text{max}}=F_{\text{max}}/F_{\text{min}}$, mean of the measured flux (densities) $\langle F \rangle$, the standard deviation $\sigma_F$, and the fractional variation,
\begin{align}
F_{\text{var}} = \sqrt{\sigma_F^2 - \Delta^2} / \langle F \rangle
\end{align}
as defined by \cite{1997ApJS..110....9R}, for the presented continuum light curves.
Here, $\Delta^2$ denotes the mean square value of the uncertainties of the individual fluxes $F_i$ of the light curve. The calculated values are tabulated in Table~\ref{tab:cont_line_stats} and show that the mean flux density, the variability amplitude and the fractional variation decrease with increasing wavelength.
%
%------------------------------------------------------------------------------
%
\begin{table}[!t]
\centering
\caption{Variability statistics of the investigated continua and emission lines with minimum (2) and maximum flux density or integrated flux (3), peak-to-peak ratio (4), mean flux density (5), standard deviation (6) and fractional variation (7).}
\begin{tabular}{lrrrrrr}
\hline 
\hline
\noalign{\smallskip}
Cont./Line &  \multicolumn{1}{c}{$F_{\text{min}}$} &  \multicolumn{1}{c}{$F_{\text{max}}$} &  \multicolumn{1}{c}{$R_{\text{max}}$} &  \multicolumn{1}{c}{$\langle F \rangle$} &  \multicolumn{1}{c}{$\sigma_F$} &  \multicolumn{1}{c}{$F_{\text{var}}$} \\ 
\multicolumn{1}{c}{(1)} & \multicolumn{1}{c}{(2)} & \multicolumn{1}{c}{(3)} & \multicolumn{1}{c}{(4)} & \multicolumn{1}{c}{(5)} & \multicolumn{1}{c}{(6)} & \multicolumn{1}{c}{(7)} \\
\hline 
\noalign{\smallskip}
Cont. 4265  & $2.78$ & $3.97$ & $1.43$ & $3.40$ & $0.37$ & $0.11$ \\
Cont. 4440  & $2.80$ & $3.78$ & $1.35$ & $3.31$ & $0.32$ & $0.10$ \\
Cont. 4775  & $2.65$ & $3.64$ & $1.37$ & $3.13$ & $0.29$ & $0.09$ \\
Cont. 5100  & $2.33$ & $3.19$ & $1.37$ & $2.68$ & $0.22$ & $0.08$ \\
Cont. 5775  & $2.10$ & $2.81$ & $1.34$ & $2.43$ & $0.22$ & $0.09$ \\
Cont. 6030  & $1.97$ & $2.66$ & $1.35$ & $2.28$ & $0.20$ & $0.09$ \\
Cont. 6250  & $1.93$ & $2.54$ & $1.32$ & $2.21$ & $0.17$ & $0.08$ \\
Cont. 6810  & $1.76$ & $2.28$ & $1.30$ & $1.99$ & $0.15$ & $0.07$ \\
Cont. 7030  & $1.78$ & $2.28$ & $1.28$ & $2.01$ & $0.14$ & $0.07$ \\
Cont. 7110  & $1.62$ & $2.16$ & $1.34$ & $1.84$ & $0.13$ & $0.07$ \\
\hline
\noalign{\smallskip}
H$\gamma$ & $130.9$ & $174.7$ & $1.33$ & $151.8$ & $10.7$ & $0.07$ \\
\text{}\ion{He}{ii}\,$\lambda4686$ & $62.5$ & $106.0$ & $1.70$ & $87.6$ & $11.8$ & $0.13$ \\
H$\beta$ & $267.2$ & $318.3$ & $1.19$ & $291.1$ & $14.7$ & $0.05$ \\
\text{}\ion{He}{i}\,$\lambda5876$ & $45.5$ & $63.3$ & $1.39$ & $53.7$ & $5.2$ & $0.10$ \\
H$\alpha$ & $973.$ & $1150.$ & $1.18$ & $1064.$ & $49.${\color{white}\_} & $0.05$ \\
\hline 
\end{tabular} 
\tablefoot{Continuum flux densities in units of $10^{-15}\,$erg$\,$cm$^{-2}$s$^{-1}\,$\AA$^{-1}$, line fluxes in units of $10^{-15}\,$erg$\,$cm$^{-2}$s$^{-1}$.}
\label{tab:cont_line_stats}
\end{table}
%
%------------------------------------------------------------------------------
%

%**********************************************************************************
%
\subsubsection{Interband continuum delays -- Univariate Analysis}\label{sec:cont_CCFs_univariate}
%
%**********************************************************************************

%
%------------------------------------------------------------------------------
%
\begin{table*}[!t] 
\centering 
\caption{Interband cross-correlation lags $\tau$ with respect to the continuum at 4265\,{\AA} (at 4425\,{\AA} in observed frame) as well as the maximum correlation coefficient $r_{\text{max}}$ and the fractional contribution $\alpha$ of the secondary component, respectively, for the univariate and bivariate model.} 
\begin{tabular}{lrrrcrr} 
\hline 
\hline 
\noalign{\smallskip}
  & \multicolumn{3}{c}{univariate model} & & \multicolumn{2}{c}{bivariate model} \\ 
\cline{2-4} 
\cline{6-7} 
\multicolumn{1}{c}{Cont.} & \multicolumn{1}{c}{$r_{\text{max}}$} & \multicolumn{1}{c}{$\tau_{\text{cent}}$} & \multicolumn{1}{c}{$\tau_{\text{peak}}$} & & \multicolumn{1}{c}{$\tau_{\text{sec}}$} & \multicolumn{1}{c}{$\alpha$} \\ 
 & & \multicolumn{1}{c}{[d]} & \multicolumn{1}{c}{[d]} & & \multicolumn{1}{c}{[d]} & \multicolumn{1}{c}{} \\ 
 \multicolumn{1}{c}{(1)} & \multicolumn{1}{c}{(2)} & \multicolumn{1}{c}{(3)} & \multicolumn{1}{c}{(4)} & & \multicolumn{1}{c}{(5)} & \multicolumn{1}{c}{(6)} \\ 
\hline 
\noalign{\smallskip}
Cont.\,4265 (ACF) & $1.00$ & $0.0_{-1.9}^{+1.9}$ & $0_{-1}^{+1{\color{white}\_}}$ & & \multicolumn{1}{c}{--} & \multicolumn{1}{c}{--}\\ 
Cont.\,4440 & $0.99$ & $0.0_{-1.4}^{+2.4}$ & $0_{-1}^{+1{\color{white}\_}}$ & & \multicolumn{1}{c}{--} & \multicolumn{1}{c}{--}\\ 
Cont.\,4775 & $0.96$ & $-0.4_{-2.1}^{+2.4}$ & $-1_{-1}^{+2{\color{white}\_}}$ & & \multicolumn{1}{c}{--} & \multicolumn{1}{c}{--}\\ 
Cont.\,5100 & $0.90$ & $-0.4_{-2.0}^{+3.3}$ & $1_{-3}^{+2{\color{white}\_}}$ & & $11.5_{-6.0}^{+4.5{\color{white}\_}}$ & $0.22_{-0.08}^{+0.09}$\\ 
V\,band (5300\,{\AA})$^{\ast}$ & $0.91$ & $5.4_{-1.9}^{+1.6}$ & $5_{-2}^{+2{\color{white}\_}}$ & & \multicolumn{1}{c}{--} & \multicolumn{1}{c}{--}\\ 
Cont.\,5775 & $0.91$ & $4.9_{-2.4}^{+2.6}$ & $3_{-1}^{+3{\color{white}\_}}$ & & $14.8_{-2.3}^{+2.8{\color{white}\_}}$ & $0.39_{-0.05}^{+0.04}$\\ 
Cont.\,6030 & $0.88$ & $5.4_{-2.4}^{+2.7}$ & $3_{0}^{+4{\color{white}\_}}$ & & $16.0_{-2.5}^{+2.1{\color{white}\_}}$ & $0.40_{-0.05}^{+0.04}$\\ 
Cont.\,6250 & $0.87$ & $6.4_{-2.0}^{+3.0}$ & $4_{-1}^{+5{\color{white}\_}}$ & & $15.5_{-2.9}^{+2.6{\color{white}\_}}$ & $0.44_{-0.05}^{+0.05}$\\ 
R\,band (6750\,{\AA})$^{\ast}$ & $0.88$ & $8.5_{-4.2}^{+1.6}$ & $6_{-3}^{+5{\color{white}\_}}$ & & \multicolumn{1}{c}{--} & \multicolumn{1}{c}{--}\\ 
Cont.\,6810 & $0.85$ & $7.0_{-3.1}^{+4.5}$ & $4_{-2}^{+8{\color{white}\_}}$ & & $14.0_{-2.1}^{+3.6{\color{white}\_}}$ & $0.49_{-0.05}^{+0.05}$\\ 
Cont.\,7030 & $0.84$ & $6.5_{-1.5}^{+6.5}$ & $4_{0}^{+10}$ & & $14.0_{-2.5}^{+4.0{\color{white}\_}}$ & $0.53_{-0.06}^{+0.08}$\\ 
Cont.\,7110 & $0.83$ & $8.0_{-4.2}^{+3.0}$ & $4_{-2}^{+8{\color{white}\_}}$ & & $13.0_{-1.6}^{+3.6{\color{white}\_}}$ & $0.51_{-0.06}^{+0.06}$\\ 
\hline 
\label{tab:continuum_lags} 
\end{tabular} 
\tablefoot{The photometric light curves marked with $^{\ast}$ were published by \citet{2018rnls.confE..57S}. We note that H$\beta$ contributes to the flux in the V\,band. The lag $\tau_{\text{sec}}$ and fractional contribution $\alpha$ of the secondary component have not been included for Cont.\,4440 and Cont.\,4775 due to its strong dependence on the chosen bin size of the correlation function.}
\end{table*}
%
%------------------------------------------------------------------------------
%
We correlate the continuum light curves with the light curve of the continuum at $4265$\,{\AA}, thereby determining interband continuum time lags with respect to the bluest continuum of this work. The univariate model underlying this analysis of interband continuum delays assumes similar, however, shifted light curves conjectured to stem from an irradiated accretion disk \citep[compare with e.g.][]{1998ApJ...500..162C}, such that
\begin{align}
    F_{\text{r, model}}(t) = F_{\text{b}}(t - \tau),
\end{align}
for two continuum light curves in a bluer band $F_{\text{b}}$ and a redder band $F_{\text{r}}$. In this equation, $\tau$ is the lag between the two light curves.

The employed correlation technique is the interpolated cross-correlation function (ICCF) described by \citet{1987ApJS...65....1G}. We calculate interband continuum lags according to both the peak $\tau_{\text{peak}}$ and the centroid $\tau_{\text{cent}}$ of the cross-correlation function. For the calculation of the centroid of the CCF, only values above $80$ percent of its peak value $r_{\text{max}}$ are considered. As demonstrated in \citet{2004ApJ...613..682P}, a threshold value of $0.8\, r_{\text{max}}$ is generally a good choice.

We derive the uncertainties by calculating the peak and the centroid lag for $\sim 20,000$ runs in a model-independent Monte-Carlo method known as flux randomisation/random subsample selection (FR/RSS). The method is described in detail by \citet{1998ApJ...501...82P}. The errors are evaluated using the centroid and peak distributions such that $68\,\%$ of the realizations yield values between the error interval.

\begin{figure}
\centering
\includegraphics[width=0.47\textwidth,angle=0]{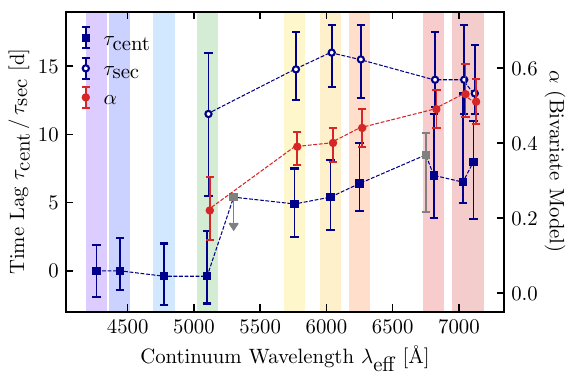}
\vspace*{-3mm} 
\caption{Mean centroid lags $\tau_{\text{cent}}$ according to the univariate model (blue squares) as well as centroid lags $\tau_{\text{sec}}$ (blue circles) and fractional contribution $\alpha$ (red circles) of the secondary component according to the bivariate model \citep{2019NatAs...3..251C} as a function of the effective wavelength of the continuum. We include the univariate lags of photometric light curves published by \citet{2018rnls.confE..57S} as grey squares. We note that H$\beta$ contributes to the flux in the V\,band. The lags and the fractional contribution are measured with respect to the continuum at $4265$\,{\AA} as a function of the effective wavelength of the continuum. The colouring of the shaded background corresponds to the light curves in Fig.~\ref{fig:WPVS48_cont_lightcurves.pdf}. The errors correspond to the $68\,\%$ ($\pm1\sigma$) confidence level after a Monte-Carlo simulation with 20,000 runs. The data of $\tau_{\text{sec}}$ and $\alpha$ have been slightly shifted wavelength-wise for better readability.} 
\label{fig:continuum_lags}
\end{figure}

The calculated lags are given in Table~\ref{tab:continuum_lags}. Fig.~\ref{fig:continuum_lags} shows the mean centroid lags as function of the effective wavelength of the respective continuum band and Fig.~\ref{fig:ContCorrelations} shows the individual correlation functions and the corresponding centroid distributions. The mean centroid time lag $\tau_{\text{cent}}$ is $\sim 0$\,days for the continua in the B\,band including the continuum at $5100$\,{\AA}. The lag jumps to $\sim 5$\,days between the continua measured at $5100$\,{\AA} and $5775$\,{\AA} and continuously increases up to $\sim 8$\,days for continua in the R\,band.

We repeated the calculation for the photometric V\,band and R\,band light curves of \citet{2018rnls.confE..57S} with respect to the continuum at $4265$\,{\AA} from this work. The derived centroid lags of the V\,band and R\,band light curves amount to $5.4^{+1.6}
_{-1.9}$ and $8.5^{+1.6}_{-4.2}$\,days, respectively, and are consistent with the interband continuum lags described above. The photometric light curves were obtained with the  VYSOS-6 and BEST-II telescopes of the Universit\"atssternwarte Bochum near Cerro Armazones in the time period between December 2013 and May 2014. Hence, the photometric light curves overlap with the spectroscopic light curves from this work. The derived lags are included in Fig.~\ref{fig:continuum_lags} as grey squares. We note that the photometric V\,band light curve may have contributions of variable line emission, especially H$\beta$ and \ion{He}{i}\,$\lambda5876$, which may shift the measured lag to larger delays. In \ref{sec:discussion_intercalibration}, we compare our red continuum light curve with the independent observations of \citet{2018rnls.confE..57S}.
%
%------------------------------------------------------------------------------
%

%**********************************************************************************
%
\subsubsection{Interband Continuum Delays -- Bivariate Analysis}
\label{sec:tau_alpha_formalism}
%
%**********************************************************************************
In addition to the univariate model, a different interpretation exists allowing an additional secondary component from a diffuse continuum originating from the BLR \citep[][]{2001ApJ...553..695K}{}{}. Therefore, we perform an additional analysis using a formalism by \citet{1994ApJ...420..806Z} adapted for a bivariate model of interband continuum delays by \citet{2013ApJ...769..124C}. The bivariate model assumes that the delayed red continuum light curve can be reconstructed with two identical superposed components of the blue continuum light curve. The primary component is not delayed, whereas the secondary is.\footnote{Note that the formalism in \citet{2013ApJ...769..124C} includes a lag $\tau_{\text{pri}}$ for the primary component as well. However, the formalism was simplified in later studies \citep[e.g.][]{2019NatAs...3..251C, 2025ApJS..276...48S}{}{} assuming a small $\tau_{\text{pri}}$ compared to both the lag of the secondary component $\tau_{\text{sec}}$ and the sampling.} This yields
\begin{align}
    F_{\text{r, model}}(t) = (1 - \alpha) F_{\text{b}}(t) + \alpha F_{\text{b}}(t - \tau_{\text{sec}}),
\end{align}
where $\alpha$ denotes the fractional contribution of the delayed component. 

We correlate the red-band light curve with a superposition of the non-delayed and delayed blue-band light curves under varying contributions and lags. We derive the fractional contributions $\alpha$ and delays $\tau_{\text{sec}}$ from the correlation with the highest correlation coefficient. The retrieved lags $\tau_{\text{sec}}$ and contributions $\alpha$ with respect to the continuum light curve at $4265$\,{\AA} ($4425$\,{\AA} in observed frame) are tabulated in columns 5 and 6 of Table~\ref{tab:continuum_lags} and plotted in Fig.~\ref{fig:continuum_lags}. We repeated the analysis in $\sim 20 000$ Monte-Carlo simulations employing both FR and RSS to evaluate the uncertainties. We note that we exclude the continua at $4440$\,{\AA} and $4775$\,{\AA}, which is discussed in Sect.~\ref{sec:discussionContinuumLags}.

The results are consistent with a contribution of the secondary delayed component with a lag of $11$ -- $15$\,d. This contribution increases with higher wavelengths from $\alpha \approx 0.2$ for the continuum at $5100$\,{\AA} to $\alpha \approx 0.5$ for the continuum at  $7030$\,{\AA}.

%**********************************************************************************
%
\subsection{Emission-Line Variability}\label{sec:line_variability_results}
%
%**********************************************************************************
%**********************************************************************************
%
\subsubsection{Broad-line Fluxes and Light Curves}\label{sec:Helium_profiles_results}
\begin{figure*}
\centering
\includegraphics[width=17cm,angle=0]{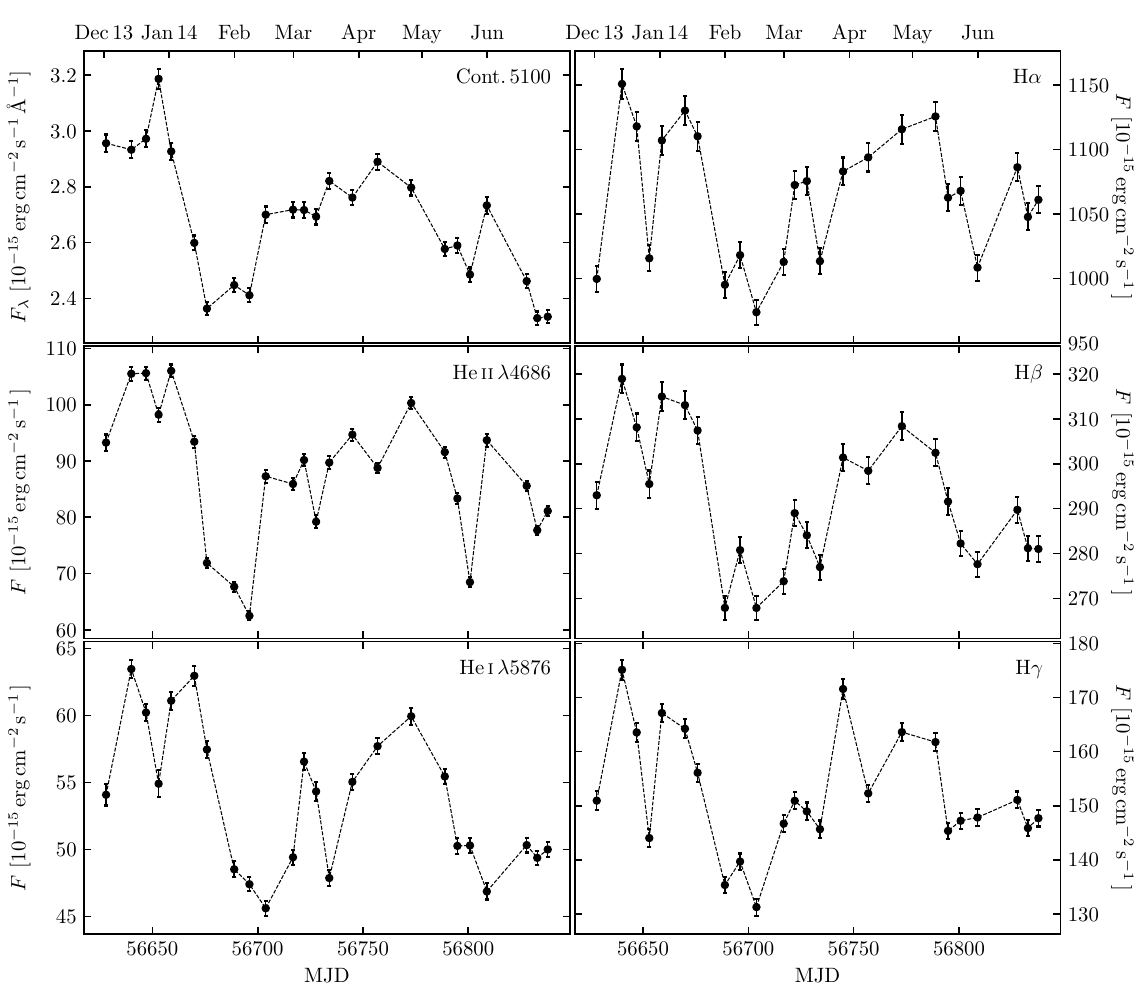}
\vspace*{-3mm} 
\caption{Emission line light curves of the Balmer lines as well as \ion{He}{i}\,$\lambda5876$ and \ion{He}{ii}$\,\lambda4686$ measured within the integration limits given in Table~\ref{tab:integrate_a_pc}. The continuum light curve at $5100\,${\AA} is included for reference.}
\label{fig:WPVS48_line_lightcurves.pdf}
\end{figure*}

%
%**********************************************************************************
We present the light curves of the integrated line fluxes of the Balmer lines as well as the \ion{He}{i}\,$\lambda5876$ and \ion{He}{ii}\,$\lambda4686$ lines in Fig.~\ref{fig:WPVS48_line_lightcurves.pdf}. We evaluate the line fluxes after the subtraction of a linear pseudo continuum. The flux integration limits as well as the regions used to interpolate the pseudo continua are given in Table~\ref{tab:integrate_a_pc}. The emission-line light curves show a similar sequence as the continuum light curves described in Sect.~\ref{sec:Results_Continuum_Light_Curves}. The flux ratio between the two periods of high flux is $\sim 1$ as observed in the light curves of the bluer continua. The fluxes of the measured lines are tabulated in \ref{tab:lineLightcurves}.
%
%------------------------------------------------------------------------------
%
%
%------------------------------------------------------------------------------
%
We derive the minimum and maximum fluxes $F_{\text{min}}$, $F_{\text{max}}$, peak-to-peak amplitudes $R_{\text{max}}=F_{\text{max}}/F_{\text{min}}$, mean of the measured flux (densities) $\langle F \rangle$, the standard deviation $\sigma_F$, and the fractional variation analogously to the continuum light curves. The calculated values for the emission-line light curves are tabulated in Table~\ref{tab:cont_line_stats}. The variability amplitudes and fractional variation are higher for \ion{He}{ii}\,$\lambda4686$ and \ion{He}{i}\,$\lambda5876$ than for the Balmer lines.

%**********************************************************************************
%
\subsubsection{Emission-line lags and BLR size}\label{sec:1D_CCFs}
%
%**********************************************************************************
We correlate light curves of the integrated emission lines with the the light curve of the continuum at $5100$\,{\AA}, thereby determining the mean light travelling time from the ionizing source to the variable broad component of the emission lines. Commonly, the continuum at $5100$\,{\AA} is used as a surrogate for the light curve of the ionizing continuum in variability studies. 

In some previous RM campaigns analysing interband continuum delays, the B\,band continuum light curves are, firstly, higher-correlated with the ionizing UV continuum \citep[e.g.][]{2019ApJ...870..123E}, and, secondly, exhibit shorter delays to the ionizing UV/X-ray continuum than the V\,band \citep[e.g.][]{2015ApJ...806..129E, 2018MNRAS.480.2881M, 2025ApJ...986..137Z}. We, therefore, correlate the emission line light curves also with the the bluest line-free continuum region at $4265$\,{\AA} in our spectra. This continuum region was specifically selected such it is free from contamination of both \ion{Fe}{ii}\,$\lambda4233$ and H$\gamma$.

\begin{figure*}[!t]
\centering
  \begin{minipage}{0.99\textwidth}
    \begin{subfigure}[t]{0.49\textwidth}
      \includegraphics[width=\textwidth]{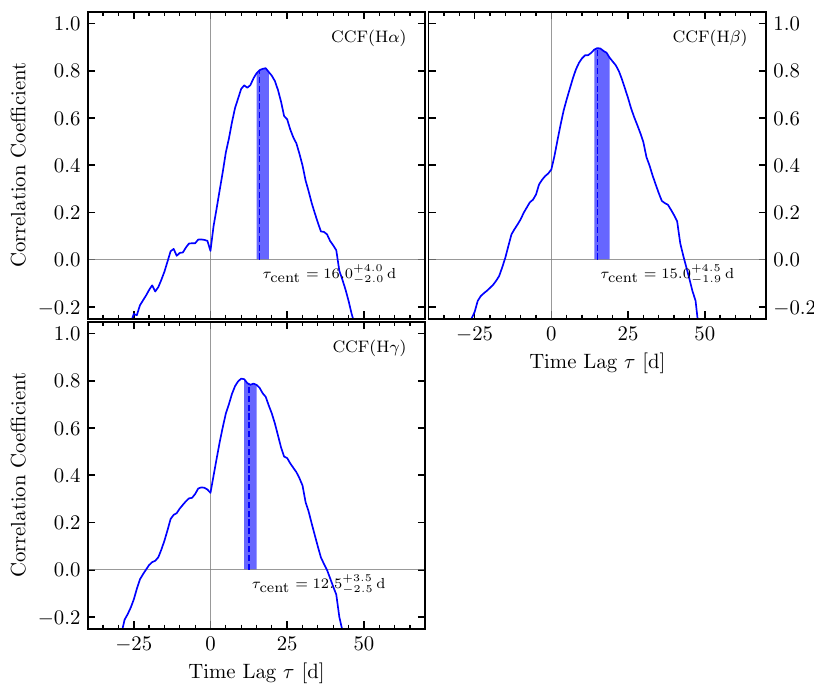}
      \caption{}
      \label{fig:correlation_balmer}
    \end{subfigure}\hfill
    \begin{subfigure}[t]{0.49\textwidth}
      \includegraphics[width=\textwidth]{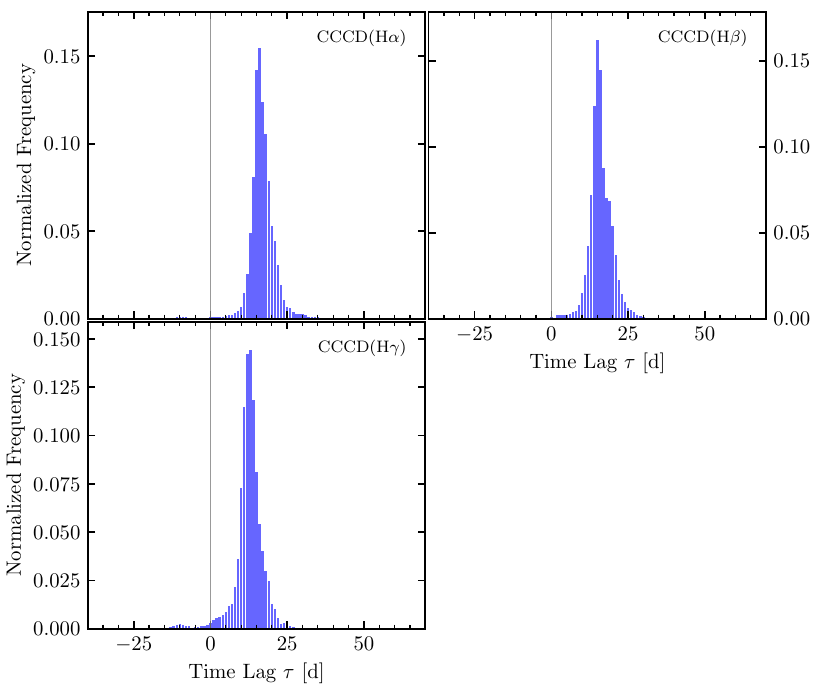}
      \caption{}
      \label{fig:distribution_balmer}
    \end{subfigure}
    \begin{subfigure}[t]{0.49\textwidth}
      \includegraphics[width=\textwidth]{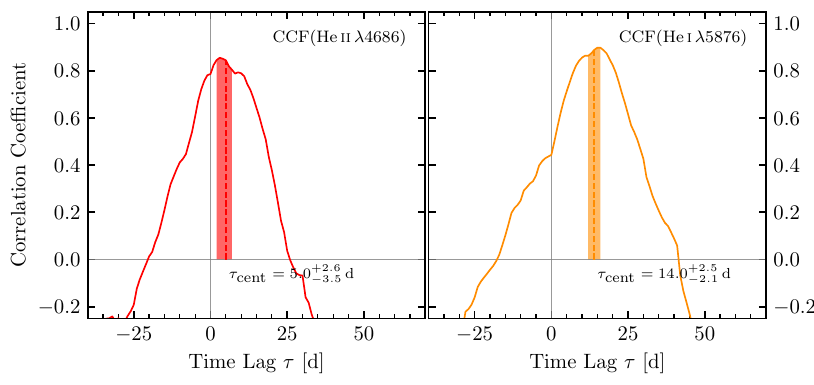}
      \caption{}
      \label{fig:correlation_helium}
    \end{subfigure}\hfill
    \begin{subfigure}[t]{0.49\textwidth}
      \includegraphics[width=\textwidth]{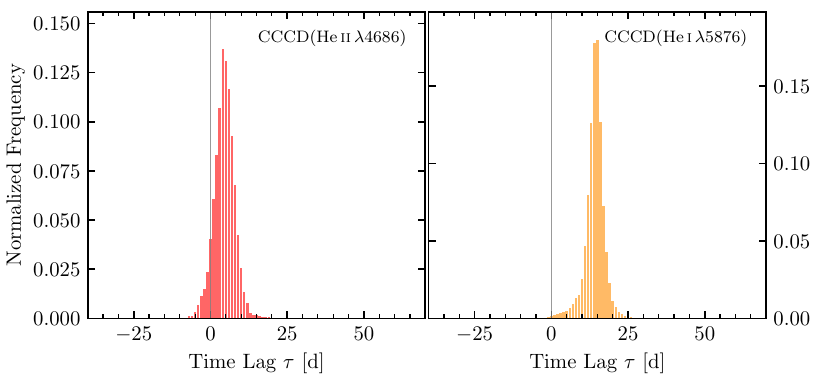}
      \caption{}
      \label{fig:distribution_helium}
    \end{subfigure}
  \caption{\textit{Left panels}: CCFs of the integrated Balmer lines in blue (a), the integrated \ion{He}{i}$\,\lambda5876$ line in orange and the integrated \ion{He}{ii}$\,\lambda4686$ line in red (c) with respect to the continuum light curve at $5100\,${\AA}. The time lag, $\tau_{\text{cent}}$, is denoted by a dashed line, with the shaded area corresponding to a $\pm 1 \sigma$ interval from the Monte-Carlo simulations. \textit{Right panels}: Cross-correlation centroid distributions after 20,000 runs employing FR and RSS of the integrated Balmer lines in blue (b) the integrated \ion{He}{i}$\,\lambda5876$ line in orange and the integrated \ion{He}{ii}$\,\lambda4686$ line in red (d) with respect to the continuum at $5100\,${\AA}.} 
  \label{fig:1DCorrelations}
  \end{minipage}
\end{figure*}

The employed correlation technique as well as the uncertainty estimation (FR/RSS) are the same as for the interband continuum delays described in Sect.~\ref{sec:cont_CCFs_univariate}. The correlation functions and the centroid distributions resulting from the Monte-Carlo method employing FR/RSS are shown in Fig.~\ref{fig:1DCorrelations}. The derived lags are tabulated in Table~\ref{tab:lags_and_masses}.
%
%------------------------------------------------------------------------------
%
%
%------------------------------------------------------------------------------
%

The light curves are well correlated with maximum correlation coefficients between $0.74$ and $0.9$ for the correlations with both continua. We determine centroid lags of the Balmer lines H$\alpha$ and H$\beta$ of $\sim 16$\,days, consistent with the H$\alpha$ lag of $\sim 18$\,days found by \citet{2018rnls.confE..57S} based on photometry with a narrow-band filter. Further we derive centroid lags of $\sim 12$\,days, $\sim 14$\,days for H$\gamma$, \ion{He}{i}\,$\lambda5876$, respectively. Notably, the lag of \ion{He}{i}\,$\lambda5876$ is identical to the lags of the Balmer lines within the errors. The corresponding centroid lags from correlations with both continua are consistent for aforementioned lines and deviate by $\lesssim 1.5$\,days. The centroid lag of \ion{He}{ii}\,$\lambda4686$ is close to the minimum sampling in this campaign and amounts to $\lesssim 5$\,days.

%**********************************************************************************
%
\subsubsection{Broad-Line Profiles}\label{sec:optical_profiles_results}
\begin{figure*}[!t]
\centering
\includegraphics[width=17cm,angle=0]{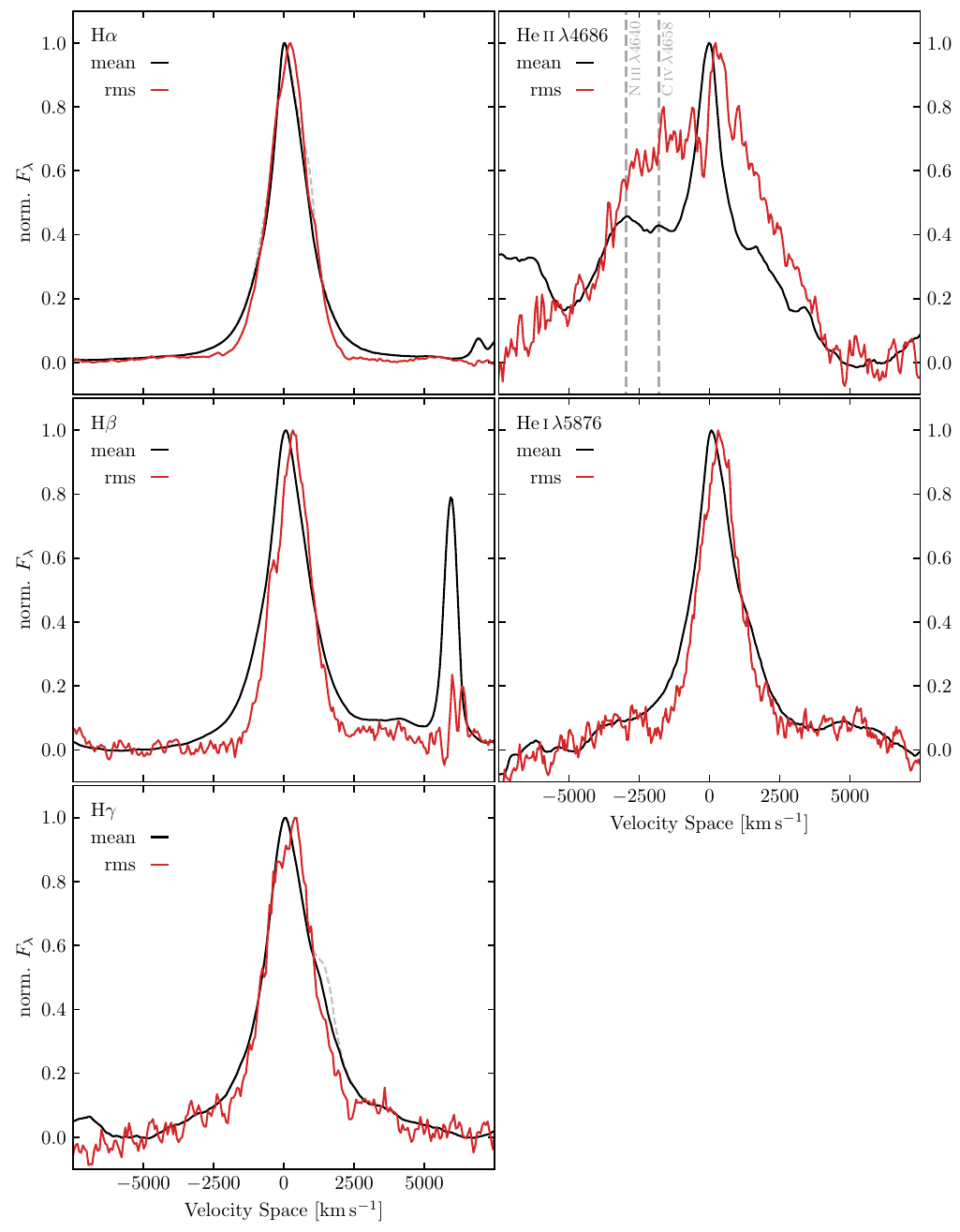}
\vspace*{-3mm} 
\caption{Normalised mean (black) and rms (red) line profiles of the Balmer and the prominent \ion{He}{i}\,$\lambda5876$ and \ion{He}{ii}\,$\lambda4686$ emission lines. While the mean profiles corrected for contributions of narrow forbidden lines are shown in red, the grey dashed line indicates the uncorrected mean profiles of H$\alpha$ and H$\gamma$. Note that the blue wing of the mean \ion{He}{ii}\,$\lambda4686$ profile blends with emission from \ion{N}{iii}\,$\lambda4640$, \ion{C}{iv}\,$\lambda4658$ and \ion{Fe}{ii}$\,(37,38)$ multiplets.}
\label{fig:WPVS48_Balmer_prominent_Helium_lineprofiles.pdf}
\end{figure*}

%
%**********************************************************************************
Fig.~\ref{fig:WPVS48_Balmer_prominent_Helium_lineprofiles.pdf} shows the normalised mean and rms profiles of the Balmer lines as well as of the \ion{He}{i}\,$\lambda5876$ and \ion{He}{ii}\,$\lambda4686$ lines after subtraction of a linear pseudo-continuum with the wavelengths indicated in Table~\ref{tab:integrate_a_pc}. In order to retrieve genuine mean line profiles without the contribution of forbidden narrow lines, we corrected the mean profiles of H$\alpha$ and H$\gamma$ for [\ion{N}{ii}]\,$\lambda\lambda6548,\,6584$ and [\ion{O}{iii}]\,$\lambda4363$, respectively, by subtracting a scaled narrow-line template of [\ion{O}{iii}]\,$\lambda5007$. The uncorrected line profiles of H$\alpha$ and H$\gamma$ are also indicated in Fig.~\ref{fig:WPVS48_Balmer_prominent_Helium_lineprofiles.pdf} with a grey dashed line. We did not subtract the narrow components of the Balmer and Helium lines, as the broad component is only slightly broader than the narrow component and the narrow components thus cannot be unambiguously separated using a narrow-line template of [\ion{O}{iii}]\,$\lambda5007$. The rms profiles reflect only the variable broad component, but not the constant narrow component of the line. Note that the closest line-free region bluewards of \ion{He}{ii}\,$\lambda4686$ is at $\approx4440$\,{\AA} beyond the \ion{Fe}{ii} emission band. Therefore, the blue wing of the \ion{He}{ii}\,$\lambda4686$ mean profile does not reach the base line.
%
%------------------------------------------------------------------------------
%
\begin{table}[!t] 
\centering 
\caption{Cross-correlation lags of the continuum light curves at 4265\,{\AA} and 5100\,{\AA} (at 4425\,{\AA} and 5290\,{\AA} in observed frame, respectively) with the light curves of the integrated lines.}
\begin{tabular}{lrrrcrrr} 
\hline 
\hline 
\noalign{\smallskip}
 & \multicolumn{3}{c}{Cont.\,4265} & & \multicolumn{3}{c}{Cont.\,5100} \\ 
\cline{2-4} 
\cline{6-8} 
\multicolumn{1}{c}{Line} & \multicolumn{1}{c}{$r_{\text{max}}$} & \multicolumn{1}{c}{$\tau_{\text{cent}}$} & \multicolumn{1}{c}{$\tau_{\text{peak}}$} & & \multicolumn{1}{c}{$r_{\text{max}}$} & \multicolumn{1}{c}{$\tau_{\text{cent}}$} & \multicolumn{1}{c}{$\tau_{\text{peak}}$} \\ 
 & & \multicolumn{1}{c}{[d]} & \multicolumn{1}{c}{[d]} & & & \multicolumn{1}{c}{[d]} & \multicolumn{1}{c}{[d]} \\
 \multicolumn{1}{c}{(1)} & \multicolumn{1}{c}{(2)} & \multicolumn{1}{c}{(3)} & \multicolumn{1}{c}{(4)} & & \multicolumn{1}{c}{(5)} & \multicolumn{1}{c}{(6)} & \multicolumn{1}{c}{(7)} \\
\hline 
\noalign{\smallskip}
H$\alpha$ & $0.76$ & $16.6_{-3.1}^{+4.5}$ & $19_{-6}^{+2}$ & & $0.81$ & $16.0_{-2.0}^{+4.0}$ & $18_{-4}^{+1}$\\ 
H$\beta$ & $0.82$ & $15.5_{-2.6}^{+4.6}$ & $15_{-2}^{+6}$ & & $0.90$ & $15.0_{-1.9}^{+4.5}$ & $15_{-2}^{+4}$\\ 
H$\gamma$ & $0.74$ & $11.0_{-3.5}^{+5.8}$ & $9_{-3}^{+9}$ & & $0.81$ & $12.5_{-2.5}^{+3.5}$ & $10_{-1}^{+6}$\\ 
\ion{He}{i} & $0.82$ & $15.1_{-3.8}^{+3.6}$ & $16_{-5}^{+4}$ & & $0.90$ & $14.0_{-2.1}^{+2.5}$ & $15_{-4}^{+1}$\\ 
\ion{He}{ii} & $0.90$ & $2.4_{-1.4}^{+4.0}$ & $2_{-1}^{+2}$ & & $0.85$ & $5.0_{-3.5}^{+2.6}$ & $3_{-1}^{+5}$\\ 
\hline 
\label{tab:lags_and_masses} 
\end{tabular} 
\end{table}
%
%------------------------------------------------------------------------------
%
%
%------------------------------------------------------------------------------
%
%
%------------------------------------------------------------------------------
%
\begin{figure*}[!t]
\includegraphics[width=17cm, angle=0]{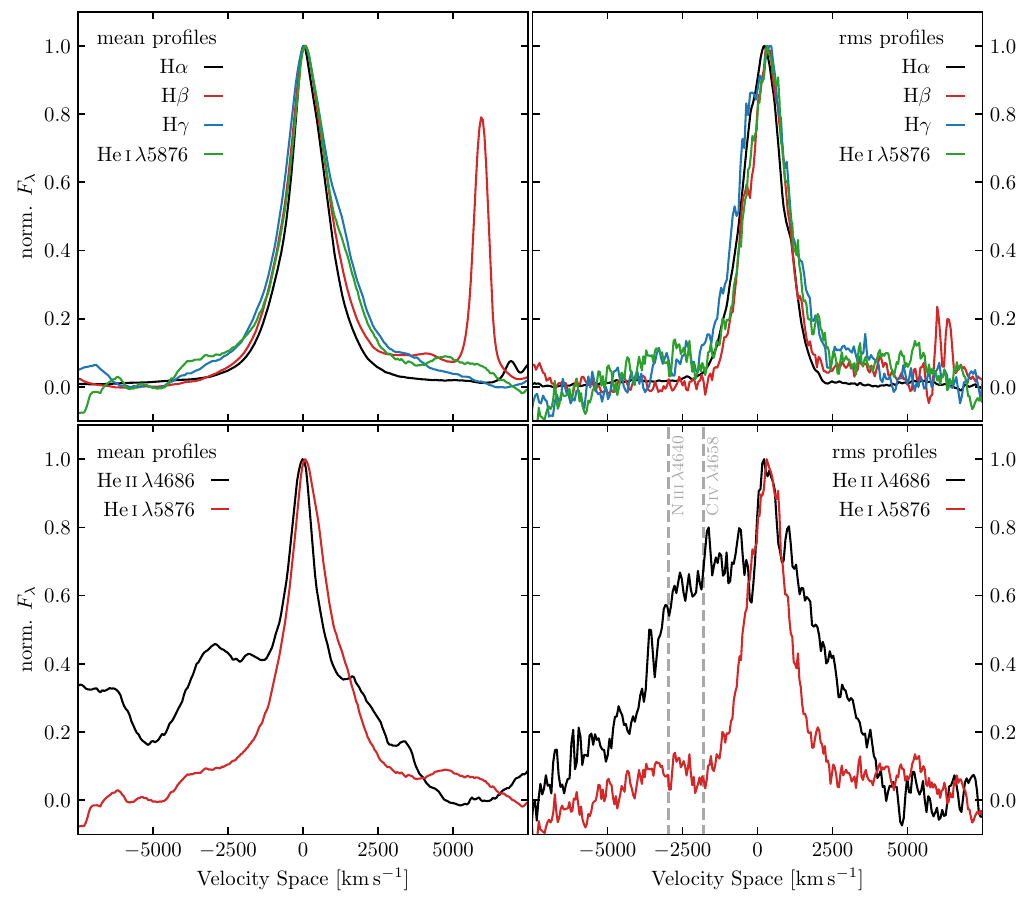}
\caption{\textit{Top panels:} Comparison of normalised mean (left panel) and rms (right panel) line shapes of the Balmer emission lines with the \ion{He}{i}\,$\lambda5876$ emission line. \textit{Bottom panels:} Comparison of normalised mean (left panel) and rms (right panel) line shapes of the \ion{He}{i}\,$\lambda5876$ emission line with the \ion{He}{ii}\,$\lambda4686$ emission line. Note that the blue wing of the mean \ion{He}{ii}\,$\lambda4686$ profile blends with emission from \ion{N}{iii}\,$\lambda4640$, \ion{C}{iv}\,$\lambda4658$ and \ion{Fe}{ii}$\,(37,38)$ multiplets.}
\label{fig:WPVS48_Line_Profile_Comparisons.pdf}
\end{figure*}
%
%------------------------------------------------------------------------------
%

We compare the mean and rms profiles of the Balmer lines with each other and with \ion{He}{i}\,$\lambda5876$ as well as \ion{He}{i}\,$\lambda5876$ with \ion{He}{ii}\,$\lambda4686$ in Fig.~\ref{fig:WPVS48_Line_Profile_Comparisons.pdf}. We note nearly identical and nearly symmetrical mean profiles of the Balmer lines and \ion{He}{i}\,$\lambda5876$ with the profiles of H$\gamma$ and \ion{He}{i}\,$\lambda5876$ being slightly broader (line widths and corresponding lags are discussed in Sect.~\ref{sec:discussion_lineprofiles}). The blue wings of the Balmer lines do not reach the base line due to blends with adjacent lines. We note higher flux at the wings of the mean profile of \ion{He}{i}\,$\lambda5876$ at $\sim -4700$\,km\,s$^{-1}$ and between $\sim +2500$ and $\sim +7300$\,km\,s$^{-1}$, which is not present in the Balmer profiles.

The Balmer lines and \ion{He}{i}\,$\lambda5876$ exhibit similar rms line shapes with FWHM of $\sim 1500$\,km\,s$^{-1}$, which furthermore are nearly symmetrical with exception of the peak. Specifically, all lines show a flux deficit at the upper $20\,\%$ of the blue wing resulting in a redwards asymmetry of the rms peaks. The redshift of the peaks in the rms profiles with respect to the peaks in the mean profiles are equal and amount to $\approx300\pm100$\,km\,s$^{-1}$ with variations on the order of the spectral resolving power of RSS with a PG900 grating.

The \ion{He}{ii}\,$\lambda4686$ line with a FWHM of $\sim 4000$\,km\,s$^{-1}$ is clearly broader than the Balmer lines and the \ion{He}{i}\,$\lambda5876$ emission. Comparing the two Helium lines, \ion{He}{ii}\,$\lambda4686$ exhibits a similar redshift of its rms profile peak as \ion{He}{i}\,$\lambda5876$ (and the Balmer lines). The red wing of \ion{He}{ii}\,$\lambda4686$ declines steadily without apparent features (except the blends with [\ion{Ar}{iv}]\,$\lambda\lambda4712, 4740$ in the mean profile). The blue wing of \ion{He}{ii}\,$\lambda4686$, however, is extended in both the mean and rms profiles due to the line blends described in Sect.~\ref{sec:line_identification}.

We measured the line widths (full width at half maximum, FWHM) of the Balmer lines as well as \ion{He}{i}\,$\lambda5876$. Furthermore, we parametrised the line widths by means of the line dispersion \citep[][]{2000ApJ...533..172F, 2004ApJ...613..682P}{}{}. The measurements for H$\alpha$ and H$\gamma$ are performed for the line profiles corrected for the narrow-line contributions of [\ion{N}{ii}]\,$\lambda6548$, [\ion{N}{ii}]\,$\lambda6584$ and [\ion{O}{iii}]\,$\lambda4363$, respectively. In order to minimise the effect of further blended lines located on the extreme wings to the line dispersion, we limited the evaluation of the line dispersion only to the variable spectral ranges given in Table~\ref{tab:integrate_a_pc}. This also encompasses the majority of the flux of the mean profiles.

As a result, we measured FWHM without and line dispersions with minimal effect of line blends for the mean and rms profiles of the Balmer lines and \ion{He}{i}\,$\lambda5876$. This, however cannot be stated for \ion{He}{ii}\,$\lambda4686$: Due to the prominent narrow component in \ion{He}{ii}\,$\lambda4686$, the half maximum is likely overestimated and, thus, the measured FWHM is likely underestimated. Both the FWHM of the rms profile as well as the line dispersion of both profiles include the extended blue wing and, thus, the emission from \ion{N}{iii}\,$\lambda4640$ and \ion{C}{iv}\,$\lambda4658$. For the total line complex, we measure a FWHM of (5060$\pm$660)\,km\,s$^{-1}$. Due to the blends, we only give the width of the red wing of the rms profile alone, which amounts to 1780\,km\,s$^{-1}$. We base the estimate of the uncertainties of the line widths on the bootstrap method described in \citet{2004ApJ...613..682P}.

%
%------------------------------------------------------------------------------
%
\begin{table}[!t] 
\caption{FWHM and line dispersion $\sigma_{\text{line}}$ of the mean and rms profiles of the Balmer lines as well as of \ion{He}{i}\,$\lambda5876$.}
\centering 
\begin{tabular}{lrrcrr} 
\hline 
\hline 
\noalign{\smallskip}
 & \multicolumn{2}{c}{FWHM [km$\,$s$^{-1}$]} & & \multicolumn{2}{c}{$\sigma_{\text{line}}$ [km$\,$s$^{-1}$]} \\
\cline{2-3} 
\cline{5-6}
\multicolumn{1}{c}{Line} & \multicolumn{1}{c}{mean} & \multicolumn{1}{c}{rms} & & \multicolumn{1}{c}{mean} & \multicolumn{1}{c}{rms} \\ 
\multicolumn{1}{c}{(1)} & \multicolumn{1}{c}{(2)} & \multicolumn{1}{c}{(3)} & & \multicolumn{1}{c}{(4)} & \multicolumn{1}{c}{(5)} \\ 
\hline 
\noalign{\smallskip}
H$\alpha$ & 1360$\pm$50 & 1530$\pm${\color{white}\_}90 & & 990$\pm$50 & 790$\pm$50 \\
H$\beta$  & 1660$\pm$60 & 1540$\pm$290 & & 950$\pm$50 & 740$\pm$80 \\
H$\gamma$ & 2050$\pm$70 & 1920$\pm$260 & & 1110$\pm$50 & 1010$\pm$80 \\
\ion{He}{i} & 1660$\pm$50 & 1560$\pm$190 & & 1150$\pm$50 & 1040$\pm$50 \\
\hline 
\label{tab:lineWidths} 
\end{tabular} 
\end{table}
%
%------------------------------------------------------------------------------
%

%**********************************************************************************
%
\subsubsection{Narrow-Line Fluxes}\label{sec:narrowLineFlux_and_errorEstimation}
%
%**********************************************************************************

We measure the constant fluxes of the narrow forbidden lines [\ion{O}{iii}]\,$\lambda4959$, [\ion{O}{iii}]\,$\lambda5007$, [\ion{O}{i}]\,$\lambda6300$, [\ion{O}{i}]\,$\lambda6364$ as well as the combined flux of [\ion{S}{ii}]\,$\lambda\lambda6716,\,6731$, in the intercalibrated extinction-corrected spectra. We present mean values of the measured fluxes alongside with the standard deviations as errors in Table~\ref{tab:narrowlinesIntensities}. We determine the fluxes of the narrow lines [\ion{O}{iii}]\,$\lambda4363$, [\ion{N}{ii}]\,$\lambda6548$ and [\ion{N}{ii}\,$\lambda6584$] based on the flux of [\ion{O}{iii}]\,$\lambda5007$ and the respective factors ($0.021$, $0.063$ and $0.201$) from the decomposition of the line blends with H$\alpha$ and H$\gamma$ in their mean profiles as explained in Sect.~\ref{sec:optical_profiles_results}. Here, we assume an uncertainty in the multiplicative factor of $0.01$, which translates to a flux uncertainty of $2.5\times10^{-15}\,$erg$\,$cm$^{-2}$s$^{-1}$.
%
%------------------------------------------------------------------------------
%
\begin{table}[!t]
\caption{Line intensities of the narrow forbidden lines.}
\centering 
\small
\begin{tabular}{lr|lr}
\hline 
\hline
\noalign{\smallskip}
\multicolumn{1}{c}{Line} & \multicolumn{1}{c}{Flux} & \multicolumn{1}{c}{Line} & \multicolumn{1}{c}{Flux} \\ 
\hline 
\vspace{-7pt}
 & & & \\
\text{}[\ion{O}{iii}]\,$\lambda4363^{\ast}$ & $5.1\pm2.5$ & [\ion{O}{i}]\,$\lambda6364$ & $21.6\pm1.5$ \\
\text{}[\ion{O}{iii}]\,$\lambda4959$ & $79.8\pm0.9$ & [\ion{N}{ii}]\,$\lambda6548^{\ast}$ & $15.4\pm2.5$ \\
\text{}[\ion{O}{iii}]\,$\lambda5007$ & $244.2\pm0.7$ & [\ion{N}{ii}]$\,\lambda6584^{\ast}$ & $49.1\pm2.5$ \\
\text{}[\ion{O}{i}]$\,\lambda6300$ & $10.4\pm0.4$ & [\ion{S}{ii}]$\,\lambda\lambda6716$,\,$6731$ & $43.2\pm0.8$ \\
\hline 
\end{tabular} 
\label{tab:narrowlinesIntensities}
\tablefoot{Line fluxes in units of $10^{-15}\,$erg$\,$cm$^{-2}$\,s$^{-1}$, standard deviations of the fluxes are assumed as errors. The [\ion{O}{i}]\,$\lambda6364$ line is blended with [\ion{Fe}{X}]\,$\lambda6375$. Line fluxes indicated by $^{\ast}$ have been estimated based on the decomposition of the H$\alpha$ and H$\gamma$ profile employing a [\ion{O}{iii}]\,$\lambda5007$ template (see Sect.~\ref{sec:optical_profiles_results})}.
\end{table}
%
%------------------------------------------------------------------------------
%

%**********************************************************************************
%
\subsection{Black hole mass estimation and Bolometric Luminosity}
\label{sec:black_hole_mass_results}
%
%**********************************************************************************
%
%------------------------------------------------------------------------------
%
\begin{table}[!t] 
\caption{Virial products based on reverberation mapping.} 
\centering 
\begin{tabular}{lccccc} 
\hline 
\hline 
\noalign{\smallskip}
Width & H$\alpha$ & H$\beta$ & H$\gamma$ & \ion{He}{i}\,$\lambda5876$ & mean \\  
FWHM & $7.3_{-1.6}^{+2.9}$ & $6.9_{-2.9}^{+5.8}$ & $9.0_{-3.6}^{+5.8}$ & $6.7_{-2.3}^{+3.2}$ & $7.3_{-2.5}^{+5.7}$ \\
$\sigma_{\text{line}}$ & $1.9_{-0.5}^{+0.8}$ & $1.6_{-0.5}^{+1.0}$ & $2.5_{-0.8}^{+1.2}$ & $3.0_{-0.7}^{+0.9}$ & $2.0_{-0.9}^{+1.7}$ \\
\hline 
\label{tab:BH_masses} 
\end{tabular} 
\tablefoot{The virial products $c\,\tau\,\Delta v^2\,G^{-1}$ are given in units of $10^6M_{\odot}$. The estimates employ both the FWHM and $\sigma_{\text{line}}$ of the respective line and the lag $\tau_{\text{cent}}$ between line emission and the continuum emission at $5100$\,{\AA}. Derived masses are discussed in Sect.~\ref{sec:discussion_BH_mass_Ledd}.}
\end{table}
%
%------------------------------------------------------------------------------
%
We estimate the BH mass from reverberation mapping data using the following equation: 
\begin{align} \label{eq:RM_BH_mass}
    M_{\text{BH}} = f\,c\,\tau\,\Delta v^2\,G^{-1}.
\end{align}
This estimate is based on the assumption that the gas dynamics are dominated by the central black hole. In our study, we use $\tau_{\text{cent}}$ with respect to the continuum at $5100$\,{\AA} (see Sect.~\ref{sec:1D_CCFs}) to estimate the mean distance of the origin of the variable line emission to the ionizing continuum source and the FWHM of the rms profiles to parametrise the mean velocities in the variable gas (see Tab.~\ref{tab:lineWidths}). The scaling factor $f$ accounts for the geometrical distribution of the gas to the line of sight and various values on the order of unity have been discussed in literature \citep[e.g.][]{2004ApJ...615..645O, 2011MNRAS.412.2211G, 2013ApJ...773...90G} for samples of AGN. The value for individual AGN, however, may differ from the values derived from these samples. The narrow width of the broad line components may originate from either a low-mass BH or a face-on geometry of WPVS\,48. A face-on geometry would likely lead to a high $f$ value \citep[e.g.][]{2001ApJ...551...72K, 2006MNRAS.373..551L, 2008MNRAS.387.1237D}. Therefore, we give only the virial product $c\,\tau\,\Delta v^2\,G^{-1}$ in this section and discuss the scaling factor $f$ and the BH mass in Sect.~\ref{sec:discussion_BH_mass_Ledd}.

The virial products for the individual emission lines based on both the FWHM and $\sigma_{\text{line}}$ and the lag to the continuum at $5100$\,{\AA} are given in Tab.~\ref{tab:BH_masses}. Here, we exclude \ion{He}{ii}\,$\lambda4686$ from the estimate, as the deblended line profile is difficult to obtain. The virial products based on the FWHM are consistent with each other, their weighted mean of the amounts to
\begin{align}
    c\,\tau\,\Delta v^2\,G^{-1} = (7.3_{-2.4}^{+5.8})\times10^6M_{\odot}.
\end{align}
The errors are propagated from both the high errors in the estimate of the line width as well as the lag from the CCF.

We, furthermore, used the mean continuum flux density of $F_{\lambda} = (2.68 \pm 0.22) \times 10^{-15}$\,erg\,cm$^{-2}$\,s$^{-1}$\,{\AA}$^{-1}$ at $5100$\,{\AA} to infer the optical luminosity of $\lambda L_{5100} = (4.74 \pm 0.39) \times 10^{43}$\,erg\,s$^{-1}$. Here, we adopt the bolometric correction factor of $\sim 19.9$ from \citet[][see their Equation 3 and Table 1]{2019MNRAS.488.5185N}
\begin{align}
    k_{\text{bol}} = 40 \times [\lambda L_{5100} / 10^{42}\,\text{erg}\,\text{s}^{-1}]^{-0.2},
\end{align}
and derive a bolometric luminosity of $L_{\text{bol}} = (8.76 \pm 0.72) \times 10^{44}$\,erg\,s$^{-1}$.

%**********************************************************************************
%
\section{Discussion}\label{sec:discussion}
%
%**********************************************************************************
%**********************************************************************************
%
\subsection{Optical variability}
%
%**********************************************************************************
%**********************************************************************************
%
\subsubsection{Comparison to the photometric campaign}\label{sec:discussion_intercalibration}
\begin{figure}[!t]
\centering
\includegraphics[width=0.47\textwidth,angle=0]{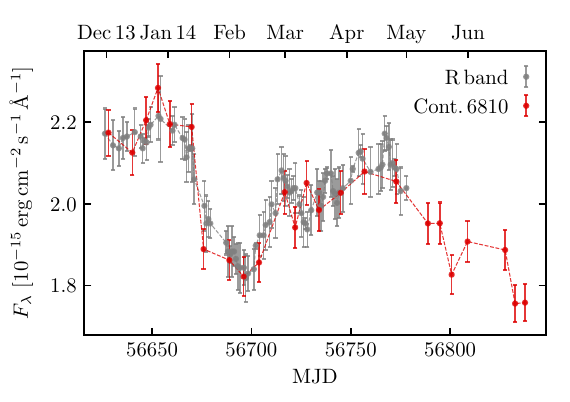}
\vspace*{-3mm} 
\caption{Comparison of the spectroscopic continuum light curve at $6810$\,{\AA} and the photometric R\,band light curve from \citet{2018rnls.confE..57S} (effective wavelength in the rest frame of WPVS\,48 $\lambda_{\text{eff}}=6750$\,{\AA}), shifted by $-1.57 \times 10^{-15}$\,erg\,cm$^{-2}$\,s$^{-1}$\,{\AA}$^{-1}$.} 
\label{fig:comparison_R_lightcurves}
\end{figure}

%
%**********************************************************************************
We compare the intercalibration of the spectroscopic data redwards of $5645$\,{\AA} with the independent calibration by \citet{2018rnls.confE..57S}. Figure~\ref{fig:comparison_R_lightcurves} shows the continuum light curve at $6810$\,{\AA} of this work as well as the R\,band light curve from \citet{2018rnls.confE..57S} after the subtraction of a constant flux density of $1.57 \times 10^{-15}$\,erg\,cm$^{-2}$\,s$^{-1}$\,{\AA}$^{-1}$. The continuum at $6810$\,{\AA} was chosen for the comparison, as it is closest to both the [\ion{S}{ii}]\,$\lambda\lambda6716,6731$ emission and the effective wavelength of the employed R\,band filter in the rest frame of WPVS\,48. We are able to reproduce the photometric light curve of \citet{2018rnls.confE..57S} well with our spectroscopic light curve. Most importantly, we are able to reproduce the variability amplitude without employing a multiplicative factor to our calibrated light curve. The necessary subtraction of flux density is likely be explained with the larger aperture with a diameter of of 7\arc5 that was used in the photometric observations.
%
%------------------------------------------------------------------------------
%
%
%------------------------------------------------------------------------------
%

%**********************************************************************************
%
\subsubsection{Spectral continuum slope}
%
%**********************************************************************************
We now test whether the spectral index of the optical continuum is consistent with the theoretical predictions from accretion disk models -- specifically, the spectral energy distribution of the UV/optical continuum from \citet{1973A&A....24..337S}, which is commonly used as first-order approximation in AGN \citep[e.g.][]{2007ApJ...668..682D, 2008Natur.454..492K}, although the details are more complex \citep[e.g.][and references therein]{2006ApJ...645.1402B, 2023Galax..11..102A}. The Shakura \& Sunyaev model predicts a flux density scaling with frequency like $F_{\nu} \propto \nu^{1/3}$ for the spectral region unaffected by contributions from the inner and outer AD that radiate in the UV and infrared, respectively. A spectral index of $\alpha = 1/3$ translates to a spectral slope of $\beta = - 7/3$, as $F_{\lambda} = F_{\nu}\text{d}\nu / \text{d}\lambda \sim \nu^{1/3}\nu^{2} \sim \lambda^{-7/3}$.

\begin{figure}
\centering
\includegraphics[width=0.47\textwidth,angle=0]{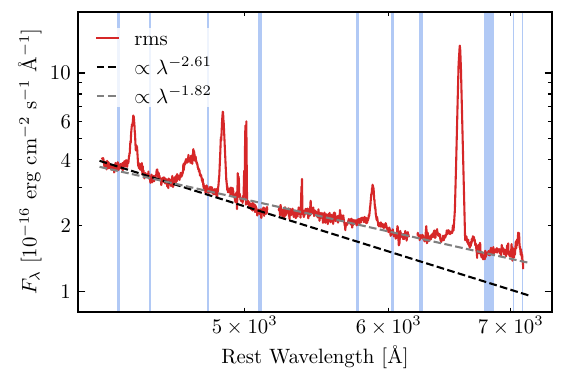}
\vspace*{-3mm} 
\caption{The rms spectrum of the SALT campaign corrected for Galactic foreground extinction (red). Two power-law models are fitted to the continuum slope. The first considers only the continua at $5100$\,{\AA} and bluewards giving a spectral index $\beta = 2.61 \pm 0.06$ (black), the second considers all measured continua giving a spectral index $\beta = 1.82 \pm 0.03$ (grey).}
\label{fig:continuum_slope}
\end{figure}

Fig.~\ref{fig:continuum_slope} shows a logarithmic plot of the observed rms spectrum corrected for Galactic foreground extinction. As the bivariate model for the interband continuum delays (see Sect.~\ref{sec:tau_alpha_formalism}) suggests a major secondary contributor other than the AD to the continuum at $5760$\,{\AA} and higher wavelengths, we fit the spectral continuum slope twice: The first fit considers the blue continua only, i.e. at $4265$\,{\AA}, $4400$\,{\AA}, $4775$\,{\AA} and $5100$\,{\AA}, whereas the second fit includes all continuum regions discussed in this work: The first fit gives $\beta = 2.61\pm0.06$ and is fairly consistent with the theoretical prediction of the Shakura \& Sunyaev model given that we test this relation over a limited wavelength range only. However, the inclusion of the continua at higher wavelengths results in a redder continuum slope, which yields $\beta = 1.86\pm0.02$. The break between the two power laws is close to H$\beta$ and and coincides with the jump in the measured univariate interband continuum delays. While the presence of a secondary continuum emitter is a possible explanation of this break in power law, we cannot exclude the host contribution being responsible for the break.

%
%------------------------------------------------------------------------------
%
%
%------------------------------------------------------------------------------
%

%**********************************************************************************
%
\subsubsection{Interband continuum delays} \label{sec:discussionContinuumLags}
%
%**********************************************************************************

In Sect.~\ref{sec:cont_CCFs_univariate}, we present interband continuum delays that increase with wavelength. Specifically, we calculate lags of $2$ -- $3$\,days between continua corresponding to the V and R\,band consistent with the lags $\lesssim 3$\,days discussed in other studies that include these wavelength bands \citep[e.g.][]{2016ApJ...821...56F, 2018ApJ...857...53C, 2019NatAs...3..251C, 2020MNRAS.498.5399H, 2021ApJ...922..151K, 2021MNRAS.504.4337V}{}{}. The calculated lag between the B and the R\,band of $\sim 7$\,days in this work is rather high compared to the lags of $\sim 1.5$ -- $4.1$ measured in other campaigns \citep[e.g.][]{2016ApJ...821...56F, 2020MNRAS.498.5399H, 2021ApJ...922..151K, 2021MNRAS.504.4337V, 2025ApJ...986..137Z}{}{}. Given the good agreement of our spectroscopic light curve with the photometric light curve of \citet{2018rnls.confE..57S} (see Sect.~\ref{sec:discussion_intercalibration}), we argue that the intercalibration performed in this work did not lead to the high lag. We, however, point out that the referenced studies that include observations in both the B and the R band are based on photometric broad-band data. The separation of continuum and line contribution is not straightforward in photometric data, whereas for the spectroscopic data of this work a separation of continuum and line flux is achieved. Therefore, the line contribution in the previous studies might have led to higher lags in the B band as discussed in \citet{2014ApJ...785..140C}{}{}, thereby decreasing the lag between the B and the R band.

In Sect.~\ref{sec:tau_alpha_formalism}, we probed the scenario of a secondary variable continuum emitter located further away to the central source that -- similar to the emission lines in the B band -- might result in longer delays: Such diffuse continuum emission from the BLR has been proposed for instance by \citet{2001ApJ...553..695K, 2019NatAs...3..251C}. We repeated the model calculations with various ICCF samplings of $0.5$, $2$ and $4$\,days. For continuum light curves whose univariate lag is comparable to the sampling of the campaign -- i.e. the continua at wavelengths $\geq 5100$\,{\AA} --  this test results in the same lags $\tau_{\text{sec}}$ and contribution factors $\alpha$ for the secondary component. However, the test struggles to find consistent delays for the continua with univariate delays far below the sampling, i.e. Cont.\,4440 and Cont\,4775. For this reason, we exclude the latter continua from the bivariate analysis. 

For the remaining continua, we are able to model the measured delays with a contribution of a secondary component with a lag of $11$ -- $15$\,days. This contribution increases with wavelength to up to $\sim 50$ percent. Intriguingly, the lag of the secondary continuum component matches the lags of H$\alpha$, H$\beta$, H$\gamma$ and \ion{He}{i}\,$\lambda5876$, pointing at a diffuse continuum component that originates in the Balmer BLR. In their continuum RM study on Mrk\,279, \citet{2019NatAs...3..251C} measured the continuum flux with intermediate-band filters ($\simeq 100$\,{\AA}-wide) at $4300$\,{\AA}, $5700$\,{\AA}, $6200$\,{\AA} and $7000$\,{\AA}. They interpreted the continuum delays with the bivariate model, however, the measured lag of the secondary continuum component in Mrk\,279 indicate an origin that is five times closer to the central source than the Balmer BLR. Furthermore, the contribution of the secondary component is higher throughout all examined intermediate-band filters ranging between $70$ and $90$ percent.

Comparing the two campaigns, we note a much higher sampling rate of $\lesssim 1$\,day for the observations of Mrk\,279. In case of WPVS\,48, lags that are five times smaller than the Balmer lags, i.e. $2$ -- $3$\,days, are in fact below the average sampling rate of our campaign. This hinders the ability of the bivariate model to distinguish two continuum signals $2$ -- $3$\,days apart delay-wise, as is described for the continua at $4440$\,{\AA} and $4775$\,{\AA} above. Furthermore, the light curves have a different variability pattern in both AGN: Mrk\,279 exhibits non-monotonic continuum variability on timescales of $\sim 5$\,days, whereas the continuum light curves of WPVS\,48 show non-monotonic continuum variability on timescales of $\sim 20$\,days. This observation of variability on different time scales cannot be accounted for the lower sampling rate of our campaign alone, as the same holds true for the higher-sampled photometric light curves of \citet[see Fig.~\ref{fig:comparison_R_lightcurves}]{2018rnls.confE..57S}. Thus, different dominant secondary continuum components may be observed due to different sampling rates or due to distinct variability behaviours of the AGN.

In addition to optical interband continuum lags, lags between the highly-ionizing continuum in the UV/X-ray and optical continua were measured in other galaxies: Specifically, the RM studies on NGC\,4151 \citep[][]{2019ApJ...870..123E}{}{}, NGC\,4593 \citep[][]{2018ApJ...857...53C, 2018MNRAS.480.2881M, 2019ApJ...870..123E}{}{}, NGC\,5548 \citep[][]{2015ApJ...806..129E, 2016ApJ...821...56F, 2019ApJ...870..123E}{}{}, Fairall\,9 \citep[][]{2020MNRAS.498.5399H, 2024ApJ...973..152E}{}{}, Mrk\,110 \citep[][]{2021MNRAS.504.4337V}{}{}, Mrk\,509 \citep[][]{2019ApJ...870..123E}{}{} and Mrk\,817 \citep[][]{2021ApJ...922..151K}{}{} include wavelength bands of the highly ionizing continuum in the UV or X-ray. They found typical lags between UV/X-ray and V\,band continua ranging from $0.48$ to $4.3$\,days.\footnote{NGC\,5548 exceeded the upper limit of the given range with a lag of $\sim 5.9$\,days between hard X-rays ($1.5$ -- $10$\,keV) and the V\,band \citep{2019ApJ...870..123E}.}

%**********************************************************************************
%
\subsubsection{Emission Line Variability}\label{sec:discussionLineVariability}
%
%**********************************************************************************

In Sect.~\ref{sec:1D_CCFs}, we deduce the BLR size for various emission lines from the CCFs with two optical continuum light curves. While the inferred sizes for the Balmer lines and the \ion{He}{i}\,$\lambda5876$ line (ranging between $\sim 11$ and $\sim 16$\,days) from the CCFs with both continua agree with each other, the CCFs for \ion{He}{ii}\,$\lambda4686$ result in BLR sizes that -- depending on the chosen continuum -- differ by a factor of $\sim 2$. Specifically, the derived BLR sizes for \ion{He}{ii}\,$\lambda4686$ range between $2$ and $5$\,days. Here, a similar argument can be made as with the continuum light curves in Sect.~\ref{sec:discussionContinuumLags}: As the found lag is below the sampling rate, the delays resultant from the cross correlations differ.

The \ion{He}{ii}\,$\lambda4686$ line is furthermore blended with \ion{N}{iii}\,$\lambda4640$ and \ion{C}{iv}\,$\lambda4658$. The extended blue wing of \ion{He}{ii}\,$\lambda4686$ points at variability in either or both of the blended lines. Variable \ion{N}{iii}\,$\lambda4640$ emission have been observed in flaring AGN so far \citep[][and references therein]{2019NatAs...3..242T, 2023ApJ...953...32M}. However, no flaring in WPVS\,48 is known to the authors. We hence show that the Bowen fluorescence line \ion{N}{iii}\,$\lambda4640$ is also present in non-outburst spectra.

%**********************************************************************************
%
\subsubsection{Balmer Decrement Variability}\label{sec:discussionBalmerVariability}
%
%**********************************************************************************
%
%------------------------------------------------------------------------------
%
\begin{figure}
\centering
\includegraphics[width=0.47\textwidth,angle=0]{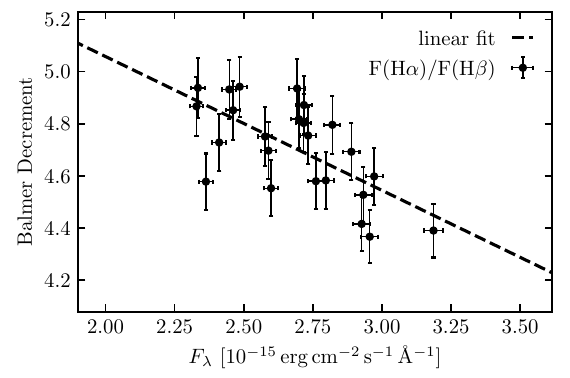}
\vspace*{-3mm} 
\caption{Balmer decrement, F(H$\alpha$)/F(H$\beta$), of the broad-line components versus continuum intensity at 5100\,{\AA}. The dashed line on the graph represents the linear regression.}
\label{fig:Balmer_decrement_Cont4440}
\end{figure}
%
%------------------------------------------------------------------------------
%
\begin{figure}
\centering
\includegraphics[width=0.47\textwidth,angle=0]{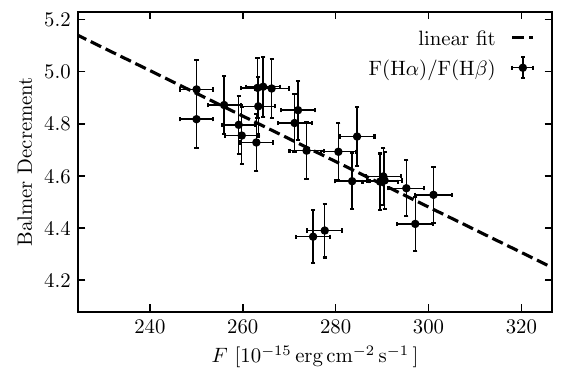}
\vspace*{-3mm} 
\caption{Balmer decrement, F(H$\alpha$)/F(H$\beta$), of the broad-line components versus broad-line H$\beta$ intensity. The dashed line on the graph represents the linear regression.}
\label{fig:Balmer_decrement_Hbeta}
\end{figure}
%
%------------------------------------------------------------------------------
%
We calculate the Balmer decrement H$\alpha$ / H$\beta$ for all epochs. Figures~\ref{fig:Balmer_decrement_Cont4440} and \ref{fig:Balmer_decrement_Hbeta} show the relation to the measured continuum flux density at 5100\,{\AA} and the H$\beta$ flux, respectively. We subtracted the flux of the narrow-line component before the evaluation of the Balmer decrement. The narrow-line fluxes ($17.8$\,erg\,cm$^{-2}$\,s$^{-1}$ and $39.3$\,erg\,cm$^{-2}$\,s$^{-1}$ for H$\beta$ and H$\alpha$, respectively) have been estimated with the help of a higher-resolved X-shooter spectrum (PI: Benny Trakhtenbrot, Program ID: 109.22YE) from 2022, i.e. 8 years after the SALT campaign. We thereby assume that the broad line profiles remain largely unchanged over this time period, which we test with an overplot of the line profiles from the X-shooter spectrum and the mean SALT spectrum. Both line profiles previously have been corrected for the narrow component using [\ion{O}{iii}]\,$\lambda5007$ as a template. We repeat the process with for the SALT spectrum employing different scaling factors for the narrow-line correction and conclude that for flux differences of $\pm 2.5$\,erg\,cm$^{-2}$\,s$^{-1}$ clear residuals of the narrow components emerge in the difference spectrum. We, therefore, adopt this as the uncertainty of the narrow component.

The values of the Balmer decrement are between $4.3$ and $5$ and show a roughly linear relation with the continuum flux density and the H$\beta$ flux, precisely, the Balmer decrement decreases with increasing flux (density). Similar Balmer decrements ranging between $\sim 3$ and $\sim 5$ have been measured in other AGN like, e.g. NGC\,7603 \citep[][]{2000A&A...361..901K}, HE\,1136-2304 \citep[][]{2018A&A...619A.168K}{}{} and Mrk\,926 \citep[][]{2022A&A...657A.122K}{}{}. The slope of the Balmer decrement in WPVS\,48 is similar to, for instance, that of NGC\,7603 in epochs of higher H$\beta$ flux ($F \gtrsim 250$\,erg\,cm$^{-2}$\,s$^{-1}$), whereas the slope in NGC\,7603 becomes steeper and the Balmer decrement reaches $\sim 7$ for epochs with lower H$\beta$ flux ($F \lesssim 250$\,erg\,cm$^{-2}$\,s$^{-1}$). \citet{2004ApJ...606..749K} simulated the responsivities of the optical emission lines prominent in AGN in relation to the incident flux and suggested that the found behaviour of the Balmer decrement as well as the apparent stratification of the BLR may be explained with optical depth effects.

%**********************************************************************************
%
\subsection{Emission Line Profiles}\label{sec:discussion_lineprofiles}
%
%**********************************************************************************

In Sect.~\ref{sec:optical_profiles_results}, we extract the broad rms profiles of the Balmer lines and \ion{He}{i}\,$\lambda5876$. The extracted rms profiles of H$\beta$, H$\alpha$ and \ion{He}{i}\,$\lambda5876$ measure almost equal line widths of $\sim 1500$\,km\,s$^{-1}$. While the bootstrap method employed to assess the uncertainties of the line widths accounts for noise in the line profiles, it, however, does not account for systematic biases from the presence of a potential narrow component. Since a robust line-profile decomposition is hard to achieve for NLS1s -- due to blending of the narrow and broad components --, we, instead, tested the influence of a possible narrow-line residuum in the upper 10\% of the rms peaks by measuring the width at the flux level of $0.45$ of the maximum. A narrow-line residuum on the order of 10\% of the peak flux would broaden the FWHM of the rms profiles by $\sim 130$\,km\,s$^{-1}$. Such a residuum, however, is not conspicuous in the rms profiles and we, therefore, conclude that such bias is smaller than $\sim 130$\,km\,s$^{-1}$ and the uncertainties already given in Sect.~\ref{sec:optical_profiles_results}.

The similar width suggests that all four lines originate from roughly the same distance to the central source, which is corroborated by the lag measurements in Sect.~\ref{sec:1D_CCFs}. Line widths and lags consistently reflect small differences in the distance from the central source, as H$\gamma$ and \ion{He}{i}\,$\lambda5876$ have both smaller lags and slightly higher widths than H$\beta$ and H$\alpha$. All four lines furthermore show a similar line shape, which includes the same feature of a slight redwards asymmetry in their rms peaks (upper $20\%$). A redwards asymmetry is furthermore present in the rms peak of \ion{He}{ii}\,$\lambda4686$. The appearance of this feature in all five lines suggests that this line feature is genuine and not a result from the reduction or intercalibration process. 

Redwards asymmetries have been observed in multiple AGN and have been reproduced using different models, for instance: \citet{2024A&A...686A..17O} fitted the double-peaked \ion{Ca}{II}\,$\lambda8662$ with a redwards asymmetry profile of NGC\,1566 with the model by \citet{1995ApJ...438..610E} of the emission of an elliptical accretion disk at an inclination angle of $i \approx (8.1\pm3.0)^{\circ}$. The single-peaked H$\beta$ profile showing a redwards asymmetry has then been replicated from the \ion{Ca}{II}\,$\lambda8662$ profile by adding turbulence. \citet{2003A&A...412L..61K} suggested gravitational redshift as an explanation for the redshifted rms peaks in the NLS1 Mrk\,110 and inferred an inclination angle of $i \approx (21\pm5)^{\circ}$. \citet{2012MNRAS.426.3086G} presented a series of line profiles expected from their disk model and linked redwards asymmetries to the combined effect of gravitational redshift, transverse Doppler shift and turbulence. Specifically, they concluded that these effects on line profiles are apparent only in near face-on systems ($i \lesssim 20^{\circ}$). While the combination of relatively narrow widths in the BLR components and a redwards asymmetry in the rms profile is consistent with a low inclination angle in these models, we note that there is no estimate of the inclination angle based on the presence of both features alone. The scaling factor $f$, therefore, is unconstrained by this observation and both the low-inclination and the low-mass scenario of the AGN in WPVS\,48 are equally possible.

We now infer (projected) rotational $v_{\text{rot}}$ and turbulent velocity components $v_{\text{turb}}$ in the BLR from the FWHM and the line dispersion $\sigma_{\text{line}}$ following the model of \citet{2011Natur.470..366K, 2013A&A...549A.100K}. The model parametrises rotationally broadened Lorentzian profiles, which are associated with turbulent motion \citep[see also][]{2012MNRAS.426.3086G}{}{} in terms of the FWHM and the FWHM / $\sigma_{\text{line}}$ ratio. This way, \citet{2011Natur.470..366K, 2013A&A...549A.100K} were able to identify characteristic turbulent velocities in various emission lines of AGN. Based on the rotational and turbulent velocities, we calculate the ratio of the height above the midplane to the radius of the line emitting regions $H/R$ as presented in \citet{2011Natur.470..366K, 2013A&A...549A.100K}
\begin{align}
    H/R = \alpha^{-1} (v_{\text{turb}} / v_{\text{rot}}),
\end{align}
where the viscosity parameter is assumed to be in the range between $0.1$ and $1$ \citep[e.g.][]{2002apa..book.....F}{}{}. For simplicity, we adopt $\alpha=1$. The calculated values for $v_{\text{turb}}$,  $v_{\text{rot}}$ and $H/R$ are given in Table~\ref{tab:rot_and_turb_velos}. We, however, note that the model of \citet{2011Natur.470..366K, 2013A&A...549A.100K} does not account for a low inclination angle or relativistic effects. In fact, \citet{2013A&A...549A.100K} point out that NLS1s tend to have lower FWHM while having FWHM / $\sigma_{\text{line}}$ ratios similar to broad-line Seyfert galaxies. A lower estimate of the line dispersion implies a lower turbulent velocity, and, thus, a lower scale height.
%
%------------------------------------------------------------------------------
%
\begin{table}[!t]
\centering
\caption{Inferred velocities of (projected) rotational and turbulent motion as well as scale height of the prominent emission lines.}
\begin{tabular}{lrrr}
\hline 
\hline
\noalign{\smallskip}
Line & \multicolumn{1}{c}{$v_{\text{rot}}$} & \multicolumn{1}{c}{$v_{\text{turb}}$} & \multicolumn{1}{c}{$H/R$} \\
 &  \multicolumn{1}{c}{[km\,s$^{-1}$]} & \multicolumn{1}{c}{[km\,s$^{-1}$]} &  \\
\multicolumn{1}{c}{(1)} & \multicolumn{1}{c}{(2)} & \multicolumn{1}{c}{(3)} & \multicolumn{1}{c}{(4)} \\
\hline 
\noalign{\smallskip}
H$\alpha$  & $925_{-80}^{+145}$ & $140_{-55}^{+80{\color{white}\_}}$ & $0.15$ \\
H$\beta$  & $950_{-135}^{+120}$ & $110_{-65}^{+90{\color{white}\_}}$ & $0.12$ \\
H$\gamma$  & $1165_{-135}^{+110}$ & $215_{-80}^{+150}$ & $0.18$ \\
\ion{He}{i}$\,\lambda$5876  & $965_{-105}^{+95}$ & $255_{-70}^{+85{{\color{white}\_}}}$ & $0.26$ \\
\hline 
\end{tabular} 
\tablefoot{The velocities were inferred based on the model of \citet{2011Natur.470..366K, 2013A&A...549A.100K}.}
\label{tab:rot_and_turb_velos}
\end{table}
%
%------------------------------------------------------------------------------
%

\subsection{Inclination Angle, Central Black Hole Mass and Eddington Ratio $L/L_{\text{Edd}}$} \label{sec:discussion_BH_mass_Ledd}

Since the discovery of NLS1s, the two scenarios of true low dispersion velocities in low-mass, high-Eddington systems and projected low dispersion velocities in low-inclination systems are primarily discussed to explain the narrow emission lines. Evidence have been found for both scenarios in different galaxies exhibiting narrow BLR lines: For instance, some surveys indicate that low-mass SMBHs accreting at a high Eddington rate are amongst the sub-class of NLS1 \citep[e.g.][and references therein]{2002ApJ...565...78B, 2004AJ....127.1799G, 2012AJ....143...83X}{}{}, but also low inclination angles have been inferred in the NLS1 Mrk\,110 \citep[][]{2003A&A...412L..61K} and in NGC\,1566\footnote{\citet{2024A&A...686A..17O} measured a FWHM of $\sim 2200$\,km\,s$^{-1}$ for H$\beta$ in NGC\,1566. While exhibiting relatively narrow BLR lines, NGC\,1566 formally is not a NLS1.} \citep[][]{2024A&A...686A..17O}. 

The principal argument for a low-inclination system in WPVS\,48 is given by \citet{2018rnls.confE..57S} in their analysis of a photometric RM campaign with various optical and IR filters, specifically, broad-band B, V, R, J and K filters as well as a narrow-band H$\alpha$ filter: They noted the sharp peak of the dust in the cross-correlation function of the IR light curves with the B\,band light curve in comparison with the smeared-out correlation function of H$\alpha$ and B\,band light curve. This may be interpreted as the quasi-simultaneous response of the upper layer of a bowl-shaped torus observed from almost face-on, as simultaneous responses should originate from the elliptic (or bowl-shaped) iso-delay surfaces. In Sect.~\ref{sec:discussion_lineprofiles}, we argue that the line profiles of WPVS\,48 are consistent with this scenario, however cannot differentiate from the low-mass scenario in the NLS1 WPVS\,48.

We, therefore, give mass estimates for both scenarios based on the virial products calculated in Sect.~\ref{sec:black_hole_mass_results}: In order to infer a BH mass, we account for the geometrical distribution and kinematics of the BLR gas by means of the scaling factor $f$. Various values for the scaling factor are discussed in literature, for instance $f=5.5$ \citep[][]{2004ApJ...615..645O}{}{}, $f=4.31$ \citep[][]{2013ApJ...773...90G}{}{} or $f=3.6$ \citep[][]{2011MNRAS.412.2211G}{}{}. 

For the estimate in the low-mass scenario, thereby assuming that inclination has little effect on the measured line width, we adopt the value of \citet{2011MNRAS.412.2211G}. We note that the referenced values for $f$ assume the line dispersion $\sigma_{\text{line}}$ as reparametrisation of the gas velocity. We, however, use the FWHM for the mass estimate and, thus, adapt the value of \citet{2011MNRAS.412.2211G} to $f = 1.8$ taking into account the ratio of FWHM/$\sigma_{\text{line}} \approx 2$ found by \citet{2004ApJ...613..682P}. This ratio of FWHM/$\sigma_{\text{line}}$ also applies to the measured widths of the rms profiles in WPVS\,48 (see Table~\ref{tab:lineWidths}). Employing this adapted scaling factor to the virial product of H$\beta$, the BH mass amounts to
\begin{align}
    M_{\text{BH,\,RM}} = (1.3_{-0.6}^{+1.1})\times10^7M_{\odot}.
\end{align}
We, furthermore, give the BH mass estimate deduced from the line dispersion $\sigma_{\text{line}}$ and the according scaling factor $f=3.6$ by \citet{2011MNRAS.412.2211G}{}{}, which amounts to $M_{\text{BH,\,RM}} = (5.8_{-1.8}^{+3.4})\times10^6M_{\odot}$.

This is in good agreement with the mass estimate following the $M_{\text{BH}}$ -- FWHM(H$\beta$)/$L_{5100}$ scaling relation of \citet[see their Equation\,5]{2006ApJ...641..689V}. Employing this scaling relation with the measured FWHM of the rms profile of H$\beta$ of $(1540\pm290)$\,km\,s$^{-1}$ (see Table~\ref{tab:lineWidths}) and the luminosity at $5100$\,{\AA} of $\lambda L_{5100} = 4.74 \times 10^{43}$\,erg\,s$^{-1}$ (see Sect.~\ref{sec:black_hole_mass_results}), we infer a BH mass of $M_{\text{BH,\,FWHM(H$\beta$),L$_{5100}$}} = (1.2_{-0.5}^{+0.6})\times10^7M_{\odot}$.

From the mass estimate of $M_{\text{BH,\,RM}} = (1.3_{-0.6}^{+1.1})\times10^7M_{\odot}$, we derive an Eddington luminosity of $L_{\text{Edd}} = 1.9\times10^{45}$\,erg\,s$^{-1}$. Given the derived bolometric luminosity of $L_{\text{bol}}=8.76 \times 10^{44}$\,erg\,s$^{-1}$ (see Sect.~\ref{sec:black_hole_mass_results}), this places the Eddington ratio at $L/L_{\text{Edd}} \approx 0.39$.

For the second scenario of WPVS\,48 observed face-on, the scaling factor $f$ can be higher than the value found by \citet{2011MNRAS.412.2211G}. In the simplest model of the BLR as a circular disk, the scaling factor scales with the inclination angle like $f \sim \sin^{-2} i$ \citep[][]{2001ApJ...551...72K, 2003A&A...412L..61K}. Following the scaling relation of $f \sim \sin^{-2} i$ and assuming an inclination angle between $5^{\circ}$ and $25^{\circ}$, we derive BH masses between $3.6\times10^7M_{\odot}$ and $8.5\times10^8M_{\odot}$. In near face-on systems, the BH mass therefore may be underestimated by up to $1.9$\,dex. The derived Eddington ratio in this scenario would be of only a few percent.

\subsection{Comparison with variability campaigns of other
(narrow-line) Seyfert galaxies}
%
%------------------------------------------------------------------------------
%
\begin{table*}[!t]
\centering
\caption{Key parameters of NLS1 galaxies with existing reverberation mapping data: Flux density (2) and variability amplitude (3) of the continuum at $5100$\,{\AA}, optical luminosity (4), FWHM of the rms profile of H$\beta$ (5), H$\beta$ lag to the optical continuum (6), BH mass (7) and Eddington luminosity (8).}
\begin{tabular}{lcccclccc}
\hline 
\hline
\noalign{\smallskip}
\multicolumn{1}{c}{NLS1 Galaxy} & \multicolumn{1}{c}{$F_{\text{5100}}$} & \multicolumn{1}{c}{$R_{\text{5100}}$} & \multicolumn{1}{c}{log $\lambda L_{5100}$} & \multicolumn{1}{c}{FWHM$_{\text{H}\beta}$} & \multicolumn{1}{c}{$\tau_{\text{H}\beta}$} & \multicolumn{1}{c}{log $M_{\text{BH}}$} & \multicolumn{1}{c}{$L/L_{\text{Edd}}$} & \multicolumn{1}{c}{Ref.} \\
% &  \multicolumn{1}{c}{} & \multicolumn{1}{c}{} & \multicolumn{1}{c}{[erg\,s$^{-1}$]} & \multicolumn{1}{c}{[km\,s$^{-1}$]} & \multicolumn{1}{c}{[days]} & \multicolumn{1}{c}{[$M_{\odot}$]} \\
\multicolumn{1}{c}{(1)} & \multicolumn{1}{c}{(2)} & \multicolumn{1}{c}{(3)} & \multicolumn{1}{c}{(4)} & \multicolumn{1}{c}{(5)} & \multicolumn{1}{c}{(6)} & \multicolumn{1}{c}{(7)} & \multicolumn{1}{c}{(8)} & \multicolumn{1}{c}{(9)} \\
\hline 
\noalign{\smallskip}
WPVS\,48   & $2.68$ & $1.37$ & $43.68$ & $1540$ & $15.0{\color{white}\_}^{+4.5}_{-1.9}$ & $7.04^{+0.12}_{-0.06}$ & $0.39^{+0.06}_{-0.10}$ & this work \\
Mrk\,110  & $2.79$ & $2.07$ & $43.62$ & $1515$ & $24.2{\color{white}\_}^{+3.7{\color{white}\_}}_{-3.3}$ & $7.29^{+0.07}_{-0.06}$ & $0.27^{+0.05}_{-0.04}$ & (a),\,(b) \\
Mrk\,335  & $5.84$ & $1.57$ & $43.68$ & $1025$ & $14.3{\color{white}\_}^{+0.7}_{-0.7}$ & $6.72^{+0.03}_{-0.03}$ & $1.12^{+0.06}_{-0.06}$ & (a),\,(c) \\
NGC\,4051 & $4.93$ & $1.69$ & $41.96$ & $1034$ & ${\color{white}\_}1.85^{+0.54}_{-0.50}$ & $5.84^{+0.12}_{-0.11}$ & $0.36^{+0.14}_{-0.09}$ & (a),\,(d) \\
NGC\,4748 & $1.15$ & $1.33$ & $42.49$ & $1212$ & ${\color{white}\_}6.30^{+1.82}_{-1.44}$ & $6.51^{+0.12}_{-0.09}$ & $0.20^{+0.06}_{-0.05}$ & (a),\,(e) \\
Ark\,564  & $5.96^{\ast}$ & $1.60$ & $43.56$ & ${\color{white}\_}960$ & ${\color{white}\_}3.56^{+27.44}_{-3.56}$ & $6.06^{+0.94}_{-0.31}$ & $4.1{\color{white}\_}_{-3.63}$ & (f) \\
SDSS\,J113913.91+335551  & - & $1.57$ & $42.85^{\dagger}$ & $1450$ & $12.5{\color{white}\_}^{+0.5}_{-1.5}$ & $6.97^{+0.02}_{-0.05}$ & $0.14^{+0.02}_{-0.01}$ & (g) \\
PG\,0934+013 & - & $1.53^{\ddagger}$ & $43.63$ & $1030$ & ${\color{white}\_}8.46^{+2.08}_{-2.14}$ & $6.50^{+0.10}_{-0.10}$ & $1.70^{+0.58}_{-0.34}$ & (h) \\
\hline 
\end{tabular} 
\tablefoot{Continuum flux densities are given in units of $10^{15}$\,erg\,cm$^{-2}$\,s$^{-1}$\,\AA$^{-1}$, optical luminosities are given in units of erg\,s$^{-1}$, FWHM are given in units of km\,s$^{-1}$, BH masses are given in units of $M_{\odot}$ and are inferred from H$\beta$ widths and lags employing uniformly the scaling factor of $f = 1.8$, thereby ignoring individual BLR geometries. Therefore, BH masses may be underestimated and Eddington ratios may be overestimated. References: (a) \citet{2013ApJ...767..149B}, (b) \citet{2001A&A...379..125K}, (c) \citet{2012ApJ...755...60G}, (d) \citet{2009ApJ...702.1353D}, (e) \citet{2009ApJ...705..199B}, (f) \citet{2012ApJS..202...10S}, (g) \citet{2013ApJ...773...24R}, (h) \citet{2017ApJ...847..125P}. $^{\ast}$Corrected for Galactic extinction adopting $A_\text{V} = 0.166$ from \citep{schlafly11} (NED). $^{\dagger}$Measured from an additional SDSS spectrum. $\ddagger$Variability amplitude taken from simultaneous photometric B\,band observations.}
\label{tab:NLS1_key_figures}
\end{table*}
%
%------------------------------------------------------------------------------
%
Several optical variability campaigns have been conducted on NLS1 galaxies, for instance on Mrk\,110 \citep[][]{2001A&A...379..125K, 2003A&A...407..461K}, NGC\,4051 \citep[][]{2009ApJ...702.1353D, 2017ApJ...840...97F}, NGC\,4748 \citep[][]{2009ApJ...705..199B, 2010ApJ...716..993B}, Ark\,564 \citep[][]{2012ApJS..202...10S}, SDSS\,J113913.91+335551 \citep[][]{2013ApJ...773...24R}{}{}, Mrk\,335 \citep[][]{2012ApJ...755...60G}, PG\,0934+013 \citep[][]{2017ApJ...847..125P}. We compile key parameters from these campaigns in Table~\ref{tab:NLS1_key_figures}. These parameters are the mean flux density $F_{5100}$ and the variability amplitude $R_{5100}$ of the optical continuum, the optical luminosity $\lambda F_{5100}$, the measured FWHM of the rms profile of H$\beta$ and the lag of H$\beta$ to the optical continuum $\tau_{\text{H}\beta}$. Optical continuum flux densities corrected for galactic foreground extinction are taken from \citet{2013ApJ...767..149B} where applicable.

We furthermore deduce homogeneously the BH mass and the Eddington ratio from these key parameters as is done in \ref{sec:discussion_BH_mass_Ledd}. For this comparison, we disregard individual geometries and adopt always a scaling factor of $1.8$ as well as the bolometric correction factor by \citet[][see \ref{sec:black_hole_mass_results}]{2019MNRAS.488.5185N}. Therefore, derived BH masses in Table~\ref{tab:NLS1_key_figures} are likely to be underestimated and Eddington ratios are likely to be overestimated. The errors of the BH masses and Eddington ratios are propagated from the uncertainties of the H$\beta$ lag. Note that the estimated error for Ark\,564 is large as several gaps of $\sim 3$\,months interrupted the variability campaign.

\begin{figure}[!t]
\centering
\includegraphics[width=0.47\textwidth,angle=0]{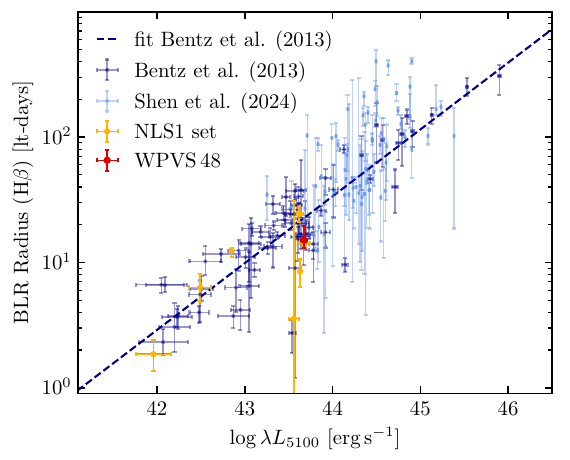}
\vspace*{-3mm} 
\caption{Optical continuum luminosity and H$\beta$ lags for WPVS\,48 and other AGN. Included are the data sets of \citet{2013ApJ...767..149B} (dark blue) and \citet{2024ApJS..272...26S} (light blue) as well as the B-L relation $\text{log} (R_{\text{BLR}} / 1 \text{lt-day}) = 1.527 + 0.533 \text{log} (\lambda L_{5100} / 10^{44} L_{\odot})$ found by \citet{2013ApJ...767..149B}. Furthermore, we highlighted a set of NLS1s, namely Mrk\,110 \citep[][]{2001A&A...379..125K, 2003A&A...407..461K}, Mrk\,335 \citep[][]{2012ApJ...755...60G}, NGC\,4051 \citep[][]{2009ApJ...702.1353D, 2017ApJ...840...97F}, NGC\,4748 \citep[][]{2009ApJ...705..199B, 2010ApJ...716..993B}, SDSS\,J113913.91+335551 \citep[][]{2013ApJ...773...24R}{}{}, Ark\,564 \citep[][]{2012ApJS..202...10S} and PG\,0934+013 \citep[][]{2017ApJ...847..125P} in orange.}
\label{fig:R-L-relation}
\end{figure}

Fig.~\ref{fig:R-L-relation} compares the measured optical luminosity $L_{5100}$ and the size of the Balmer BLR of WPVS\,48 with those inferred from other variability campaigns on both NLS1s and other AGN in the samples of \citet{2013ApJ...767..149B} and \citet{2024ApJS..272...26S}. In particular, we highlight the NLS1s from above for which optical luminosities and BLR sizes have been estimated. WPVS\,48 is in good agreement with the R-L relation found by \citet{2013ApJ...767..149B}.
%
%------------------------------------------------------------------------------
%
%
%------------------------------------------------------------------------------
%

WPVS\,48 is similar in terms of BLR size, luminosity and derived BH mass to the NLS1 galaxies Mrk\,110, Mrk\,335, PG\,0934+013 and SDSS\,J113913.91+335551. The optical luminosities and the BLR radii of the sample of reverberation-mapped NLS1 galaxies fall between the lower end and the middle of the R-L range of the AGN samples of \citet{2013ApJ...767..149B} and \citet{2024ApJS..272...26S}. The optical variability amplitude of $1.37$ in WPVS\,48 is similar to typical variability amplitudes observed in other AGN -- NLS1 galaxies and broad-line Seyfert galaxies alike. For most AGN, the variability amplitudes range from $\sim 1.2$ to $\sim 2$ \citep[e.g.][]{1998ApJ...501...82P, 2009ApJ...705..199B, 2012ApJ...755...60G}.

WPVS\,48 furthermore shares some of the spectral and variability features with the other NLS1s for which optical mean and rms spectra are published: \ion{Fe}{ii} emission is present in the mean spectra of WPVS\,48, Mrk\,110 \citep[][see their Fig.~5]{2001A&A...379..125K}, Mrk\,335 \citep[][see their Fig.~1]{2012ApJ...755...60G}, NGC\,4748 \citep[][see their Fig.~7]{2010ApJ...716..993B} and NGC\,4051 \citep[][see their Fig.~4]{2017ApJ...840...97F}, however, \ion{Fe}{ii} seems to be absent or very weak in the corresponding rms spectra (see the same figures except for Mrk\,110, there see Fig.~7), which indicates low variability in the \ion{Fe}{ii} lines. Moreover, we identified several low-intensity \ion{He}{i} lines in both the mean and rms spectrum of WPVS\,48 as was done in Mrk\,110.

%**********************************************************************************
%
\section{Conclusions}\label{sec:conclusions}
%
%**********************************************************************************
We performed reverberation mapping of both continuum and emission lines 
in WPVS\,48, using the SALT 10\,m telescope. In conclusion, our key findings can be summarized as follows:

\begin{enumerate}
    \item The BLR of WPVS\,48 is stratified with respect to the distance of the line emitting regions. The integrated emission line intensities of H$\alpha$, H$\beta$, H$\gamma$ and \ion{He}{i}\,$\lambda5876$ originate at distances of $16.0^{+4.0}_{-2.0}$, 15.0$^{+4.5}_{-1.9}$, $12.5^{+3.5}_{-2.5}$ and $14.0^{+2.5}_{-2.1}$ light-days with respect to the optical continuum at 5100\,\AA{}. The \ion{He}{ii}\,$\lambda 4686$ lag of $\lesssim 5$\,days is not perfectly-resolved with the minimum sampling of $5$\,days.
    \item WPVS\,48 shows variability similar to other NLS1s and varied by a factor of $\sim 1.4$ in the optical continuum, by a factor of $\sim 1.2$ in the Balmer lines, by a factor of $\sim 1.4$ in \ion{He}{i}\,$\lambda5876$ and by a factor of $\sim 1.7$ in \ion{He}{ii}\,$\lambda4686$. The variability amplitudes of the integrated emission lines decrease in the stratified BLR of WPVS\,48 with distance to the ionizing continuum source. The Balmer decrement in WPVS\,48 furthermore increases as a function of decreasing continuum luminosity.
    \item We identify variable emission of the Bowen line \ion{N}{iii}\,$\lambda4640$ in the mean and rms spectrum of WPVS\,48.
    \item We derive interband continuum lags increasing with wavelength in WPVS\,48. Assuming two continuum components, we obtain coherent lags of $\sim 15$\,days for the secondary component. We, therefore, suggest a continuum source that is co-located with the Balmer-emitting region.
    \item We derive a central BH mass of $(1.3_{-0.6}^{+1.1})\times10^7M_{\odot}$ and an Eddington ratio of $L/L_{\text{Edd}} \approx 0.39$ based on the lag as well as the width of the H$\beta$ rms profile. We address a potential bias due to the presence of a narrow-line H$\beta$ residuum. Such narrow-line residuum, however, is not conspicuous in the rms profile. We, furthermore, discuss the correction for a near face-on geometry as suggested by \citet{2018rnls.confE..57S}.
\end{enumerate}

Very few NLS1 galaxies have been subject to a detailed analysis by means of reverberation mapping campaigns so far. We now have contributed key parameters of the BLR for one specimen of this AGN class.

\begin{acknowledgements}
The authors greatly acknowledge support by the DFG grants KO 857/35-1 and KO 857/35-2. MWO gratefully acknowledges the support German Aerospace Center (DLR) within the framework of the `Verbundforschung Astronomie und Astrophysik' through grant 50OR2305 with funds from the German Federal Ministry for Economic Affairs and Climate Action (BMWK).

The spectra reported in this paper were obtained with the Southern African Large Telescope (SALT under proposal code 2013-2-GU-001, PI: Kollatschny).

\end{acknowledgements}

\bibliographystyle{aa} % style aa.bst
\bibliography{literature} % your references Yourfile.bib

\begin{thebibliography}{96}
\expandafter\ifx\csname natexlab\endcsname\relax\def\natexlab#1{#1}\fi

\bibitem[{{Antonucci}(2023)}]{2023Galax..11..102A}
{Antonucci}, R. R.~J. 2023, Galaxies, 11, 102

\bibitem[{{Bentz} {et~al.}(2013){Bentz}, {Denney}, {Grier}, {Barth},
  {Peterson}, {Vestergaard}, {Bennert}, {Canalizo}, {De Rosa}, {Filippenko},
  {Gates}, {Greene}, {Li}, {Malkan}, {Pogge}, {Stern}, {Treu}, \&
  {Woo}}]{2013ApJ...767..149B}
{Bentz}, M.~C., {Denney}, K.~D., {Grier}, C.~J., {et~al.} 2013, \apj, 767, 149

\bibitem[{{Bentz} {et~al.}(2009){Bentz}, {Walsh}, {Barth}, {Baliber},
  {Bennert}, {Canalizo}, {Filippenko}, {Ganeshalingam}, {Gates}, {Greene},
  {Hidas}, {Hiner}, {Lee}, {Li}, {Malkan}, {Minezaki}, {Sakata}, {Serduke},
  {Silverman}, {Steele}, {Stern}, {Street}, {Thornton}, {Treu}, {Wang}, {Woo},
  \& {Yoshii}}]{2009ApJ...705..199B}
{Bentz}, M.~C., {Walsh}, J.~L., {Barth}, A.~J., {et~al.} 2009, \apj, 705, 199

\bibitem[{{Bentz} {et~al.}(2010){Bentz}, {Walsh}, {Barth}, {Yoshii}, {Woo},
  {Wang}, {Treu}, {Thornton}, {Street}, {Steele}, {Silverman}, {Serduke},
  {Sakata}, {Minezaki}, {Malkan}, {Li}, {Lee}, {Hiner}, {Hidas}, {Greene},
  {Gates}, {Ganeshalingam}, {Filippenko}, {Canalizo}, {Bennert}, \&
  {Baliber}}]{2010ApJ...716..993B}
{Bentz}, M.~C., {Walsh}, J.~L., {Barth}, A.~J., {et~al.} 2010, \apj, 716, 993

\bibitem[{{Blaes} {et~al.}(2006){Blaes}, {Davis}, {Hirose}, {Krolik}, \&
  {Stone}}]{2006ApJ...645.1402B}
{Blaes}, O.~M., {Davis}, S.~W., {Hirose}, S., {Krolik}, J.~H., \& {Stone},
  J.~M. 2006, \apj, 645, 1402

\bibitem[{{Blandford} \& {McKee}(1982)}]{blandford82}
{Blandford}, R.~D. \& {McKee}, C.~F. 1982, \apj, 255, 419

\bibitem[{{Boroson}(2002)}]{2002ApJ...565...78B}
{Boroson}, T.~A. 2002, \apj, 565, 78

\bibitem[{{Brown} {et~al.}(2017){Brown}, {Holoien}, {Auchettl}, {Stanek},
  {Kochanek}, {Shappee}, {Prieto}, \& {Grupe}}]{2017MNRAS.466.4904B}
{Brown}, J.~S., {Holoien}, T.~W.~S., {Auchettl}, K., {et~al.} 2017, \mnras,
  466, 4904

\bibitem[{{Brown} {et~al.}(2018){Brown}, {Kochanek}, {Holoien}, {Stanek},
  {Auchettl}, {Shappee}, {Prieto}, {Morrell}, {Falco}, {Strader}, {Chomiuk},
  {Post}, {Villanueva}, {Mathur}, {Dong}, {Chen}, \&
  {Bose}}]{2018MNRAS.473.1130B}
{Brown}, J.~S., {Kochanek}, C.~S., {Holoien}, T.~W.~S., {et~al.} 2018, \mnras,
  473, 1130

\bibitem[{{Cackett} {et~al.}(2021){Cackett}, {Bentz}, \&
  {Kara}}]{2021iSci...24j2557C}
{Cackett}, E.~M., {Bentz}, M.~C., \& {Kara}, E. 2021, iScience, 24, 102557

\bibitem[{{Cackett} {et~al.}(2018){Cackett}, {Chiang}, {McHardy}, {Edelson},
  {Goad}, {Horne}, \& {Korista}}]{2018ApJ...857...53C}
{Cackett}, E.~M., {Chiang}, C.-Y., {McHardy}, I., {et~al.} 2018, \apj, 857, 53

\bibitem[{{Cardelli} {et~al.}(1989){Cardelli}, {Clayton}, \&
  {Mathis}}]{cardelli89}
{Cardelli}, J.~A., {Clayton}, G.~C., \& {Mathis}, J.~S. 1989, \apj, 345, 245

\bibitem[{{Chelouche} {et~al.}(2019){Chelouche}, {Pozo Nu{\~n}ez}, \&
  {Kaspi}}]{2019NatAs...3..251C}
{Chelouche}, D., {Pozo Nu{\~n}ez}, F., \& {Kaspi}, S. 2019, Nature Astronomy,
  3, 251

\bibitem[{{Chelouche} {et~al.}(2014){Chelouche}, {Shemmer}, {Cotlier}, {Barth},
  \& {Rafter}}]{2014ApJ...785..140C}
{Chelouche}, D., {Shemmer}, O., {Cotlier}, G.~I., {Barth}, A.~J., \& {Rafter},
  S.~E. 2014, \apj, 785, 140

\bibitem[{{Chelouche} \& {Zucker}(2013)}]{2013ApJ...769..124C}
{Chelouche}, D. \& {Zucker}, S. 2013, \apj, 769, 124

\bibitem[{{Choloniewski}(1981)}]{1981AcA....31..293C}
{Choloniewski}, J. 1981, \actaa, 31, 293

\bibitem[{{Collier} {et~al.}(1998){Collier}, {Horne}, {Kaspi}, {Netzer},
  {Peterson}, {Wanders}, {Alexander}, {Bertram}, {Comastri}, {Gaskell},
  {Malkov}, {Maoz}, {Mignoli}, {Pogge}, {Pronik}, {Sergeev}, {Snedden},
  {Stirpe}, {Bochkarev}, {Burenkov}, {Shapovalova}, \&
  {Wagner}}]{1998ApJ...500..162C}
{Collier}, S.~J., {Horne}, K., {Kaspi}, S., {et~al.} 1998, \apj, 500, 162

\bibitem[{{Davis} {et~al.}(2007){Davis}, {Woo}, \&
  {Blaes}}]{2007ApJ...668..682D}
{Davis}, S.~W., {Woo}, J.-H., \& {Blaes}, O.~M. 2007, \apj, 668, 682

\bibitem[{{Decarli} {et~al.}(2008){Decarli}, {Labita}, {Treves}, \&
  {Falomo}}]{2008MNRAS.387.1237D}
{Decarli}, R., {Labita}, M., {Treves}, A., \& {Falomo}, R. 2008, \mnras, 387,
  1237

\bibitem[{{Denney} {et~al.}(2009){Denney}, {Watson}, {Peterson}, {Pogge},
  {Atlee}, {Bentz}, {Bird}, {Brokofsky}, {Comins}, {Dietrich}, {Doroshenko},
  {Eastman}, {Efimov}, {Gaskell}, {Hedrick}, {Klimanov}, {Klimek}, {Kruse},
  {Lamb}, {Leighly}, {Minezaki}, {Nazarov}, {Petersen}, {Peterson},
  {Poindexter}, {Schlesinger}, {Sakata}, {Sergeev}, {Tobin}, {Unterborn},
  {Vestergaard}, {Watkins}, \& {Yoshii}}]{2009ApJ...702.1353D}
{Denney}, K.~D., {Watson}, L.~C., {Peterson}, B.~M., {et~al.} 2009, \apj, 702,
  1353

\bibitem[{{Edelson} {et~al.}(2019){Edelson}, {Gelbord}, {Cackett}, {Peterson},
  {Horne}, {Barth}, {Starkey}, {Bentz}, {Brandt}, {Goad}, {Joner}, {Korista},
  {Netzer}, {Page}, {Uttley}, {Vaughan}, {Breeveld}, {Cenko}, {Done}, {Evans},
  {Fausnaugh}, {Ferland}, {Gonzalez-Buitrago}, {Gropp}, {Grupe}, {Kaastra},
  {Kennea}, {Kriss}, {Mathur}, {Mehdipour}, {Mudd}, {Nousek}, {Schmidt},
  {Vestergaard}, \& {Villforth}}]{2019ApJ...870..123E}
{Edelson}, R., {Gelbord}, J., {Cackett}, E., {et~al.} 2019, \apj, 870, 123

\bibitem[{{Edelson} {et~al.}(2015){Edelson}, {Gelbord}, {Horne}, {McHardy},
  {Peterson}, {Ar{\'e}valo}, {Breeveld}, {De Rosa}, {Evans}, {Goad}, {Kriss},
  {Brandt}, {Gehrels}, {Grupe}, {Kennea}, {Kochanek}, {Nousek}, {Papadakis},
  {Siegel}, {Starkey}, {Uttley}, {Vaughan}, {Young}, {Barth}, {Bentz},
  {Brewer}, {Crenshaw}, {Dalla Bont{\`a}}, {De Lorenzo-C{\'a}ceres}, {Denney},
  {Dietrich}, {Ely}, {Fausnaugh}, {Grier}, {Hall}, {Kaastra}, {Kelly},
  {Korista}, {Lira}, {Mathur}, {Netzer}, {Pancoast}, {Pei}, {Pogge},
  {Schimoia}, {Treu}, {Vestergaard}, {Villforth}, {Yan}, \&
  {Zu}}]{2015ApJ...806..129E}
{Edelson}, R., {Gelbord}, J.~M., {Horne}, K., {et~al.} 2015, \apj, 806, 129

\bibitem[{{Edelson} {et~al.}(2024){Edelson}, {Peterson}, {Gelbord}, {Horne},
  {Goad}, {McHardy}, {Vaughan}, \& {Vestergaard}}]{2024ApJ...973..152E}
{Edelson}, R., {Peterson}, B.~M., {Gelbord}, J., {et~al.} 2024, \apj, 973, 152

\bibitem[{{Eracleous} {et~al.}(1995){Eracleous}, {Livio}, {Halpern}, \&
  {Storchi-Bergmann}}]{1995ApJ...438..610E}
{Eracleous}, M., {Livio}, M., {Halpern}, J.~P., \& {Storchi-Bergmann}, T. 1995,
  \apj, 438, 610

\bibitem[{{Fausnaugh} {et~al.}(2016){Fausnaugh}, {Denney}, {Barth}, {Bentz},
  {Bottorff}, {Carini}, {Croxall}, {De Rosa}, {Goad}, {Horne}, {Joner},
  {Kaspi}, {Kim}, {Klimanov}, {Kochanek}, {Leonard}, {Netzer}, {Peterson},
  {Schn{\"u}lle}, {Sergeev}, {Vestergaard}, {Zheng}, {Zu}, {Anderson},
  {Ar{\'e}valo}, {Bazhaw}, {Borman}, {Boroson}, {Brandt}, {Breeveld}, {Brewer},
  {Cackett}, {Crenshaw}, {Dalla Bont{\`a}}, {De Lorenzo-C{\'a}ceres},
  {Dietrich}, {Edelson}, {Efimova}, {Ely}, {Evans}, {Filippenko}, {Flatland},
  {Gehrels}, {Geier}, {Gelbord}, {Gonzalez}, {Gorjian}, {Grier}, {Grupe},
  {Hall}, {Hicks}, {Horenstein}, {Hutchison}, {Im}, {Jensen}, {Jones},
  {Kaastra}, {Kelly}, {Kennea}, {Kim}, {Korista}, {Kriss}, {Lee}, {Lira},
  {MacInnis}, {Manne-Nicholas}, {Mathur}, {McHardy}, {Montouri}, {Musso},
  {Nazarov}, {Norris}, {Nousek}, {Okhmat}, {Pancoast}, {Papadakis}, {Parks},
  {Pei}, {Pogge}, {Pott}, {Rafter}, {Rix}, {Saylor}, {Schimoia}, {Siegel},
  {Spencer}, {Starkey}, {Sung}, {Teems}, {Treu}, {Turner}, {Uttley},
  {Villforth}, {Weiss}, {Woo}, {Yan}, \& {Young}}]{2016ApJ...821...56F}
{Fausnaugh}, M.~M., {Denney}, K.~D., {Barth}, A.~J., {et~al.} 2016, \apj, 821,
  56

\bibitem[{{Fausnaugh} {et~al.}(2017){Fausnaugh}, {Grier}, {Bentz}, {Denney},
  {De Rosa}, {Peterson}, {Kochanek}, {Pogge}, {Adams}, {Barth}, {Beatty},
  {Bhattacharjee}, {Borman}, {Boroson}, {Bottorff}, {Brown}, {Brown},
  {Brotherton}, {Coker}, {Crawford}, {Croxall}, {Eftekharzadeh}, {Eracleous},
  {Joner}, {Henderson}, {Holoien}, {Horne}, {Hutchison}, {Kaspi}, {Kim},
  {King}, {Li}, {Lochhaas}, {Ma}, {MacInnis}, {Manne-Nicholas}, {Mason},
  {Montuori}, {Mosquera}, {Mudd}, {Musso}, {Nazarov}, {Nguyen}, {Okhmat},
  {Onken}, {Ou-Yang}, {Pancoast}, {Pei}, {Penny}, {Poleski}, {Rafter},
  {Romero-Colmenero}, {Runnoe}, {Sand}, {Schimoia}, {Sergeev}, {Shappee},
  {Simonian}, {Somers}, {Spencer}, {Starkey}, {Stevens}, {Tayar}, {Treu},
  {Valenti}, {Van Saders}, {Villanueva}, {Villforth}, {Weiss}, {Winkler}, \&
  {Zhu}}]{2017ApJ...840...97F}
{Fausnaugh}, M.~M., {Grier}, C.~J., {Bentz}, M.~C., {et~al.} 2017, \apj, 840,
  97

\bibitem[{{Frank} {et~al.}(2002){Frank}, {King}, \&
  {Raine}}]{2002apa..book.....F}
{Frank}, J., {King}, A., \& {Raine}, D.~J. 2002, {Accretion Power in
  Astrophysics: Third Edition}

\bibitem[{{Fromerth} \& {Melia}(2000)}]{2000ApJ...533..172F}
{Fromerth}, M.~J. \& {Melia}, F. 2000, \apj, 533, 172

\bibitem[{{Gaskell} \& {Peterson}(1987)}]{1987ApJS...65....1G}
{Gaskell}, C.~M. \& {Peterson}, B.~M. 1987, \apjs, 65, 1

\bibitem[{{Gaskell} {et~al.}(2022){Gaskell}, {Thakur}, {Tian}, \&
  {Saravanan}}]{2022AN....34310112G}
{Gaskell}, M., {Thakur}, N., {Tian}, B., \& {Saravanan}, A. 2022, Astronomische
  Nachrichten, 343, e210112

\bibitem[{{Gezari} {et~al.}(2015){Gezari}, {Chornock}, {Lawrence}, {Rest},
  {Jones}, {Berger}, {Challis}, \& {Narayan}}]{2015ApJ...815L...5G}
{Gezari}, S., {Chornock}, R., {Lawrence}, A., {et~al.} 2015, \apjl, 815, L5

\bibitem[{{Goad} {et~al.}(2012){Goad}, {Korista}, \&
  {Ruff}}]{2012MNRAS.426.3086G}
{Goad}, M.~R., {Korista}, K.~T., \& {Ruff}, A.~J. 2012, \mnras, 426, 3086

\bibitem[{{Goodrich}(1989)}]{1989ApJ...342..224G}
{Goodrich}, R.~W. 1989, \apj, 342, 224

\bibitem[{{Graham} {et~al.}(2011){Graham}, {Onken}, {Athanassoula}, \&
  {Combes}}]{2011MNRAS.412.2211G}
{Graham}, A.~W., {Onken}, C.~A., {Athanassoula}, E., \& {Combes}, F. 2011,
  \mnras, 412, 2211

\bibitem[{{Grier} {et~al.}(2013{\natexlab{a}}){Grier}, {Martini}, {Watson},
  {Peterson}, {Bentz}, {Dasyra}, {Dietrich}, {Ferrarese}, {Pogge}, \&
  {Zu}}]{2013ApJ...773...90G}
{Grier}, C.~J., {Martini}, P., {Watson}, L.~C., {et~al.} 2013{\natexlab{a}},
  \apj, 773, 90

\bibitem[{{Grier} {et~al.}(2013{\natexlab{b}}){Grier}, {Peterson}, {Horne},
  {Bentz}, {Pogge}, {Denney}, {De Rosa}, {Martini}, {Kochanek}, {Zu},
  {Shappee}, {Siverd}, {Beatty}, {Sergeev}, {Kaspi}, {Araya Salvo}, {Bird},
  {Bord}, {Borman}, {Che}, {Chen}, {Cohen}, {Dietrich}, {Doroshenko}, {Efimov},
  {Free}, {Ginsburg}, {Henderson}, {King}, {Mogren}, {Molina}, {Mosquera},
  {Nazarov}, {Okhmat}, {Pejcha}, {Rafter}, {Shields}, {Skowron}, {Szczygiel},
  {Valluri}, \& {van Saders}}]{2013ApJ...764...47G}
{Grier}, C.~J., {Peterson}, B.~M., {Horne}, K., {et~al.} 2013{\natexlab{b}},
  \apj, 764, 47

\bibitem[{{Grier} {et~al.}(2012){Grier}, {Peterson}, {Pogge}, {Denney},
  {Bentz}, {Martini}, {Sergeev}, {Kaspi}, {Minezaki}, {Zu}, {Kochanek},
  {Siverd}, {Shappee}, {Stanek}, {Araya Salvo}, {Beatty}, {Bird}, {Bord},
  {Borman}, {Che}, {Chen}, {Cohen}, {Dietrich}, {Doroshenko}, {Drake},
  {Efimov}, {Free}, {Ginsburg}, {Henderson}, {King}, {Koshida}, {Mogren},
  {Molina}, {Mosquera}, {Nazarov}, {Okhmat}, {Pejcha}, {Rafter}, {Shields},
  {Skowron}, {Szczygiel}, {Valluri}, \& {van Saders}}]{2012ApJ...755...60G}
{Grier}, C.~J., {Peterson}, B.~M., {Pogge}, R.~W., {et~al.} 2012, \apj, 755, 60

\bibitem[{{Grupe}(2004)}]{2004AJ....127.1799G}
{Grupe}, D. 2004, \aj, 127, 1799

\bibitem[{{Haas} {et~al.}(2011){Haas}, {Chini}, {Ramolla}, {Pozo Nu{\~n}ez},
  {Westhues}, {Watermann}, {Hoffmeister}, \& {Murphy}}]{2011A&A...535A..73H}
{Haas}, M., {Chini}, R., {Ramolla}, M., {et~al.} 2011, \aap, 535, A73

\bibitem[{{Hern{\'a}ndez Santisteban} {et~al.}(2020){Hern{\'a}ndez
  Santisteban}, {Edelson}, {Horne}, {Gelbord}, {Barth}, {Cackett}, {Goad},
  {Netzer}, {Starkey}, {Uttley}, {Brandt}, {Korista}, {Lohfink}, {Onken},
  {Page}, {Siegel}, {Vestergaard}, {Bisogni}, {Breeveld}, {Cenko}, {Dalla
  Bont{\`a}}, {Evans}, {Ferland}, {Gonzalez-Buitrago}, {Grupe}, {Joner},
  {Kriss}, {LaPorte}, {Mathur}, {Marshall}, {Mehdipour}, {Mudd}, {Peterson},
  {Schmidt}, {Vaughan}, \& {Valenti}}]{2020MNRAS.498.5399H}
{Hern{\'a}ndez Santisteban}, J.~V., {Edelson}, R., {Horne}, K., {et~al.} 2020,
  \mnras, 498, 5399

\bibitem[{{Kara} {et~al.}(2021){Kara}, {Mehdipour}, {Kriss}, {Cackett}, {Arav},
  {Barth}, {Byun}, {Brotherton}, {De Rosa}, {Gelbord}, {Hern{\'a}ndez
  Santisteban}, {Hu}, {Kaastra}, {Landt}, {Li}, {Miller}, {Montano},
  {Partington}, {Aceituno}, {Bai}, {Bao}, {Bentz}, {Brink}, {Chelouche},
  {Chen}, {Colmenero}, {Dalla Bont{\`a}}, {Dehghanian}, {Du}, {Edelson},
  {Ferland}, {Ferrarese}, {Fian}, {Filippenko}, {Fischer}, {Goad},
  {Gonz{\'a}lez Buitrago}, {Gorjian}, {Grier}, {Guo}, {Hall}, {Ho},
  {Homayouni}, {Horne}, {Ili{\'c}}, {Jiang}, {Joner}, {Kaspi}, {Kochanek},
  {Korista}, {Kynoch}, {Li}, {Liu}, {McHardy}, {McLane}, {Mitchell}, {Netzer},
  {Olson}, {Pogge}, {Popovi{\'c}}, {Proga}, {Storchi-Bergmann}, {Strasburger},
  {Treu}, {Vestergaard}, {Wang}, {Ward}, {Waters}, {Williams}, {Yang}, {Yao},
  {Zastrocky}, {Zhai}, \& {Zu}}]{2021ApJ...922..151K}
{Kara}, E., {Mehdipour}, M., {Kriss}, G.~A., {et~al.} 2021, \apj, 922, 151

\bibitem[{{Kishimoto} {et~al.}(2008){Kishimoto}, {Antonucci}, {Blaes},
  {Lawrence}, {Boisson}, {Albrecht}, \& {Leipski}}]{2008Natur.454..492K}
{Kishimoto}, M., {Antonucci}, R., {Blaes}, O., {et~al.} 2008, \nat, 454, 492

\bibitem[{{Kobulnicky} {et~al.}(2003){Kobulnicky}, {Nordsieck}, {Burgh},
  {Smith}, {Percival}, {Williams}, \& {O'Donoghue}}]{kobulnicky03}
{Kobulnicky}, H.~A., {Nordsieck}, K.~H., {Burgh}, E.~B., {et~al.} 2003, in
  Society of Photo-Optical Instrumentation Engineers (SPIE) Conference Series,
  Vol. 4841, Instrument Design and Performance for Optical/Infrared
  Ground-based Telescopes, ed. M.~{Iye} \& A.~F.~M. {Moorwood}, 1634--1644

\bibitem[{{Kollatschny}(2003{\natexlab{a}})}]{2003A&A...407..461K}
{Kollatschny}, W. 2003{\natexlab{a}}, \aap, 407, 461

\bibitem[{{Kollatschny}(2003{\natexlab{b}})}]{2003A&A...412L..61K}
{Kollatschny}, W. 2003{\natexlab{b}}, \aap, 412, L61

\bibitem[{{Kollatschny} {et~al.}(2000){Kollatschny}, {Bischoff}, \&
  {Dietrich}}]{2000A&A...361..901K}
{Kollatschny}, W., {Bischoff}, K., \& {Dietrich}, M. 2000, \aap, 361, 901

\bibitem[{{Kollatschny} {et~al.}(2001){Kollatschny}, {Bischoff}, {Robinson},
  {Welsh}, \& {Hill}}]{2001A&A...379..125K}
{Kollatschny}, W., {Bischoff}, K., {Robinson}, E.~L., {Welsh}, W.~F., \&
  {Hill}, G.~J. 2001, \aap, 379, 125

\bibitem[{{Kollatschny} {et~al.}(2022){Kollatschny}, {Ochmann}, {Kaspi},
  {Schumacher}, {Behar}, {Chelouche}, {Horne}, {M{\"u}ller}, {Rafter}, {Chini},
  {Haas}, \& {Probst}}]{2022A&A...657A.122K}
{Kollatschny}, W., {Ochmann}, M.~W., {Kaspi}, S., {et~al.} 2022, \aap, 657,
  A122

\bibitem[{{Kollatschny} {et~al.}(2018){Kollatschny}, {Ochmann}, {Zetzl},
  {Haas}, {Chelouche}, {Kaspi}, {Pozo Nu{\~n}ez}, \&
  {Grupe}}]{2018A&A...619A.168K}
{Kollatschny}, W., {Ochmann}, M.~W., {Zetzl}, M., {et~al.} 2018, \aap, 619,
  A168

\bibitem[{{Kollatschny} {et~al.}(2014){Kollatschny}, {Ulbrich}, {Zetzl},
  {Kaspi}, \& {Haas}}]{2014A&A...566A.106K}
{Kollatschny}, W., {Ulbrich}, K., {Zetzl}, M., {Kaspi}, S., \& {Haas}, M. 2014,
  \aap, 566, A106

\bibitem[{{Kollatschny} \& {Zetzl}(2011)}]{2011Natur.470..366K}
{Kollatschny}, W. \& {Zetzl}, M. 2011, \nat, 470, 366

\bibitem[{{Kollatschny} \& {Zetzl}(2013)}]{2013A&A...549A.100K}
{Kollatschny}, W. \& {Zetzl}, M. 2013, \aap, 549, A100

\bibitem[{{Korista} \& {Goad}(2001)}]{2001ApJ...553..695K}
{Korista}, K.~T. \& {Goad}, M.~R. 2001, \apj, 553, 695

\bibitem[{{Korista} \& {Goad}(2004)}]{2004ApJ...606..749K}
{Korista}, K.~T. \& {Goad}, M.~R. 2004, \apj, 606, 749

\bibitem[{{Kova{\v{c}}evi{\'c}} {et~al.}(2010){Kova{\v{c}}evi{\'c}},
  {Popovi{\'c}}, \& {Dimitrijevi{\'c}}}]{2010ApJS..189...15K}
{Kova{\v{c}}evi{\'c}}, J., {Popovi{\'c}}, L.~{\v{C}}., \& {Dimitrijevi{\'c}},
  M.~S. 2010, \apjs, 189, 15

\bibitem[{{Krolik}(2001)}]{2001ApJ...551...72K}
{Krolik}, J.~H. 2001, \apj, 551, 72

\bibitem[{{Labita} {et~al.}(2006){Labita}, {Treves}, {Falomo}, \&
  {Uslenghi}}]{2006MNRAS.373..551L}
{Labita}, M., {Treves}, A., {Falomo}, R., \& {Uslenghi}, M. 2006, \mnras, 373,
  551

\bibitem[{{Liu} {et~al.}(2022){Liu}, {Gebhardt}, {Cooper}, {Davis},
  {Schneider}, {Ciardullo}, {Farrow}, {Finkelstein}, {Gronwall}, {Guo}, {Hill},
  {House}, {Jeong}, {Jogee}, {Kollatschny}, {Krumpe}, {Landriau}, {Chavez
  Ortiz}, {Zhang}, \& {HETDEX Collaboration}}]{2022ApJS..261...24L}
{Liu}, C., {Gebhardt}, K., {Cooper}, E.~M., {et~al.} 2022, \apjs, 261, 24

\bibitem[{{Makrygianni} {et~al.}(2023){Makrygianni}, {Trakhtenbrot}, {Arcavi},
  {Ricci}, {Lam}, {Horesh}, {Sfaradi}, {Bostroem}, {Hosseinzadeh}, {Howell},
  {Pellegrino}, {Fender}, {Green}, {Williams}, \&
  {Bright}}]{2023ApJ...953...32M}
{Makrygianni}, L., {Trakhtenbrot}, B., {Arcavi}, I., {et~al.} 2023, \apj, 953,
  32

\bibitem[{{McHardy} {et~al.}(2018){McHardy}, {Connolly}, {Horne}, {Cackett},
  {Gelbord}, {Peterson}, {Pahari}, {Gehrels}, {Goad}, {Lira}, {Arevalo},
  {Baldi}, {Brandt}, {Breedt}, {Chand}, {Dewangan}, {Done}, {Elvis},
  {Emmanoulopoulos}, {Fausnaugh}, {Kaspi}, {Kochanek}, {Korista}, {Papadakis},
  {Rao}, {Uttley}, {Vestergaard}, \& {Ward}}]{2018MNRAS.480.2881M}
{McHardy}, I.~M., {Connolly}, S.~D., {Horne}, K., {et~al.} 2018, \mnras, 480,
  2881

\bibitem[{{Mullaney} {et~al.}(2013){Mullaney}, {Alexander}, {Fine}, {Goulding},
  {Harrison}, \& {Hickox}}]{2013MNRAS.433..622M}
{Mullaney}, J.~R., {Alexander}, D.~M., {Fine}, S., {et~al.} 2013, \mnras, 433,
  622

\bibitem[{{Netzer}(2019)}]{2019MNRAS.488.5185N}
{Netzer}, H. 2019, \mnras, 488, 5185

\bibitem[{{Netzer}(2022)}]{2022MNRAS.509.2637N}
{Netzer}, H. 2022, \mnras, 509, 2637

\bibitem[{{Netzer} {et~al.}(1985){Netzer}, {Elitzur}, \&
  {Ferland}}]{1985ApJ...299..752N}
{Netzer}, H., {Elitzur}, M., \& {Ferland}, G.~J. 1985, \apj, 299, 752

\bibitem[{{Ochmann} {et~al.}(2024){Ochmann}, {Kollatschny}, {Probst},
  {Romero-Colmenero}, {Buckley}, {Chelouche}, {Chini}, {Grupe}, {Haas},
  {Kaspi}, {Komossa}, {Parker}, {Santos-Lleo}, {Schartel}, \&
  {Famula}}]{2024A&A...686A..17O}
{Ochmann}, M.~W., {Kollatschny}, W., {Probst}, M.~A., {et~al.} 2024, \aap, 686,
  A17

\bibitem[{{Oh} {et~al.}(2022){Oh}, {Koss}, {Ueda}, {Stern}, {Ricci},
  {Trakhtenbrot}, {Powell}, {den Brok}, {Lamperti}, {Mushotzky}, {Ricci},
  {B{\"a}r}, {Rojas}, {Ichikawa}, {Riffel}, {Treister}, {Harrison}, {Urry},
  {Bauer}, \& {Schawinski}}]{2022ApJS..261....4O}
{Oh}, K., {Koss}, M.~J., {Ueda}, Y., {et~al.} 2022, \apjs, 261, 4

\bibitem[{{Onken} {et~al.}(2004){Onken}, {Ferrarese}, {Merritt}, {Peterson},
  {Pogge}, {Vestergaard}, \& {Wandel}}]{2004ApJ...615..645O}
{Onken}, C.~A., {Ferrarese}, L., {Merritt}, D., {et~al.} 2004, \apj, 615, 645

\bibitem[{{Osterbrock} \& {Cohen}(1982)}]{1982ApJ...261...64O}
{Osterbrock}, D.~E. \& {Cohen}, R.~D. 1982, \apj, 261, 64

\bibitem[{{Osterbrock} \& {Pogge}(1985)}]{1985ApJ...297..166O}
{Osterbrock}, D.~E. \& {Pogge}, R.~W. 1985, \apj, 297, 166

\bibitem[{{Park} {et~al.}(2022){Park}, {Barth}, {Ho}, \&
  {Laor}}]{2022ApJS..258...38P}
{Park}, D., {Barth}, A.~J., {Ho}, L.~C., \& {Laor}, A. 2022, \apjs, 258, 38

\bibitem[{{Park} {et~al.}(2017){Park}, {Woo}, {Romero-Colmenero}, {Crawford},
  {Park}, {Cho}, {Jeon}, {Choi}, {Barth}, {Pei}, {Hickox}, {Sung}, \&
  {Im}}]{2017ApJ...847..125P}
{Park}, S., {Woo}, J.-H., {Romero-Colmenero}, E., {et~al.} 2017, \apj, 847, 125

\bibitem[{{Pei} {et~al.}(2017){Pei}, {Fausnaugh}, {Barth}, {Peterson}, {Bentz},
  {De Rosa}, {Denney}, {Goad}, {Kochanek}, {Korista}, {Kriss}, {Pogge},
  {Bennert}, {Brotherton}, {Clubb}, {Dalla Bont{\`a}}, {Filippenko}, {Greene},
  {Grier}, {Vestergaard}, {Zheng}, {Adams}, {Beatty}, {Bigley}, {Brown},
  {Brown}, {Canalizo}, {Comerford}, {Coker}, {Corsini}, {Croft}, {Croxall},
  {Deason}, {Eracleous}, {Fox}, {Gates}, {Henderson}, {Holmbeck}, {Holoien},
  {Jensen}, {Johnson}, {Kelly}, {Kim}, {King}, {Lau}, {Li}, {Lochhaas}, {Ma},
  {Manne-Nicholas}, {Mauerhan}, {Malkan}, {McGurk}, {Morelli}, {Mosquera},
  {Mudd}, {Muller Sanchez}, {Nguyen}, {Ochner}, {Ou-Yang}, {Pancoast}, {Penny},
  {Pizzella}, {Poleski}, {Runnoe}, {Scott}, {Schimoia}, {Shappee}, {Shivvers},
  {Simonian}, {Siviero}, {Somers}, {Stevens}, {Strauss}, {Tayar}, {Tejos},
  {Treu}, {Van Saders}, {Vican}, {Villanueva}, {Yuk}, {Zakamska}, {Zhu},
  {Anderson}, {Ar{\'e}valo}, {Bazhaw}, {Bisogni}, {Borman}, {Bottorff},
  {Brandt}, {Breeveld}, {Cackett}, {Carini}, {Crenshaw}, {De
  Lorenzo-C{\'a}ceres}, {Dietrich}, {Edelson}, {Efimova}, {Ely}, {Evans},
  {Ferland}, {Flatland}, {Gehrels}, {Geier}, {Gelbord}, {Grupe}, {Gupta},
  {Hall}, {Hicks}, {Horenstein}, {Horne}, {Hutchison}, {Im}, {Joner}, {Jones},
  {Kaastra}, {Kaspi}, {Kelly}, {Kennea}, {Kim}, {Kim}, {Klimanov}, {Lee},
  {Leonard}, {Lira}, {MacInnis}, {Mathur}, {McHardy}, {Montouri}, {Musso},
  {Nazarov}, {Netzer}, {Norris}, {Nousek}, {Okhmat}, {Papadakis}, {Parks},
  {Pott}, {Rafter}, {Rix}, {Saylor}, {Schn{\"u}lle}, {Sergeev}, {Siegel},
  {Skielboe}, {Spencer}, {Starkey}, {Sung}, {Teems}, {Turner}, {Uttley},
  {Villforth}, {Weiss}, {Woo}, {Yan}, {Young}, \& {Zu}}]{2017ApJ...837..131P}
{Pei}, L., {Fausnaugh}, M.~M., {Barth}, A.~J., {et~al.} 2017, \apj, 837, 131

\bibitem[{{Peterson} {et~al.}(2002){Peterson}, {Berlind}, {Bertram},
  {Bischoff}, {Bochkarev}, {Borisov}, {Burenkov}, {Calkins}, {Carrasco},
  {Chavushyan}, {Chornock}, {Dietrich}, {Doroshenko}, {Ezhkova}, {Filippenko},
  {Gilbert}, {Huchra}, {Kollatschny}, {Leonard}, {Li}, {Lyuty}, {Malkov},
  {Matheson}, {Merkulova}, {Mikhailov}, {Modjaz}, {Onken}, {Pogge}, {Pronik},
  {Qian}, {Romano}, {Sergeev}, {Sergeeva}, {Shapovalova}, {Spiridonova}, {Tao},
  {Tokarz}, {Valdes}, {Vlasiuk}, {Wagner}, \& {Wilkes}}]{2002ApJ...581..197P}
{Peterson}, B.~M., {Berlind}, P., {Bertram}, R., {et~al.} 2002, \apj, 581, 197

\bibitem[{{Peterson} {et~al.}(2004){Peterson}, {Ferrarese}, {Gilbert}, {Kaspi},
  {Malkan}, {Maoz}, {Merritt}, {Netzer}, {Onken}, {Pogge}, {Vestergaard}, \&
  {Wandel}}]{2004ApJ...613..682P}
{Peterson}, B.~M., {Ferrarese}, L., {Gilbert}, K.~M., {et~al.} 2004, \apj, 613,
  682

\bibitem[{{Peterson} {et~al.}(1998){Peterson}, {Wanders}, {Bertram}, {Hunley},
  {Pogge}, \& {Wagner}}]{1998ApJ...501...82P}
{Peterson}, B.~M., {Wanders}, I., {Bertram}, R., {et~al.} 1998, \apj, 501, 82

\bibitem[{{Planck Collaboration} {et~al.}(2020){Planck Collaboration},
  {Aghanim}, {Akrami}, {Ashdown}, {Aumont}, {Baccigalupi}, {Ballardini},
  {Banday}, {Barreiro}, {Bartolo}, {Basak}, {Battye}, {Benabed}, {Bernard},
  {Bersanelli}, {Bielewicz}, {Bock}, {Bond}, {Borrill}, {Bouchet}, {Boulanger},
  {Bucher}, {Burigana}, {Butler}, {Calabrese}, {Cardoso}, {Carron},
  {Challinor}, {Chiang}, {Chluba}, {Colombo}, {Combet}, {Contreras}, {Crill},
  {Cuttaia}, {de Bernardis}, {de Zotti}, {Delabrouille}, {Delouis}, {Di
  Valentino}, {Diego}, {Dor{\'e}}, {Douspis}, {Ducout}, {Dupac}, {Dusini},
  {Efstathiou}, {Elsner}, {En{\ss}lin}, {Eriksen}, {Fantaye}, {Farhang},
  {Fergusson}, {Fernandez-Cobos}, {Finelli}, {Forastieri}, {Frailis},
  {Fraisse}, {Franceschi}, {Frolov}, {Galeotta}, {Galli}, {Ganga},
  {G{\'e}nova-Santos}, {Gerbino}, {Ghosh}, {Gonz{\'a}lez-Nuevo}, {G{\'o}rski},
  {Gratton}, {Gruppuso}, {Gudmundsson}, {Hamann}, {Handley}, {Hansen},
  {Herranz}, {Hildebrandt}, {Hivon}, {Huang}, {Jaffe}, {Jones}, {Karakci},
  {Keih{\"a}nen}, {Keskitalo}, {Kiiveri}, {Kim}, {Kisner}, {Knox},
  {Krachmalnicoff}, {Kunz}, {Kurki-Suonio}, {Lagache}, {Lamarre}, {Lasenby},
  {Lattanzi}, {Lawrence}, {Le Jeune}, {Lemos}, {Lesgourgues}, {Levrier},
  {Lewis}, {Liguori}, {Lilje}, {Lilley}, {Lindholm}, {L{\'o}pez-Caniego},
  {Lubin}, {Ma}, {Mac{\'\i}as-P{\'e}rez}, {Maggio}, {Maino}, {Mandolesi},
  {Mangilli}, {Marcos-Caballero}, {Maris}, {Martin}, {Martinelli},
  {Mart{\'\i}nez-Gonz{\'a}lez}, {Matarrese}, {Mauri}, {McEwen}, {Meinhold},
  {Melchiorri}, {Mennella}, {Migliaccio}, {Millea}, {Mitra},
  {Miville-Desch{\^e}nes}, {Molinari}, {Montier}, {Morgante}, {Moss}, {Natoli},
  {N{\o}rgaard-Nielsen}, {Pagano}, {Paoletti}, {Partridge}, {Patanchon},
  {Peiris}, {Perrotta}, {Pettorino}, {Piacentini}, {Polastri}, {Polenta},
  {Puget}, {Rachen}, {Reinecke}, {Remazeilles}, {Renzi}, {Rocha}, {Rosset},
  {Roudier}, {Rubi{\~n}o-Mart{\'\i}n}, {Ruiz-Granados}, {Salvati}, {Sandri},
  {Savelainen}, {Scott}, {Shellard}, {Sirignano}, {Sirri}, {Spencer},
  {Sunyaev}, {Suur-Uski}, {Tauber}, {Tavagnacco}, {Tenti}, {Toffolatti},
  {Tomasi}, {Trombetti}, {Valenziano}, {Valiviita}, {Van Tent}, {Vibert},
  {Vielva}, {Villa}, {Vittorio}, {Wandelt}, {Wehus}, {White}, {White},
  {Zacchei}, \& {Zonca}}]{2020A&A...641A...6P}
{Planck Collaboration}, {Aghanim}, N., {Akrami}, Y., {et~al.} 2020, \aap, 641,
  A6

\bibitem[{{Rafter} {et~al.}(2013){Rafter}, {Kaspi}, {Chelouche}, {Sabach},
  {Karl}, \& {Behar}}]{2013ApJ...773...24R}
{Rafter}, S.~E., {Kaspi}, S., {Chelouche}, D., {et~al.} 2013, \apj, 773, 24

\bibitem[{{Rodr{\'\i}guez-Pascual} {et~al.}(1997){Rodr{\'\i}guez-Pascual},
  {Alloin}, {Clavel}, {Crenshaw}, {Horne}, {Kriss}, {Krolik}, {Malkan},
  {Netzer}, {O'Brien}, {Peterson}, {Reichert}, {Wamsteker}, {Alexander},
  {Barr}, {Blandford}, {Bregman}, {Carone}, {Clements}, {Courvoisier}, {De
  Robertis}, {Dietrich}, {Dottori}, {Edelson}, {Filippenko}, {Gaskell},
  {Huchra}, {Hutchings}, {Kollatschny}, {Koratkar}, {Korista}, {Laor},
  {MacAlpine}, {Martin}, {Maoz}, {McCollum}, {Morris}, {Perola}, {Pogge},
  {Ptak}, {Recondo-Gonz{\'a}lez}, {Rodr{\'\i}guez-Espinoza}, {Rokaki},
  {Santos-Lle{\'o}}, {Sekiguchi}, {Shull}, {Snijders}, {Sparke}, {Stirpe},
  {Stoner}, {Sun}, {Wagner}, {Wanders}, {Wilkes}, {Winge}, \&
  {Zheng}}]{1997ApJS..110....9R}
{Rodr{\'\i}guez-Pascual}, P.~M., {Alloin}, D., {Clavel}, J., {et~al.} 1997,
  \apjs, 110, 9

\bibitem[{{Sakata} {et~al.}(2010){Sakata}, {Minezaki}, {Yoshii}, {Kobayashi},
  {Koshida}, {Aoki}, {Enya}, {Tomita}, {Suganuma}, {Katsuno Uchimoto}, \&
  {Sugawara}}]{2010ApJ...711..461S}
{Sakata}, Y., {Minezaki}, T., {Yoshii}, Y., {et~al.} 2010, \apj, 711, 461

\bibitem[{{Schlafly} \& {Finkbeiner}(2011)}]{schlafly11}
{Schlafly}, E.~F. \& {Finkbeiner}, D.~P. 2011, \apj, 737, 103

\bibitem[{{Shakura} \& {Sunyaev}(1973)}]{1973A&A....24..337S}
{Shakura}, N.~I. \& {Sunyaev}, R.~A. 1973, \aap, 24, 337

\bibitem[{{Shapovalova} {et~al.}(2010){Shapovalova}, {Popovi{\'c}}, {Burenkov},
  {Chavushyan}, {Ili{\'c}}, {Kollatschny}, {Kova{\v{c}}evi{\'c}}, {Bochkarev},
  {Carrasco}, {Le{\'o}n-Tavares}, {Mercado}, {Valdes}, {Vlasuyk}, \& {de La
  Fuente}}]{2010A&A...517A..42S}
{Shapovalova}, A.~I., {Popovi{\'c}}, L.~{\v{C}}., {Burenkov}, A.~N., {et~al.}
  2010, \aap, 517, A42

\bibitem[{{Shapovalova} {et~al.}(2012){Shapovalova}, {Popovi{\'c}}, {Burenkov},
  {Chavushyan}, {Ili{\'c}}, {Kova{\v{c}}evi{\'c}}, {Kollatschny},
  {Kova{\v{c}}evi{\'c}}, {Bochkarev}, {Valdes}, {Torrealba},
  {Le{\'o}n-Tavares}, {Mercado}, {Ben{\'\i}tez}, {Carrasco}, {Dultzin}, \& {de
  la Fuente}}]{2012ApJS..202...10S}
{Shapovalova}, A.~I., {Popovi{\'c}}, L.~{\v{C}}., {Burenkov}, A.~N., {et~al.}
  2012, \apjs, 202, 10

\bibitem[{{Shen} {et~al.}(2024){Shen}, {Grier}, {Horne}, {Stone}, {Li}, {Yang},
  {Homayouni}, {Trump}, {Anderson}, {Brandt}, {Hall}, {Ho}, {Jiang},
  {Petitjean}, {Schneider}, {Tao}, {Donnan}, {AlSayyad}, {Bershady}, {Blanton},
  {Bizyaev}, {Bundy}, {Chen}, {Davis}, {Dawson}, {Fan}, {Greene},
  {Gr{\"o}ller}, {Guo}, {Ibarra-Medel}, {Jiang}, {Keenan}, {Kollmeier},
  {Lejoly}, {Li}, {de la Macorra}, {Moe}, {Nie}, {Rossi}, {Smith}, {Tee},
  {Weijmans}, {Xu}, {Yue}, {Zhou}, {Zhou}, \& {Zou}}]{2024ApJS..272...26S}
{Shen}, Y., {Grier}, C.~J., {Horne}, K., {et~al.} 2024, \apjs, 272, 26

\bibitem[{{Sobrino Figaredo} {et~al.}(2025){Sobrino Figaredo}, {Chelouche},
  {Haas}, {Ramolla}, {Kaspi}, {Panda}, {Ochmann}, {Zucker}, {Chini}, {Probst},
  {Kollatschny}, \& {Murphy}}]{2025ApJS..276...48S}
{Sobrino Figaredo}, C., {Chelouche}, D., {Haas}, M., {et~al.} 2025, \apjs, 276,
  48

\bibitem[{{Sobrino Figaredo} {et~al.}(2018){Sobrino Figaredo}, {Nu{\~n}ez},
  {Ramolla}, {Haas}, {Chini}, {Hodapp}, {Willner}, \&
  {Ashby}}]{2018rnls.confE..57S}
{Sobrino Figaredo}, C., {Nu{\~n}ez}, F.~P., {Ramolla}, M., {et~al.} 2018, in
  Revisiting Narrow-Line Seyfert 1 Galaxies and their Place in the Universe, 57

\bibitem[{{Trakhtenbrot} {et~al.}(2019){Trakhtenbrot}, {Arcavi}, {Ricci},
  {Tacchella}, {Stern}, {Netzer}, {Jonker}, {Horesh}, {Mej{\'\i}a-Restrepo},
  {Hosseinzadeh}, {Hallefors}, {Howell}, {McCully}, {Balokovi{\'c}}, {Heida},
  {Kamraj}, {Lansbury}, {Wyrzykowski}, {Gromadzki}, {Hamanowicz}, {Cenko},
  {Sand}, {Hsiao}, {Phillips}, {Diamond}, {Kara}, {Gendreau}, {Arzoumanian}, \&
  {Remillard}}]{2019NatAs...3..242T}
{Trakhtenbrot}, B., {Arcavi}, I., {Ricci}, C., {et~al.} 2019, Nature Astronomy,
  3, 242

\bibitem[{{V{\'e}ron-Cetty} {et~al.}(2001){V{\'e}ron-Cetty}, {V{\'e}ron}, \&
  {Gon{\c{c}}alves}}]{2001A&A...372..730V}
{V{\'e}ron-Cetty}, M.~P., {V{\'e}ron}, P., \& {Gon{\c{c}}alves}, A.~C. 2001,
  \aap, 372, 730

\bibitem[{{Vestergaard} \& {Peterson}(2006)}]{2006ApJ...641..689V}
{Vestergaard}, M. \& {Peterson}, B.~M. 2006, \apj, 641, 689

\bibitem[{{Vincentelli} {et~al.}(2021){Vincentelli}, {McHardy}, {Cackett},
  {Barth}, {Horne}, {Goad}, {Korista}, {Gelbord}, {Brandt}, {Edelson},
  {Miller}, {Pahari}, {Peterson}, {Schmidt}, {Baldi}, {Breedt}, {Hern{\'a}ndez
  Santisteban}, {Romero-Colmenero}, {Ward}, \&
  {Williams}}]{2021MNRAS.504.4337V}
{Vincentelli}, F.~M., {McHardy}, I., {Cackett}, E.~M., {et~al.} 2021, \mnras,
  504, 4337

\bibitem[{{Winkler}(1997)}]{1997MNRAS.292..273W}
{Winkler}, H. 1997, \mnras, 292, 273

\bibitem[{{Winkler} {et~al.}(1992){Winkler}, {Glass}, {van Wyk}, {Marang},
  {Jones}, {Buckley}, \& {Sekiguchi}}]{1992MNRAS.257..659W}
{Winkler}, H., {Glass}, I.~S., {van Wyk}, F., {et~al.} 1992, \mnras, 257, 659

\bibitem[{{Wright}(2006)}]{2006PASP..118.1711W}
{Wright}, E.~L. 2006, \pasp, 118, 1711

\bibitem[{{Xu} {et~al.}(2012){Xu}, {Komossa}, {Zhou}, {Lu}, {Li}, {Grupe},
  {Wang}, \& {Yuan}}]{2012AJ....143...83X}
{Xu}, D., {Komossa}, S., {Zhou}, H., {et~al.} 2012, \aj, 143, 83

\bibitem[{{Zhou} {et~al.}(2025){Zhou}, {Sun}, {Feng}, {Li}, {Xue}, {Wang},
  {Cai}, {Bai}, {Li}, {Guo}, {Liu}, {Lu}, {Mao}, {Marculewicz}, \&
  {Wang}}]{2025ApJ...986..137Z}
{Zhou}, S., {Sun}, M., {Feng}, H.-C., {et~al.} 2025, \apj, 986, 137

\bibitem[{{Zucker} \& {Mazeh}(1994)}]{1994ApJ...420..806Z}
{Zucker}, S. \& {Mazeh}, T. 1994, \apj, 420, 806

\end{thebibliography}

\begin{appendix}
\section{Additional figures}
%
%------------------------------------------------------------------------------
%
\begin{figure}[!h]
\centering
  \begin{minipage}{0.99\textwidth}
    \begin{subfigure}[t]{0.49\textwidth}
      \includegraphics[width=\textwidth]{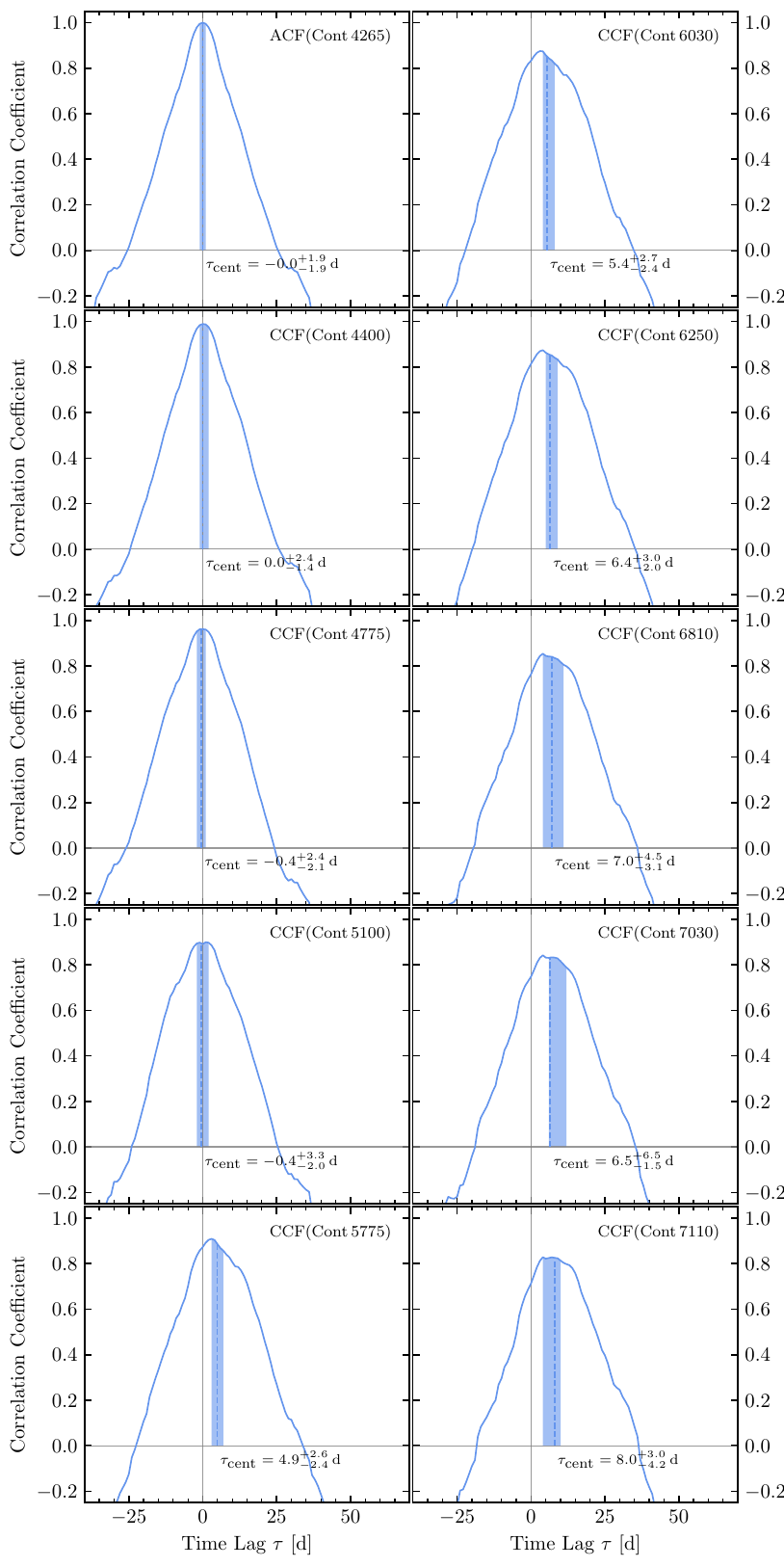}
      \caption{}
      \label{fig:correlation_balmer}
    \end{subfigure}\hfill
    \begin{subfigure}[t]{0.49\textwidth}
      \includegraphics[width=\textwidth]{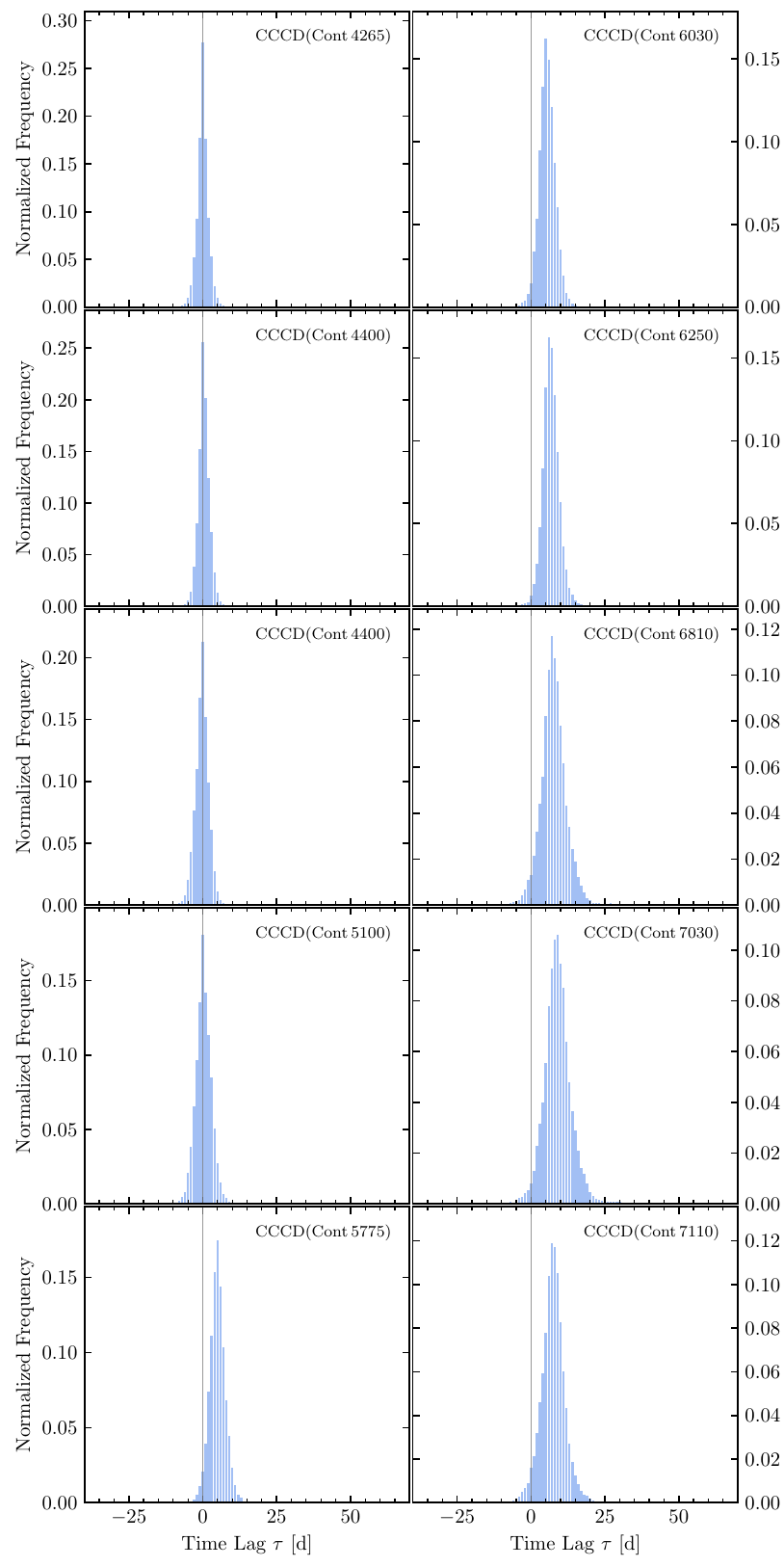}
      \caption{}
      \label{fig:distribution_balmer}
    \end{subfigure}
  \caption{\textit{Left panel}: CCFs of the interband continuum correlations (a) with respect to the continuum light curve at $4265\,${\AA}. The time lag, $\tau_{\text{cent}}$, is denoted by a dashed line, with the shaded area corresponding to a $\pm 1 \sigma$ interval. \textit{Right panel}: Cross-correlation centroid distributions after 20,000 runs employing FR and RSS of the various continua with respect to the combined continuum at $4265\,${\AA} (rest frame).} 
  \label{fig:ContCorrelations}
  \end{minipage}
\end{figure}
%
%------------------------------------------------------------------------------
%

\newpage
\clearpage
\newpage

\section{Additional tables}

\begin{sidewaystable}[!t]
\centering
\tabcolsep1.mm
\vspace{-9cm}
\caption{Continuum Flux Densities.} 
\begin{tabular}{ccccccccccc} 
\hline 
\hline
\noalign{\smallskip}
MJD  & Cont.$\,4265$ & Cont.$\,4440$ & Cont.$\,4775$ & Cont.$\,5100$ & Cont.$\,5775$ & Cont.$\,6030$ & Cont.$\,6250$ & Cont.$\,6810$ & Cont.$\,7030$ & Cont.$\,7110$ \\ 
 (1) & (2) & (3) & (4) & (5) & (6) & (7) & (8) & (9) & (10) & (11) \\ 
\hline 
\noalign{\smallskip}
$56628.03$ & $3.53 \pm 0.04$ & $3.52 \pm 0.05$ & $3.33 \pm 0.04$ & $2.96 \pm 0.03$ & $2.66 \pm 0.03$ & $2.52 \pm 0.03$ & $2.35 \pm 0.02$ & $2.18 \pm 0.02$ & $2.20 \pm 0.03$ & $1.97 \pm 0.03$ \\ 
$56640.01$ & $3.96 \pm 0.04$ & $3.74 \pm 0.04$ & $3.49 \pm 0.04$ & $2.93 \pm 0.03$ & $2.81 \pm 0.03$ & $2.61 \pm 0.03$ & $2.46 \pm 0.02$ & $2.13 \pm 0.02$ & $2.10 \pm 0.02$ & $1.94 \pm 0.02$ \\ 
$56647.00$ & $3.96 \pm 0.04$ & $3.79 \pm 0.04$ & $3.48 \pm 0.04$ & $2.97 \pm 0.03$ & $2.76 \pm 0.03$ & $2.58 \pm 0.03$ & $2.49 \pm 0.03$ & $2.21 \pm 0.02$ & $2.20 \pm 0.02$ & $2.01 \pm 0.02$ \\ 
$56652.97$ & $3.76 \pm 0.05$ & $3.72 \pm 0.04$ & $3.64 \pm 0.04$ & $3.19 \pm 0.03$ & $2.80 \pm 0.03$ & $2.66 \pm 0.03$ & $2.56 \pm 0.03$ & $2.28 \pm 0.02$ & $2.29 \pm 0.03$ & $2.16 \pm 0.03$ \\ 
$56658.95$ & $3.70 \pm 0.04$ & $3.60 \pm 0.04$ & $3.40 \pm 0.03$ & $2.93 \pm 0.03$ & $2.72 \pm 0.03$ & $2.54 \pm 0.03$ & $2.47 \pm 0.02$ & $2.20 \pm 0.02$ & $2.21 \pm 0.03$ & $2.00 \pm 0.02$ \\ 
$56669.92$ & $3.37 \pm 0.04$ & $3.23 \pm 0.04$ & $3.05 \pm 0.03$ & $2.60 \pm 0.03$ & $2.56 \pm 0.03$ & $2.42 \pm 0.02$ & $2.37 \pm 0.02$ & $2.19 \pm 0.02$ & $2.25 \pm 0.03$ & $2.02 \pm 0.03$ \\ 
$56675.90$ & $2.90 \pm 0.03$ & $2.86 \pm 0.03$ & $2.69 \pm 0.03$ & $2.36 \pm 0.02$ & $2.18 \pm 0.02$ & $2.08 \pm 0.02$ & $2.05 \pm 0.02$ & $1.89 \pm 0.02$ & $1.95 \pm 0.02$ & $1.79 \pm 0.02$ \\ 
$56688.87$ & $2.78 \pm 0.03$ & $2.80 \pm 0.03$ & $2.77 \pm 0.03$ & $2.45 \pm 0.03$ & $2.21 \pm 0.02$ & $2.14 \pm 0.02$ & $2.05 \pm 0.02$ & $1.86 \pm 0.02$ & $1.86 \pm 0.02$ & $1.69 \pm 0.02$ \\ 
$56696.08$ & $2.88 \pm 0.03$ & $2.88 \pm 0.04$ & $2.76 \pm 0.03$ & $2.41 \pm 0.02$ & $2.16 \pm 0.02$ & $2.03 \pm 0.02$ & $1.99 \pm 0.02$ & $1.82 \pm 0.02$ & $1.86 \pm 0.02$ & $1.70 \pm 0.02$ \\ 
$56703.83$ & $3.45 \pm 0.05$ & $3.22 \pm 0.04$ & $3.15 \pm 0.04$ & $2.70 \pm 0.03$ & $2.24 \pm 0.02$ & $2.13 \pm 0.02$ & $2.03 \pm 0.02$ & $1.86 \pm 0.02$ & $1.85 \pm 0.02$ & $1.71 \pm 0.02$ \\ 
$56716.78$ & $3.51 \pm 0.04$ & $3.43 \pm 0.04$ & $3.28 \pm 0.04$ & $2.72 \pm 0.03$ & $2.47 \pm 0.03$ & $2.29 \pm 0.02$ & $2.25 \pm 0.02$ & $2.03 \pm 0.02$ & $2.06 \pm 0.02$ & $1.87 \pm 0.02$ \\ 
$56722.02$ & $3.38 \pm 0.03$ & $3.30 \pm 0.04$ & $3.19 \pm 0.03$ & $2.72 \pm 0.03$ & $2.42 \pm 0.02$ & $2.27 \pm 0.02$ & $2.20 \pm 0.02$ & $1.94 \pm 0.02$ & $1.94 \pm 0.02$ & $1.77 \pm 0.02$ \\ 
$56727.77$ & $3.34 \pm 0.04$ & $3.22 \pm 0.04$ & $3.11 \pm 0.04$ & $2.69 \pm 0.03$ & $2.40 \pm 0.03$ & $2.29 \pm 0.02$ & $2.23 \pm 0.02$ & $2.05 \pm 0.02$ & $2.08 \pm 0.02$ & $1.93 \pm 0.02$ \\ 
$56734.00$ & $3.83 \pm 0.05$ & $3.62 \pm 0.04$ & $3.41 \pm 0.04$ & $2.82 \pm 0.03$ & $2.45 \pm 0.03$ & $2.29 \pm 0.02$ & $2.21 \pm 0.02$ & $1.99 \pm 0.02$ & $2.02 \pm 0.02$ & $1.85 \pm 0.02$ \\ 
$56744.97$ & $3.95 \pm 0.04$ & $3.77 \pm 0.04$ & $3.39 \pm 0.04$ & $2.76 \pm 0.03$ & $2.53 \pm 0.03$ & $2.32 \pm 0.02$ & $2.24 \pm 0.02$ & $2.03 \pm 0.02$ & $2.05 \pm 0.02$ & $1.91 \pm 0.02$ \\ 
$56756.92$ & $3.68 \pm 0.04$ & $3.57 \pm 0.04$ & $3.38 \pm 0.03$ & $2.89 \pm 0.03$ & $2.64 \pm 0.03$ & $2.44 \pm 0.02$ & $2.32 \pm 0.02$ & $2.08 \pm 0.02$ & $2.05 \pm 0.02$ & $1.87 \pm 0.02$ \\ 
$56772.89$ & $3.64 \pm 0.04$ & $3.54 \pm 0.04$ & $3.31 \pm 0.03$ & $2.80 \pm 0.03$ & $2.59 \pm 0.03$ & $2.40 \pm 0.02$ & $2.35 \pm 0.02$ & $2.05 \pm 0.02$ & $2.06 \pm 0.02$ & $1.88 \pm 0.02$ \\ 
$56788.84$ & $3.41 \pm 0.04$ & $3.31 \pm 0.03$ & $3.07 \pm 0.03$ & $2.58 \pm 0.03$ & $2.39 \pm 0.02$ & $2.24 \pm 0.02$ & $2.16 \pm 0.02$ & $1.95 \pm 0.02$ & $2.00 \pm 0.02$ & $1.82 \pm 0.02$ \\ 
$56794.81$ & $3.22 \pm 0.04$ & $3.12 \pm 0.03$ & $2.98 \pm 0.03$ & $2.59 \pm 0.03$ & $2.35 \pm 0.02$ & $2.18 \pm 0.02$ & $2.14 \pm 0.02$ & $1.95 \pm 0.02$ & $1.98 \pm 0.02$ & $1.78 \pm 0.02$ \\ 
$56800.80$ & $2.99 \pm 0.03$ & $2.96 \pm 0.03$ & $2.84 \pm 0.03$ & $2.49 \pm 0.03$ & $2.16 \pm 0.02$ & $2.07 \pm 0.02$ & $2.02 \pm 0.02$ & $1.83 \pm 0.02$ & $1.85 \pm 0.02$ & $1.69 \pm 0.02$ \\ 
$56808.80$ & $3.52 \pm 0.04$ & $3.47 \pm 0.04$ & $3.19 \pm 0.04$ & $2.73 \pm 0.03$ & $2.37 \pm 0.02$ & $2.19 \pm 0.02$ & $2.14 \pm 0.02$ & $1.91 \pm 0.02$ & $1.94 \pm 0.02$ & $1.74 \pm 0.02$ \\ 
$56827.72$ & $3.08 \pm 0.03$ & $3.04 \pm 0.03$ & $2.86 \pm 0.03$ & $2.46 \pm 0.03$ & $2.24 \pm 0.02$ & $2.12 \pm 0.02$ & $2.09 \pm 0.02$ & $1.89 \pm 0.02$ & $1.91 \pm 0.02$ & $1.74 \pm 0.02$ \\ 
$56832.73$ & $2.85 \pm 0.03$ & $2.82 \pm 0.03$ & $2.65 \pm 0.03$ & $2.33 \pm 0.02$ & $2.10 \pm 0.02$ & $1.97 \pm 0.02$ & $1.93 \pm 0.02$ & $1.76 \pm 0.02$ & $1.79 \pm 0.02$ & $1.62 \pm 0.02$ \\ 
$56837.72$ & $2.96 \pm 0.03$ & $2.92 \pm 0.03$ & $2.72 \pm 0.03$ & $2.33 \pm 0.02$ & $2.16 \pm 0.02$ & $2.02 \pm 0.02$ & $1.98 \pm 0.02$ & $1.76 \pm 0.02$ & $1.78 \pm 0.02$ & $1.63 \pm 0.02$ \\ 
 
\hline 
\label{tab:contLightcurves} 
\end{tabular} 
\tablefoot{Continuum flux densities in units of $10^{-15}\,$erg$\,$cm$^{-2}$s$^{-1}\,$\AA$^{-1}$.} 
\end{sidewaystable}

\begin{table*}[!h]
\centering
\tabcolsep1.mm
\caption{Line Fluxes.} 
\begin{tabular}{ccccccc} 
\hline 
\hline 
\noalign{\smallskip}
MJD  & H$\alpha$ & H$\beta$ & H$\gamma$ & $\textup{He\,\textsc{\lowercase{i}}}$$\,\lambda5876$ & $\textup{He\,\textsc{\lowercase{ii}}}$$\,\lambda4686$ \\ 
 (1) & (2) & (3) & (4) & (5) & (6) \\ 
\hline 
\noalign{\smallskip}
$56628.03$ & $1000. \pm 10.$ & $293.0 \pm 3.0$ & $150.9 \pm 1.7$ & $54.1 \pm 0.8$ &  ${\color{white}\_}93.3 \pm 1.5$ \\ 
$56640.01$ & $1151. \pm 12.$ & $318.9 \pm 3.2$ & $175.1 \pm 1.9$ & $63.5 \pm 0.7$ & $105.5 \pm 1.3$ \\ 
$56647.00$ & $1118. \pm 12.$ & $308.1 \pm 3.1$ & $163.5 \pm 1.7$ & $60.2 \pm 0.7$ & $105.6 \pm 1.2$ \\ 
$56652.97$ & $1016. \pm 11.$ & $295.5 \pm 3.1$ & $144.0 \pm 1.7$ & $54.9 \pm 1.0$ &  ${\color{white}\_}98.2 \pm 1.3$ \\ 
$56658.95$ & $1107. \pm 12.$ & $315.0 \pm 3.2$ & $167.1 \pm 1.8$ & $61.1 \pm 0.7$ & $106.0 \pm 1.2$ \\ 
$56669.92$ & $1130. \pm 12.$ & $313.1 \pm 3.2$ & $164.2 \pm 1.8$ & $63.0 \pm 0.8$ &  ${\color{white}\_}93.4 \pm 1.1$ \\ 
$56675.90$ & $1110. \pm 12.$ & $307.4 \pm 3.1$ & $156.1 \pm 1.7$ & $57.5 \pm 0.6$ &  ${\color{white}\_}71.9 \pm 0.9$ \\ 
$56688.87$ &  ${\color{white}\_}995. \pm 10.$ & $267.9 \pm 2.8$ & $135.3 \pm 1.5$ & $48.5 \pm 0.6$ &  ${\color{white}\_}67.7 \pm 1.0$ \\ 
$56696.08$ & $1018. \pm 11.$ & $280.8 \pm 2.9$ & $139.7 \pm 1.5$ & $47.4 \pm 0.6$ &  ${\color{white}\_}62.5 \pm 0.8$ \\ 
$56703.83$ &  ${\color{white}\_}974. \pm 10.$ & $267.9 \pm 2.8$ & $131.3 \pm 1.6$ & $45.6 \pm 0.6$ &  ${\color{white}\_}87.3 \pm 1.2$ \\ 
$56716.78$ & $1013. \pm 11.$ & $273.8 \pm 2.8$ & $146.7 \pm 1.6$ & $49.4 \pm 0.6$ &  ${\color{white}\_}85.9 \pm 1.1$ \\ 
$56722.02$ & $1073. \pm 11.$ & $289.0 \pm 3.0$ & $150.9 \pm 1.6$ & $56.6 \pm 0.7$ &  ${\color{white}\_}90.2 \pm 1.1$ \\ 
$56727.77$ & $1076. \pm 11.$ & $284.1 \pm 2.9$ & $148.9 \pm 1.7$ & $54.3 \pm 0.8$ &  ${\color{white}\_}79.2 \pm 1.2$ \\ 
$56734.00$ & $1013. \pm 11.$ & $277.0 \pm 2.9$ & $145.7 \pm 1.7$ & $47.9 \pm 0.6$ &  ${\color{white}\_}89.7 \pm 1.2$ \\ 
$56744.97$ & $1083. \pm 11.$ & $301.4 \pm 3.1$ & $171.6 \pm 1.8$ & $55.0 \pm 0.6$ &  ${\color{white}\_}94.7 \pm 1.1$ \\ 
$56756.92$ & $1094. \pm 11.$ & $298.4 \pm 3.0$ & $152.3 \pm 1.6$ & $57.7 \pm 0.7$ &  ${\color{white}\_}88.8 \pm 1.0$ \\ 
$56772.89$ & $1116. \pm 12.$ & $308.4 \pm 3.1$ & $163.6 \pm 1.7$ & $59.9 \pm 0.7$ & $100.3 \pm 1.1$ \\ 
$56788.84$ & $1126. \pm 12.$ & $302.5 \pm 3.1$ & $161.8 \pm 1.7$ & $55.5 \pm 0.6$ &  ${\color{white}\_}91.6 \pm 1.0$ \\ 
$56794.81$ & $1063. \pm 11.$ & $291.6 \pm 3.0$ & $145.3 \pm 1.6$ & $50.3 \pm 0.6$ &  ${\color{white}\_}83.3 \pm 1.0$ \\ 
$56800.80$ & $1068. \pm 11.$ & $282.2 \pm 2.9$ & $147.2 \pm 1.6$ & $50.3 \pm 0.6$ &  ${\color{white}\_}68.5 \pm 1.0$ \\ 
$56808.80$ & $1009. \pm 11.$ & $277.6 \pm 2.9$ & $147.8 \pm 1.6$ & $46.9 \pm 0.6$ &  ${\color{white}\_}93.7 \pm 1.2$ \\ 
$56827.72$ & $1086. \pm 11.$ & $289.7 \pm 3.0$ & $151.1 \pm 1.6$ & $50.3 \pm 0.6$ &  ${\color{white}\_}85.6 \pm 1.0$ \\ 
$56832.73$ & $1048. \pm 11.$ & $281.2 \pm 2.9$ & $145.9 \pm 1.5$ & $49.4 \pm 0.6$ &  ${\color{white}\_}77.7 \pm 0.9$ \\ 
$56837.72$ & $1061. \pm 11.$ & $281.0 \pm 2.9$ & $147.7 \pm 1.6$ & $50.0 \pm 0.6$ &  ${\color{white}\_}81.1 \pm 0.9$ \\ 
\hline 
\label{tab:lineLightcurves} 
\end{tabular} 
\tablefoot{Line fluxes in units of $10^{-15}\,$erg$\,$cm$^{-2}$s$^{-1}$.} 
\end{table*}

\end{appendix}

\end{document}